\documentclass[fleqn,usenatbib]{mnras}

\usepackage{newtxtext,newtxmath}

\usepackage[T1]{fontenc}

\DeclareRobustCommand{\VAN}[3]{#2}
\let\VANthebibliography\thebibliography
\def\thebibliography{\DeclareRobustCommand{\VAN}[3]{##3}\VANthebibliography}

\usepackage{graphicx}	
\usepackage{amsmath}	

\title[CMa~OB1 stellar groups contents]{Canis Major OB1 stellar groups contents revealed by {\it Gaia}}

\author[Santos-Silva et al.]{
T. Santos-Silva$^{1}$\thanks{E-mail: thaisfi@gmail.com},
H. D. Perottoni$^{1}$,
F. Almeida-Fernandes$^{1}$,
J. Gregorio-Hetem$^{1}$, 
V. Jatenco-Pereira$^{1}$,
\newauthor
C. Mendes de Oliveira$^{1}$,
T. Montmerle$^{2}$,
E. Bica$^{3}$,
C. Bonatto$^{3}$,
H. Monteiro$^{4}$,
W. S. Dias$^{4}$,
\newauthor
C.E. Barbosa$^{1}$,
B. Fernandes$^{1}$,
P.A.B. Galli$^{5}$,
M. Borges Fernandes$^{6}$,
A.~Kanaan$^{7}$,
T.~Ribeiro$^{8}$,
W.~Schoenell$^{9}$
\\
$^{1}$Universidade de S\~ao Paulo, Instituto de Astronomia, Geof\'isica e Ci\^encias Atmosf\'ericas, Departamento de Astronomia, SP 05508-090, S\~ao Paulo, Brazil\\
$^{2}$Institut d'Astrophysique de Paris, 75014, Paris, France\\
$^{3}$Departamento de Astronomia, Universidade Federal do Rio Grande do Sul, Av. Bento Gonçalves 9500, Porto Alegre 91501-970, RS, Brazil\\
$^{4}$Instituto de F\'isica e Qu\'imica, Universidade Federal de Itajub\'a, Av. BPS 1303 Pinheirinho, Itajub\'a 37500-903, MG, Brazil\\
$^{5}$Laboratoire d’Astrophysique de Bordeaux, Univ. Bordeaux, CNRS, B18N, allée Geoffroy Saint-Hillaire, F-33615 Pessac, France\\
$^{6}$Observat\'{o}rio Nacional, Rua General Jos\'{e} Cristino 77, CEP: 20921-400, S\~{a}o Crist\'{o}v\~{a}o, Rio de Janeiro, Brasil\\
$^{7}$Departamento de F\'isica, Universidade Federal de Santa Catarina, Florian\'{o}polis, SC, 88040-900, Brazil\\
$^{8}$NOAO, P.O. Box 26732, Tucson, AZ 85726\\
$^{9}$GMTO Corporation 465 N. Halstead Street, Suite 250 Pasadena, CA 91107
}

\date{Accepted 2021 August 12. Received 2021 August 12; in original form 2020 December 29}

\pubyear{2020}

\begin{document}
\label{firstpage}
\pagerange{\pageref{firstpage}--\pageref{lastpage}}
\maketitle

\begin{abstract}
Canis~Major~OB1  (CMa\,OB1) is a Galactic stellar association with a very intriguing star-formation scenario. There are more than two dozen known star clusters in its line of sight, but it is not clear which ones are physically associated with CMa\,OB1. We use a clustering code that employs 5-dimensional data from the {\it Gaia} DR2 catalogue to identify physical groups and obtain their astrometric parameters and, in addition, we use two different isochrone-fitting methods to estimate the ages of these groups. We find 15 stellar groups with distances between 570\,pc and 1650\,pc, including 10 previously known and 5 new open cluster candidates. 
Four groups, precisely the youngest ones ($<$ 20 Myr), CMa05, CMa06, CMa07 and CMa08, are confirmed to be part of CMa~OB1. We find that CMa08, a new cluster candidate, may be the progenitor cluster of runaway stars. CMa06 coincides with the well-studied CMa~R1 star-forming region. While CMa06 is still forming stars, due to the remaining material of the molecular cloud associated with the Sh 2-262 nebula, CMa05, CMa07 and CMa08 seem to be in more evolved stages of evolution, with no recent star-forming activity. 
The properties of these CMa~OB1 physical groups fit well in a monolithic scenario of star formation, with a common formation mechanism, and having suffered multiple episodes of star formation. This suggests that the hierarchical model alone, which explains the populations of other parts of the same association, is not sufficient to explain its whole formation history.

\end{abstract}




\section{Introduction}
\label{sec:intro}

Detailed studies of the stellar content of OB associations, as well as star clusters, provide insights into several important issues about Galactic structure and evolution. They bring crucial information that allows the identification of different stellar populations in the Galaxy and testing stellar models. Issues such as the origin of the binary star populations, differences in initial mass functions, identification of the dominant processes in star formation and stellar fragmentation may all be enlightened by the study of the young populations, while processes of dissolving clusters and chemical Galaxy evolution can be investigated through the study of more evolved stellar groups \citep{2001RMxAC..11...89B}.

\begin{figure*}
\begin{center}

\includegraphics[width=2.0\columnwidth, angle=0]{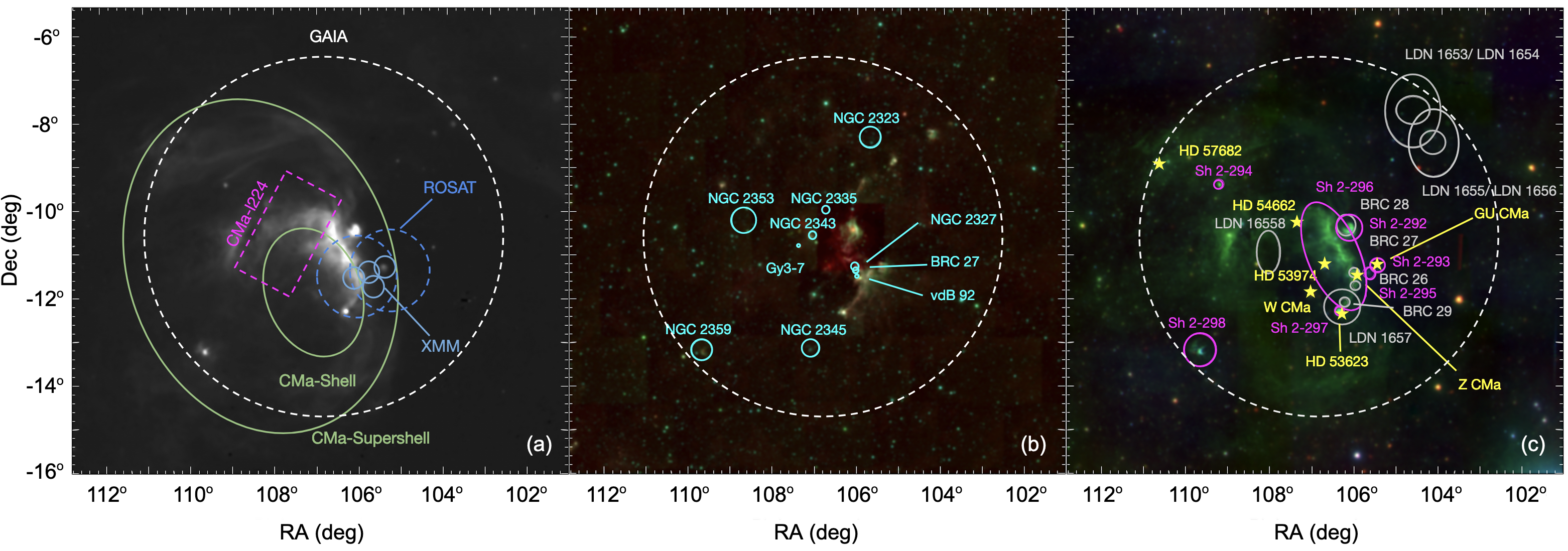}

\end{center}

\caption{(a):  H$\alpha$ image from The Southern H-Alpha Sky Survey Atlas (SHASSA) of the stellar association CMa~OB1 superposed with regions studied in previous works, including ROSAT (dashed blue circles) and {\it XMM-Newton} (solid blue circles) X-ray fields studied by \citet{2009A&A...506..711G} and \citet{2018A&A...609A.127S}, respectively (blue circles); Shell and supershell structures identified by \citet[][grey ellipses]{2019A&A...628A..44F}; CMa-l224 from \citet[][magenta rectangle]{2019ApJS..240...26S}; and with {\it Gaia} DR2 data in this work (white circle). (b):  Combined  Wide-field Infrared Survey (WISE) image in  W1 (blue), W2 (green) and W4 (red) bands, the cyan circles represent the clusters with distances similar to the CMa~OB1 association known in the literature \citep{1974A&A....37..229C,2002A&A...389..871D,2016A&A...585A.101K,2018A&A...618A..93C,2019ApJS..245...32L,2019AJ....157...12B,2019A&A...623A.108B}.  (c): Combined  Digital Sky Survey (DSS) image of the 700 nm (red), 640 nm (green) and 468 nm (blue). The HII regions (Sh 2-292, Sh 2-293, Sh 2-294, Sh 2-295, Sh 2-296, Sh 2-297 and Sh 2-298 \citep{1959ApJS....4..257S} are highlighted in magenta. The dark clouds: LDN 1653, LDN 1654, LDN 1655, LDN 1656, LDN 1657 and LDN 1658 \citep{2005PASJ...57S...1D} and bright-rimmed clouds: BRC 26, BRC 27, BRC 28 and BRC 29 \citep{1991ApJS...77...59S} are represented in grey. Stars HD 53623,  W CMa, Z CMa and GU CMa and the three runaway stars (HD 53974, HD 54662 and HD 57682) from \citet{2019A&A...628A..44F} are shown in yellow.}

\label{fig1}

\end{figure*}

It is historically accepted that most stars are born in gravitationally-bound groups, inside molecular clouds (embedded clusters) with at least 35 stars with a density greater than 1 M$_{\odot}$ pc$^{-3}$  \citep{2003ARA&A..41...57L}. Depending on the formation scenario, most of them ($\sim$ 95\%) should evolve into unbound groups that should dissolve themselves within 10 to 20 Myr  \cite[][]{2003ARA&A..41...57L,2011A&A...536A..90P, 2009A&A...498L..37P}, becoming field stars or associations, currently unbound \citep[e.g.][]{2017MNRAS.472.3887M}. In the monolithic scenario of formation, OB associations are the current configurations of systems that were originally much more compact, and that has subsequently expanded from a single (singularly monolithic) or from several clusters (multiply monolithic)  \citep[e.g.][]{1991ASPC...13....3L, 1997MNRAS.285..479B, 2001MNRAS.321..699K}. The most accepted mechanism responsible for these processes has been the expulsion of residual gas from the embedded clusters by means of stellar feedback, making the clusters super-virial \citep[e.g.][]{1980ApJ...235..986H, 2006MNRAS.373..752G, 2007MNRAS.380.1589B}.

A number of theoretical \cite[][]{2012MNRAS.419..841K, 2012MNRAS.420..613G, 2015MNRAS.451..987D} and observational studies \cite[][]{2016A&A...595A..27G, 2018MNRAS.476..381W,2020MNRAS.495..663W} have, however, disputed this general belief: they have shown that gas exhaustion has been a more efficient mechanism for the dynamical evolution of young clusters than gas expulsion \citep[see][]{2014prpl.conf..291L}. This is in agreement with a scenario in which the stars are formed through a continuous distribution of densities, following the fractal structure of the gas distribution from which they formed, thus, according to this scenario, most OB associations were never grouped \citep{2020MNRAS.495..663W}. Moreover, using {\it Gaia} data, \citet{2017MNRAS.472.3887M} and \citet{2020MNRAS.493.2339M} found that most associations are not undergoing expansion. On the other hand, \citet{2019A&A...621A.115C} suggest that although Vela OB2, as well as its stellar distribution, are expanding, and this expansion started before the formation of the stars. Vela OB2 is, therefore, an example of an OB association formed globally unbound but that nevertheless shows signs of expansion.

Our particular interest is the study of the stellar association Canis Major OB1 (henceforth CMa~OB1), located at a distance of $\sim$ 1200  pc \citep{2019ApJ...879..125Z,2020A&A...633A..51Z}. It is composed of more than 200 B stars, few late-type O stars \citep{2008hsf2.book....1G}, almost 500 young stellar objects (YSOs) \citep{2016ApJ...827...96F,2019ApJS..240...26S} and about 400 H$\alpha$ emitters \citep{2019A&A...630A..90P}, showing a low fraction of disk-bearing stars \citep{2015MNRAS.448..119F, 2016ApJ...827...96F}. This population is mostly related to a reflection nebulae association CMa~R1, including three connected HII regions \citep{1978ApJ...223..471H}: Sh 2-292, Sh 2-296, Sh 2-297 \citep{1959ApJS....4..257S}, four bright-rimmed clouds \citep[BRC from 26 to 29, see][]{1991ApJS...77...59S}, six dark clouds: LDN 1653, LDN 1654, LDN 1655, LDN 1656, LDN 1657 and LDN 1658 \citep{2005PASJ...57S...1D} and more than a dozen clusters \citep{2002A&A...389..871D,2016A&A...585A.101K,2018A&A...618A..93C,2019ApJS..245...32L,2019AJ....157...12B} are in its line of sight, which are shown in Fig. \ref{fig1}.

There are at least two open questions related to the CMa~OB1 association: one refers to membership of its clusters and the other regards its star formation history. It is well known that there are embedded clusters such as NGC 2327 and BRC 27,  with 1.5 Myr, Gy 3-7,  with $\sim$ 2Myr 
\cite[][]{2002A&A...388..172S, 2003BASBr..23..192S, 2013AJ....145...15R}, Z CMa and GU CMa,  with $<$ 5 Myr and $>$ 10 Myr respectively \citep{2009A&A...506..711G} and  vdB 92, possibly cluster in a dissolution stage, with  $\sim$5-7 Myr \cite[][]{2013AJ....145...15R, 2010A&A...516A..81B}. All of these are probably related to CMa~OB1, given their measured distances and young ages. On the other hand, there are famous clusters like NGC 2353, initially suggested to be the nucleus of CMa~OB1, as well as  NGC 2343, NGC 2335 and NGC 2323 \citep{1974A&A....37..229C, 1998A&AS..128..131C} that were discarded and the general conclusion was that these are unrelated to CMa~OB1 due to their ages ($>$ 100 Myr).  

There have been, however, other contrasting views about the plausible star formation and evolutionary scenario of the CMa~OB1 association. One of them, developed by \citet{1977ApJ...217..473H}, suggests a star formation induced by a supernova explosion (SNE) about 0.5 Myr ago. This scenario is consistent with the \citet{2016ApJ...827...96F} results on the distribution of YSOs found in the centre of CMa~OB1. On the other hand, \citet{1978ApJ...224...94R}  proposed that the star formation in the region is triggered by strong stellar winds or an expanding old ``{\it fossil}'' HII region. 

Based on the spatial distribution of YSOs  of the HII region Sh~2-297 to the west side of CMa~OB1, \citet{2012ApJ...759...48M} argue that the youngest sources in the region are distributed away from the ionising source, indicating a possible evolutionary sequence. This scenario supports the hypothesis of triggered star formation in this region, which seems to have propagated from the massive ionising star HD 53623, towards the cold dark cloud LDN 1657A. 
On the other hand, \citet{2019ApJS..240...26S} studied the other side of CMa~OB1 (east side - see the rectangle in Fig. \ref{fig1} (a)), in a region now dubbed CMa-l224 and they suggest that the most likely scenario that explains the star formation observed in this region includes the spontaneous gravitational collapse of filaments.

The star formation region CMa~R1, the most prominent feature of CMa~OB1, also has been studied by our group at several wavelengths for the last many years. \citet{2009A&A...506..711G}, based on {\it ROSAT} data, found two distinct groups: one with ages $<$ 5 Myr, immersed in the region with a high concentration of gas and dust, around Z CMa star, and another in the opposite side, near GU CMa star,  with an older young population ($>$ 10 Myr). Both populations were confirmed by \citet{2018A&A...609A.127S}, using {\it XMM-Newton} satellite. The authors not only increased the number of known sources in the region to about 400 X-ray sources, but they also, using {\it 2MASS} and {\it WISE} counterparts and CO maps, proposed a new star formation scenario including at least two episodes. The first episode occurred more slowly, within the whole studied region, at least 10 Myr ago, dispersing almost all of the present gas, while the second, ongoing episode, has been occurring for less than 5 Myr. The latter is dynamically faster, and it takes place in the region where the gas is still present, suggesting that the association is going through the final stages of star formation.

Finally, in a recent work, \citet{2019A&A...628A..44F} used images in several wavelengths (optical, IR, HI, CO, etc.) to show that CMa~OB1 consists of a shell with diameter D $\sim$ 60 pc, where the Sh 2-296 nebula is nested in a super bubble of 140 pc in diameter. They also found 3 runaway stars that were probably ejected from approximately the same location within the {\it CMa Shell}, by at least three successive SNE for $\sim$ 6 Myr, $\sim$2 Myr and $\sim$ 1 Myr. This suggests that there were more SNE than the one predicted by \citet{1977ApJ...217..473H}. In  that work, \citet{2019A&A...628A..44F} also show that the O stars in the region cannot, by themselves, be responsible for the nebula heating and they suggest, taking into account the scenario of multiple SNE, that Sh 2-296 is being heated by X-rays.

These results are in agreement with a scenario that considers a second episode of star formation, as proposed previously by \citet{2018A&A...609A.127S}. However, this cannot explain the older stellar population ($>$ 10 Myr) found by them and by \citet{2009A&A...506..711G}, which leads us to believe that the association may have an even more intriguing star-formation scenario. 

This work was done in the context of the Southern Photometric Local Universe Survey\footnote[1]{\url{www.splus.iag.usp.br}}  \citep[S-PLUS, ][]{2019MNRAS.489..241M} collaboration. In the future, we plan to apply the method developed in this paper to do a massive search for stellar associations in the S-PLUS catalogue, which will then be characterised in detail. We then intend to compute physical parameters, such as mass, age, distance, extinction and metallicity of the populations of the associations, as well as of the young star clusters in the Galaxy, using the 5 sloan-bands and 7 narrow-bands from the 12-band Javalambre system of S-PLUS.

Aiming at clarifying the complex star-forming history of CMa~OB1 and to confirm its cluster membership, we conduct a multi-dimensional study about stellar groups in the region, taking into account the positions, proper motions and parallax from the {\it Gaia} DR2 catalogue \citep{gaiadr2}. 

This work is structured as follows. In Section \ref{sec:data}, we present the data used in our analysis.  In Section \ref{sec:subs}, we describe in detail the clustering method we apply to find groups and the comparison with previous results from the literature.  In Section \ref{sec:characterization}, we describe the determination of fundamental parameters: age, distance, visual extinction and metallicity. In Section \ref{sec:comparison}, we report on the stellar populations of the region and in Section \ref{sec:discussion}  we discuss the content of the clusters of CMa~OB1 and the star-forming history of the association. Finally, in Section \ref{sec:conclusions} we present a summary of our results and conclusions.

\section{Data} 
\label{sec:data}

The {\it Gaia Mission} is an ambitious survey that aims to construct the most accurate 3D map of our Galaxy. Its second data release (hereafter {\it Gaia} DR2; \citealt{gaiadr2})  provides the five parameter astrometric solution ($\alpha$, $\delta$, $\mu_{\alpha}\cos\delta$, $\mu_{\delta}$, $\varpi$) with unprecedented precision and photometry in three bands (G, $G_{BP}$ and $G_{RP}$) for more than 1.3 billion  of stars.  In addition, there are also radial velocities, astrophysical parameters (stellar effective temperature, extinction, reddening, radius, and luminosity), and variability measurements for part of the sample.

\subsection{Sample selection}
\label{subsec:sampleselection}
 
We queried the {\it Gaia} DR2 catalogue data from {\it Gaia archive}\footnote[2]{\url{https://gea.esac.esa.int/archive/}} taking into account the astrometric and photometric constraints detailed  below.  We select objects  within a search radius of 4.1 degrees centred on the coordinates ($RA$, $DEC$) = ($106.7^{o}, -10.6^{o}$), covering the entire CMa~OB1 association (See dashed circle in Fig. \ref{fig1}).

 To ensure a good astrometric and photometric quality of the data, we apply constrains using the re-normalised unit weight error (RUWE). We still took into account the zero point (0.029 mas) provided by \citet{Lindegren2018}, and  followed their suggested criteria and also those  suggested by \citet{Arenou2018} for the photometric data.  Thus we choose only the objects that follow:
 
\begin{enumerate}[i.]
\item RUWE $\leq$ 1.4 
\item |($\varpi +0.029$ mas)/$\sigma_{\varpi}$| $<$ 5
\item {\tt phot\_g\_mean\_flux\_over\_error} $>$ 50
\item {\tt phot\_bp\_mean\_flux\_over\_error} $>$ 20
\item {\tt phot\_rp\_mean\_flux\_over\_error} $>$ 20
\item {\tt phot\_bp\_rp\_excess\_factor} $<$ 1.3 $+$ 0.06$\cdot ({\tt bp\_rp})^2$
\item {\tt phot\_bp\_rp\_excess\_factor} $>$ 1.0 $+$ 0.015$\cdot ({\tt bp\_rp})^2$
\item {\tt visibility\_periods\_used} $>$ 8
\end{enumerate}

We use, as a further selection criterion, a cut in parallax values choosing only objects with 0.4 $<$ $\varpi$ (mas) $<$ 2.0 in order to ensure that all CMa~OB1 potential members are considered and to obtain a knowledge of their close neighbourhood. We emphasise that the association has an estimated distance of around 1200 pc, using {\it Gaia} data \cite[][]{2019ApJ...879..125Z, 2020A&A...633A..51Z}, and our parallax constraints correspond to distances between 500 pc and 2500 pc. After applying all of these criteria, our final sample contains 249522 stars.

\subsection{Astrometric distances and extinction correction}
\label{subsec:extinction}

 Recent studies have determined quite accurately the distance of CMa~OB1  \citep{2020A&A...633A..51Z,2019ApJ...879..125Z,2020MNRAS.495..663W} from {\it Gaia} data. Therefore, for consistency, in this work, we also used astrometric distances (D$_{A}$) estimated by \citet{bailerjones2018}, using {\it Gaia} DR2 data,  for all stars selected in the previous section.

\begin{figure*}
\begin{center}

\includegraphics[width=2.0\columnwidth, angle=0]{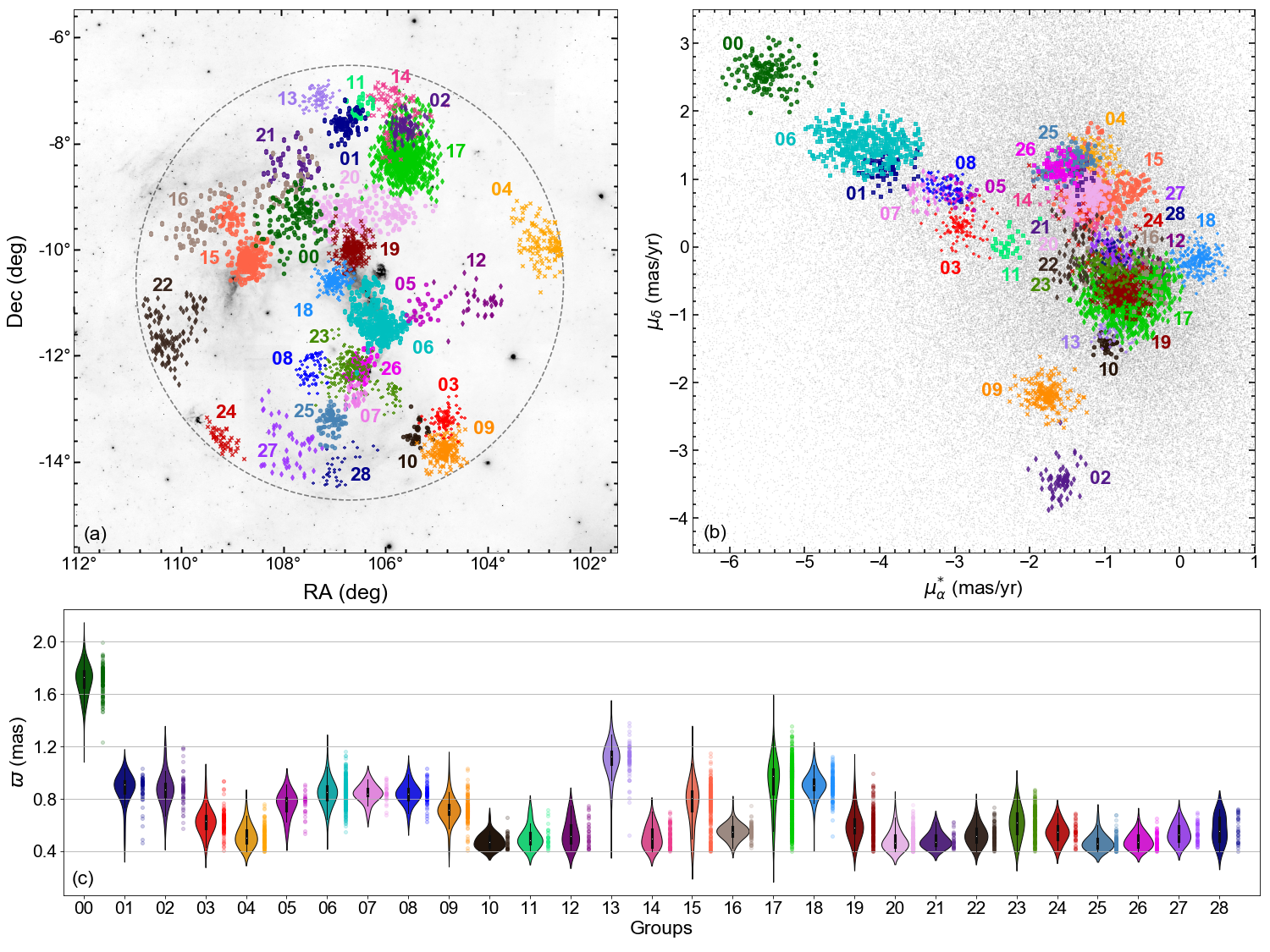}
\end{center}

\caption{ (a) Spatial distribution of groups found by \texttt{HDBSCAN} in CMa~OB1 region on Digital Sky Survey image  (640 nm) of CMa~OB1. The dashed circle delimits the selecting area from {\it Gaia} DR2; (b)  Proper Motion distribution of groups, including field stars (all sources selected in our sample that were not considered in any group - grey points).  $\mu_{\alpha}^{*}$ = $\mu_{\alpha}$ cos${\delta}$; (c) Violin histograms of parallaxes and their exact values (light points in the right side of the histograms) for each group.}

\label{fig2}

\end{figure*}

As it is well known, the association is located in a region of high extinction, and therefore the visual extinction values (A$_{V}$) can significantly vary with distance. In order to take this into account, we use the astrometric distances of each star to obtain its A$_{V}$  from the three-dimensional dust map of {\tt Bayestar19}\footnote[3]{\url{http://argonaut.skymaps.info/}} \citep{bayestar2019}. To consider the probabilistic nature of the {\tt Bayestar19}, we used it with the mode $=$ \textit{mean}. Furthermore, we applied the corrections from \citet{Schlafly2011} to the A$_{V}$ values. 

\section{Substructure search}
\label{sec:subs}

Especially after the {\it Gaia} DR2 was published, several different techniques based on multi-dimensional parameter space analysis were applied to the data to find new and/or to confirm existing stellar populations. In particular, these kinds of analysis  have been successfully applied in the search for associations \citep{2020AJ....159..105L, 2019A&A...626A..17C, 2019A&A...621A.115C, 2018ApJ...856...23G}, star-forming regions \citep{2020A&A...634A..98G,  2019A&A...628A.123Z} and open clusters \citep[see][]{2020A&A...635A..45C, 2020A&A...633A..99C, 2020A&A...640A...1C,2019ApJS..245...32L,2019A&A...628A..66L, 2018A&A...618A..93C, 2018MNRAS.481.3887D, 2018AJ....156..121G}.

Clustering  methods is the most commonly used technique of unsupervised learning and a powerful tool for data analysis. There are several clustering algorithms \citep[for instance, see][and references therein]{scikit-learn}\footnote[4]{\url{https://scikit-learn.org/}}, but one that has  shown to be powerful and efficient in different astronomy fields is the  Hierarchical Density-Based Spatial Clustering of Applications with Noise\footnote[5]{\url{https://hdbscan.readthedocs.io/en/latest/}}, \citep[HDBSCAN;][]{campello2013,campello2015}, which have been used in a variety of contexts \citet{kounkelhdbscan}, \citet{2020AJ....160..279K}, \citet{loganhdbscan}, \citet{Limberg2020} and \citet{2020arXiv201112961K}.  This algorithm is based on \texttt{DBSCAN}  \citep{Ester1996}, also used to  search for stellar clusters \citep{2018A&A...618A..59C, 2019A&A...627A..35C, 2020A&A...635A..45C}. 
This section summarises the methodology used to search for stellar populations in CMa~OB1  as well as its validation.

\subsection{Searching for groups}
\label{subsec:seachrclusters}

In this work, we opted for the use of the \texttt{HDBSCAN} code in order to search for physical groups of stars in the line of sight of CMa~OB1.
The most remarkable feature of the tool in the era of large surveys is that it can identify groups with varying densities and arbitrary shapes without the need of specifying the number of clusters in the sample, as opposed to, for example, the more commonly used k-means algorithm \citep{1967AJ.....72Q.814M}.

\texttt{HDBSCAN} can handle multidimensional data, and depends on six main parameters, including three parameters ({\tt min\_cluster\_size}, {\tt min\_samples} and {\tt cluster\_selection\_method}) that have significant effect on our work, which we discuss below.

The two primary parameters are the minimum number of objects to be classified as a cluster {\tt min\_cluster\_size} and the {\tt min\_samples} is the minimum number of samples in a neighborhood for a point to be considered as a core point. In other words, for a larger value of {\tt min\_samples}, the cluster will be reduced to a more dense area, consequently, in these cases, there will be fewer clusters and a stronger connection between objects. On the other hand, the smaller {\tt min\_samples} values could lead to fragmenting into many small clusters and the decreasing of the noise. Determining the parameters of {\tt min\_samples} and {\tt min\_cluster\_size} is data-dependent and might be difficult. The  {\tt cluster\_selection\_method} parameter is used to select the clusters from the cluster tree hierarchy. The standard approach is the Excess of Mass ({\tt EOM}) which tends to select one or two of the largest clusters and some smaller clusters. Another option is to use the method {\tt Leaf}, which selects several small and more homogeneous clusters.

We apply the \texttt{HDBSCAN} using a Python implementation to the sample selected in Sect. \ref{subsec:sampleselection} in the 5D space of astrometric parameters ($\mu_{\alpha}\cos\delta$, $\mu_{\delta}$, $\varpi$,  $\alpha$ and $\delta$). The \texttt{HDBSCAN} configuration we adopted is  {\tt min\_cluster\_size} $=$ 30, {\tt min\_samples} $=$ 60, and {\tt cluster\_selection\_method} $=$ {\tt Leaf}. These  selection parameters were used aiming at a search for more homogeneous clusters, while the value of {\tt min\_cluster\_size} was chosen in order to guarantee that the groups had at least 30 stars,  to allow their good characterisation. 

In an experimental phase, we performed tests varying  {\tt min\_samples}  from 20 to 250, in steps of 10, to choose the best value for this parameter. In this set, the number of found groups varies significantly, from 36 to 11\footnote[6]{For {\tt min\_samples} $<$ 30 the number of groups increases rapidly for smaller values of this parameter, finding about 80 groups for {\tt min\_samples}~$=$~10. On the other hand, for {\tt min\_samples} $>$ 250 the number of groups fluctuates between 9 and 11, up to {\tt min\_samples} $=$ 350.}, and the number of groups with distances compatible with CMa~OB1 falls from 13 to 4. Farther, for 40 $<$ {\tt min\_samples} $<$ 80, the number of groups with distances consistent with the association varies from 8 to 12, thus, the choice for {\tt min\_samples} $=$ 60 ensures that we are not discarding groups that could possibly be members of CMa~OB1. 

Using this configuration, the code found 29 groups whose spatial and proper motion distributions, as well as violin parallaxes histogram, are shown in Fig. \ref{fig2}.

 To validate the clusters identified by \texttt{HDBSCAN} and estimate the membership probability of each star belonging to a specific group, we have computed 400 bootstrap repetitions taking into the account the uncertainties of the stars astrometric parameters. For each repetition, we have used the \texttt{HDBSCAN} function {\tt approximate\_predict} to evaluate to which cluster the star belongs according to the original hierarchical cluster tree. We attribute a membership probability (P) of each star belonging to a specific cluster according to the percentage of assignment to it. We considered that a star is a cluster member if it is assigned to a specific cluster in at least 50\% of the realisations (P $\geq$ 50\%). This method is similar to that one described by \citet{Limberg2020}.

\begin{figure}
\begin{center}

\includegraphics[width=0.77\columnwidth, angle=0]{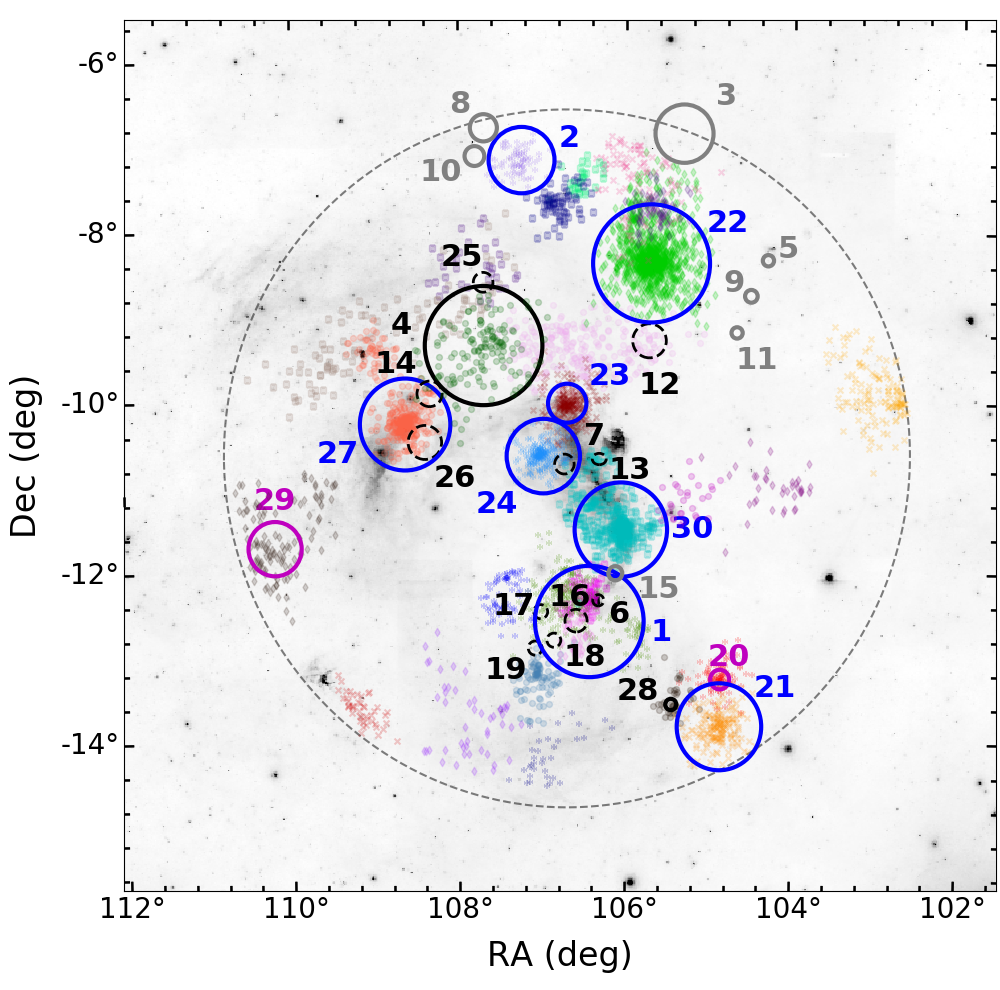}
\includegraphics[width=0.21\columnwidth, angle=0]{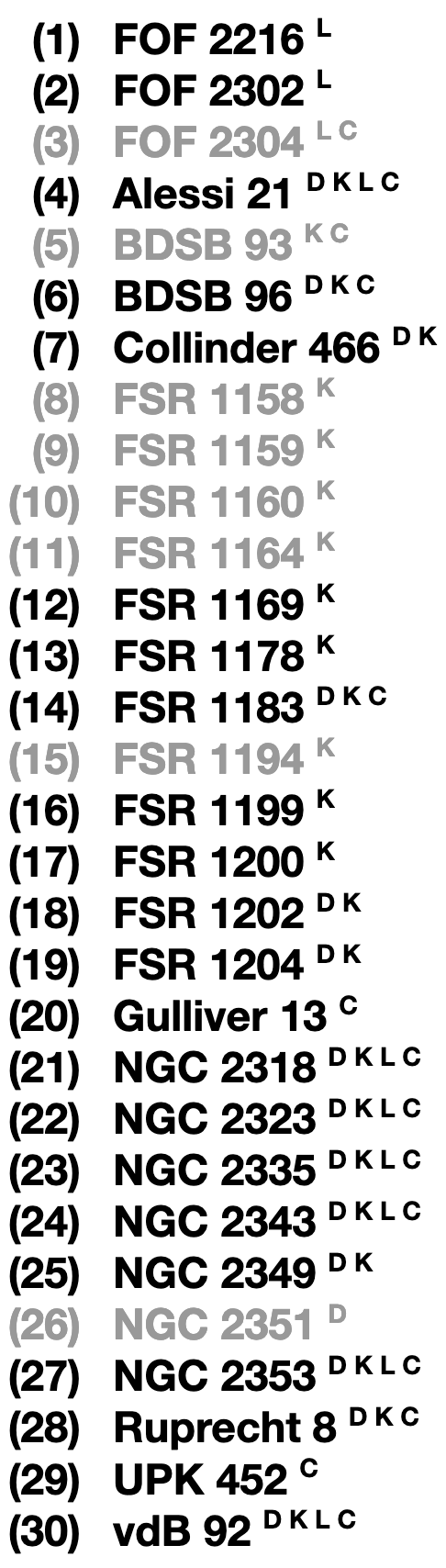}

\end{center}

\caption{Spatial distribution of 29 groups found by \texttt{HDBSCAN}, following Fig.~\ref{fig2} colours,  on a Digital Sky Survey image (640 nm) and 30 clusters from the literature. Solid contours present the angular dimension of clusters known in the literature that coincide with the groups detected in this work.  
In black are clusters from (D) \citet{2002A&A...389..871D} and (K) \citet{2016A&A...585A.101K}  with  angular dimensions from \citet{2019AJ....157...12B},  in blue are clusters from (L) \citet{2019ApJS..245...32L}  and the magenta circles are radii of 50\% of cluster members from (C) \citet{2020A&A...640A...1C}.
Dashed contours present objects that can be related to a given group and grey circles are clusters from these authors that were not detected by us. The letters indicate in which catalogues each cluster is present.
}

\label{fig3}

\end{figure}

\subsection{Validation}
\label{subsec:validatio}

The method used here to search for stellar groups was evaluated and validated by us by performing two tests that aim at proving that the code can find real structures, already known in the literature and that it is able to select the correct membership for the groups.  First,  we compare the spatial distribution of our groups with clusters known in the literature. For the second test, we compare membership of two groups with  CMa~R1 star-forming region members selected using a Bayesian method. 

\subsubsection{Cross-matching with groups previously known in the literature}
\label{subsec:lit_clusters}

In the first test we compared the spatial distribution of our 29 groups with angular dimensions of clusters, present in four large stellar cluster catalogues: two before {\it Gaia} era \citep{2002A&A...389..871D,2016A&A...585A.101K} and two based on {\it Gaia} data \citep{2019ApJS..245...32L,2020A&A...640A...1C}, see Fig. \ref{fig3}. 

 There are, respectively, 24, 14, 10 and 14 clusters with the parallaxes or distances within the same range adopted here (see Sect.~\ref{subsec:sampleselection}) in each one of these catalogues. 
Seven clusters are present in the four catalogues,  other three are present in all but in one  \citet{2019ApJS..245...32L}, four are found only in the two older catalogues, those before publication of the {\it Gaia} catalogues\footnote[7]{ FSR 1163, FSR 1172, FSR 1180, FSR 1207, FSR 1212 and Berkeley 76 from \citet{2016A&A...585A.101K}, NGC 2345 from \citet{2002A&A...389..871D} and FSR 1170 and Ivanov 4 from both, even having distances smaller than 2500 pc in these catalogues, were excluded of these comparisons because they have parallaxes estimated by \citet{2020A&A...633A..99C} using {\it Gaia} data, out of ranges adopted in this work (see Sect. \ref{subsec:sampleselection}).} and BDSB 93 is present only in \citet{2016A&A...585A.101K} and \citet{2020A&A...640A...1C}. Finally, other  15 clusters are present in only one these catalogue: 9 in \citet{2016A&A...585A.101K},  3 in \citet{2019ApJS..245...32L}, 2 in \citet{2020A&A...640A...1C} and NGC 2351  is in \citet{2002A&A...389..871D}.

 Our comparisons were performed by means of a visual inspection of Fig. \ref{fig3}, in which we  prioritise the use of more recent estimates of the angular dimensions of clusters. So that for objects present only in \citet{2002A&A...389..871D} and/or \citet{2016A&A...585A.101K} we use values available in \citet{2019AJ....157...12B}.  
 For Gulliver 13 and UPK 452,  present only in \citet{2020A&A...640A...1C}, we use {\it r50} (radius containing half of the members) provided by the authors because they have no values for the total sizes of the clusters. For all the other objects, we use the {\it rmax} (maximum cluster member's distance to average position) provided by \citet{2019ApJS..245...32L}\footnote[8]{Only for Alessi 21, also present in this catalogue, we adopt the radius provided by \citet{2019AJ....157...12B} to make it clearer in Fig. \ref{fig3}, because the {\it rmax} of this cluster is too large ($ 1.83^{o}$) and it occupies the same area as many groups and clusters.}.

We found 12 clusters whose angular dimensions are, as a whole, projected onto at least one of our groups and, therefore, they are probably the same population. These are the clusters highlighted by solid black, blue and magenta contours in Fig.~\ref{fig3}.  A total of 11 clusters have a few members of our groups  projected within their angular dimension (dashed contours).  Other seven clusters are projected onto regions with no overlap with any members of our groups (grey lines). On the other hand, 13 of our groups are located in places where there are no clusters present in the catalogues used here, indicating that they could be new cluster candidates. 

Among the clusters having no groups projected in their direction, five of them (FSR~1158, FSR 1159, FSR 1160, FSR 1164 and FSR 1194) present only in \citet{2016A&A...585A.101K}  were not detected by \texttt{HDBSCAN} probably because they are asterisms according to \citet{2018A&A...618A..93C} and they were not found by catalogues based on {\it Gaia} data. Other two clusters, FOF 2304 and BDSB 93, have no groups detected by the code in their lines of sight due likely to our sample selection criteria (Sect. \ref{subsec:sampleselection}) and our choice of parameters for \texttt{HDBSCAN} (Sect. \ref{subsec:seachrclusters}). A more detailed discussion on our groups that are clusters known in the literature is are presented in Sect.~\ref{subsec:knowclusters}, and clusters not found by \texttt{HDBSCAN} are shown in Appendix \ref{sec:cluster_not_found}.

\begin{figure}
\begin{center}

\includegraphics[width=1.0\columnwidth, angle=0]{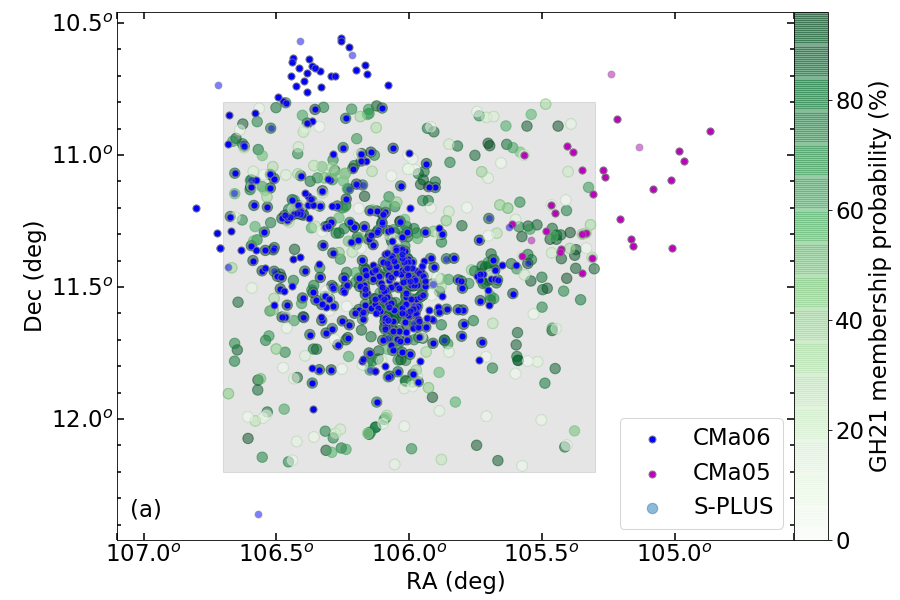}
\includegraphics[width=1.0\columnwidth, angle=0]{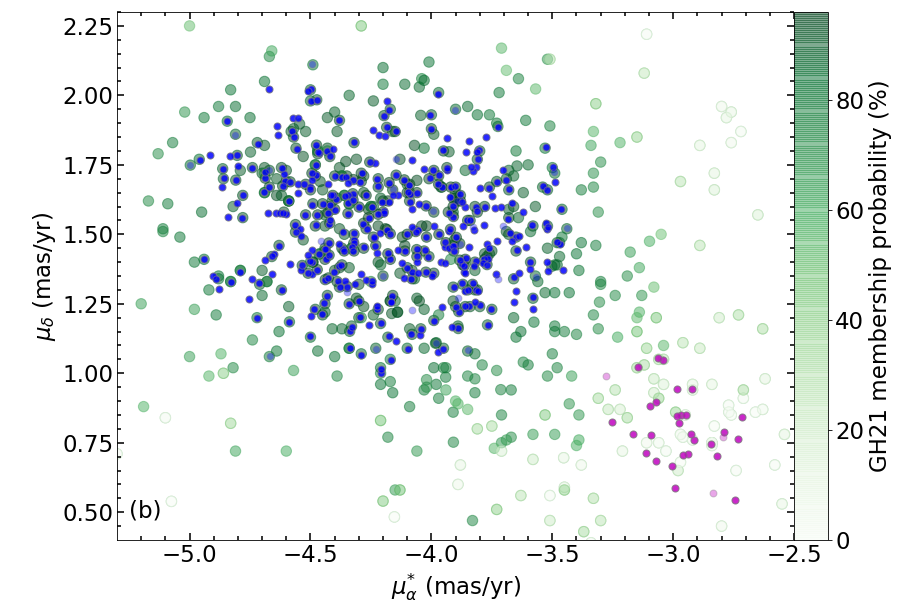}
\end{center}

\caption{ (a) Spatial and  (b) proper motion distributions of CMa06 and CMa05 (blue and pink points, respectively) compared with CMa~R1 members from \citet{2021AJ....161..133G} shown by green circles, according to their membership probability.  CMa05 and CMa06 members with P $<$ 50\% are presented by light points. The hatched area in panel (a) represents the region observed by T80-S telescope.}

\label{fig4}

\end{figure}

\subsubsection{Bayesian method}
\label{subsec:spluscount}

The second test to validate our results is based on the comparison with a recent study of the young population related to our groups. \citet[][hereafter GH21]{2021AJ....161..133G} used multi-band photometry from {\it T80S} telescope of the S-PLUS collaboration \citep{2019MNRAS.489..241M}, covering an area of 2 deg$^{2}$ in the direction of CMa~R1. The sample selection was performed by combining a Bayesian model and the Cross Entropy technique \citep[see][]{2019MNRAS.490.2521H} using astrometric data ($\mu_{\alpha}\ cos\  \delta$, $\mu_{\delta}$) from {\it Gaia} DR2\footnote[9]{The data query was limited to 0.8 $<$ $\varpi$ (mas) $<$ 1.25,  $\varpi/\sigma_{\varpi}$ $>$ 3 and RUWE $<$ 1.4.}. Their entire sample with {\it Gaia} DR2 data contains 669 stars, of which 155 are already known in the literature, 395 are new CMa~R1 members (with membership probability P~$\geq$~50\%, determined by them) and 119 member candidates (P~$<$~50\%). 

We performed a coherent cross-matching between CMa~R1 members and our groups, selecting 501 stars from GH21 that comply with our selection criteria described in Sect. \ref{subsec:sampleselection}.  From our sample, we selected 275 members of CMa06 and 12 from CMa05 within {\it T80S} FOV that fulfil the author's selection criteria\footnote[10]{CMa06 has 31 members out of {\it T80S} FOV and other 98 stars with parallax lower than the limit of 0.8 mas adopted by GH21 to select their sample and CMa05 has 14 objects out of {\it T80S} FOV and 5 with lower parallax.}.
It is important to note that the {\it T80S} survey only covers, spatially, a fraction of our sources identified in CMa05 and CMa06 (see hatched area in Fig.~\ref{fig4}~(a)). 

\begin{table*}
\caption{Astrometric parameters of 15 groups found in CMa~OB1 region.}
\begin{center}
{
\small
\begin{tabular}{lcccccccccccccc}
\hline 

Group	&	N$^{(a)}$	& N$_{50}^{(b)}$ & P$_{50}^{(c)}$	&	R.A.$^{(d)}$	&	 Dec.$^{(e)}$ 	&	$\mu_{\alpha}\cos\delta^{(f)}$  &		$\mu_{\delta}^{(g)}$	&   $\varpi^{(h)}$	&	D$_{A}^{(i)}$ &	A$_{V}^{(j)}$ & V$_{t}^{(k)}$ & $\sigma_{V2D}^{(l)}$	\\

	&	(stars)	& (stars)	&  (\%)	& (deg)	&	(deg) 	&	(mas yr$^{-1}$)&	(mas yr$^{-1}$)	&  (mas)   &	(pc)	& (mag) & (km/s) & (km/s)	\\ \hline \hline

CMa00 & 152 & 148 & 97 & 107.70$^{+0.25}_{-0.35}$  &  -9.36$^{+0.37}_{-0.27}$  &  -5.52$^{+0.19}_{-0.26}$  &  2.59$^{+0.22}_{-0.19}$  &  1.73$^{+0.14}_{-0.07}$  &  571$^{+23}_{-51}$  &  0.27$^{+0.12}_{-0.13}$  &  16.6 & 1.2\\
CMa01 & 66 & 64 & 97 & 106.80$^{+0.18}_{-0.14}$  &  -7.66$^{+0.16}_{-0.11}$  &  -3.94$^{+0.14}_{-0.18}$  &  1.13$^{+0.10}_{-0.15}$  &  0.91$^{+0.09}_{-0.03}$  &  1066$^{+32}_{-134}$  &  0.47$^{+0.10}_{-0.33}$  &  20.8 & 4.2\\
CMa02 & 71 & 65 & 92 & 105.68$^{+0.13}_{-0.16}$  &  -7.74$^{+0.13}_{-0.13}$  &  -1.59$^{+0.13}_{-0.09}$  &  -3.45$^{+0.21}_{-0.14}$  &  0.87$^{+0.08}_{-0.08}$  &  1120$^{+87}_{-129}$  &  0.53$^{+0.11}_{-0.09}$  &  20.1 & 2.6\\
CMa03 & 103 & 101 & 98 & 104.86$^{+0.09}_{-0.13}$  &  -13.25$^{+0.10}_{-0.14}$  &  -2.92$^{+0.13}_{-0.16}$  &  0.31$^{+0.15}_{-0.18}$  &  0.63$^{+0.08}_{-0.06}$  &  1526$^{+120}_{-211}$  &  0.94$^{+0.31}_{-0.31}$  &  21.5 & 3.3\\
CMa05 & 31 & 28 & 90 & 105.32$^{+0.29}_{-0.14}$  &  -11.20$^{+0.15}_{-0.21}$  &  -2.98$^{+0.11}_{-0.15}$  &  0.79$^{+0.08}_{-0.14}$  &  0.78$^{+0.10}_{-0.05}$  & 1224$^{+55}_{-139}$  &  1.08$^{+0.38}_{-0.35}$  &  18.5 & 1.9\\
CMa06 & 404 & 377 & 93 & 106.11$^{+0.14}_{-0.29}$  &  -11.45$^{+0.17}_{-0.29}$  &  -4.20$^{+0.35}_{-0.38}$  &  1.52$^{+0.19}_{-0.21}$  &  0.85$^{+0.09}_{-0.07}$  &  1147$^{+77}_{-133}$  &  1.18$^{+0.46}_{-1.09}$  &  24.3 & 3.4\\
CMa07 & 34 & 26 & 76 & 106.60$^{+0.11}_{-0.15}$  &  -12.79$^{+0.11}_{-0.18}$  &  -3.36$^{+0.20}_{-0.08}$  &  0.79$^{+0.13}_{-0.13}$  &  0.84$^{+0.04}_{-0.06}$  &  1159$^{+55}_{-69}$  &  0.70$^{+0.13}_{-0.19}$  &  19.2 & 1.6\\
CMa08 & 76 & 64 & 84 & 107.43$^{+0.11}_{-0.14}$  &  -12.30$^{+0.20}_{-0.23}$  &  -3.13$^{+0.19}_{-0.17}$  &  0.87$^{+0.13}_{-0.13}$  &  0.84$^{+0.05}_{-0.06}$  &  1162$^{+62}_{-62}$  &  0.97$^{+0.22}_{-0.26}$  &  18.2 & 1.6\\
CMa09 & 180 & 173 & 96 & 104.87$^{+0.16}_{-0.19}$  &  -13.81$^{+0.16}_{-0.17}$  &  -1.78$^{+0.14}_{-0.13}$  &  -2.18$^{+0.14}_{-0.14}$  &  0.71$^{+0.05}_{-0.08}$  &  1356$^{+100}_{-117}$  &  0.74$^{+0.19}_{-0.14}$  &  18.1 & 2.6\\
CMa13 & 80 & 68 & 85 & 107.27$^{+0.11}_{-0.17}$  &  -7.19$^{+0.15}_{-0.16}$  &  -0.91$^{+0.11}_{-0.12}$  &  -1.22$^{+0.12}_{-0.11}$  &  1.13$^{+0.11}_{-0.07}$  &  869$^{+31}_{-51}$  &  0.40$^{+0.10}_{-0.12}$  &  6.3 & 0.9\\
CMa15 & 384 & 365 & 95 & 108.66$^{+0.13}_{-0.20}$  &  -10.23$^{+0.13}_{-0.39}$  &  -1.09$^{+0.18}_{-0.29}$  &  0.77$^{+0.18}_{-0.19}$  &  0.80$^{+0.22}_{-0.08}$   &  1204$^{+96}_{-440}$  &  0.46$^{+0.17}_{-0.31}$  &  7.6 & 3.6\\
CMa17 & 1096 & 1039 & 95 & 105.70$^{+0.21}_{-0.19}$  &  -8.33$^{+0.20}_{-0.30}$  &  -0.71$^{+0.26}_{-0.33}$  &  -0.66$^{+0.28}_{-0.21}$  &  0.97$^{+0.25}_{-0.08}$  &  1003$^{+68}_{-317}$  &  0.65$^{+0.14}_{-0.21}$  &  5.0 & 2.1\\
CMa18 & 227 & 218 & 96 & 107.01$^{+0.15}_{-0.10}$  &  -10.62$^{+0.10}_{-0.10}$  &  0.26$^{+0.18}_{-0.13}$  &  -0.18$^{+0.15}_{-0.12}$  &  0.91$^{+0.06}_{-0.06}$ &  1076$^{+56}_{-74}$  &  0.59$^{+0.16}_{-0.27}$  &  1.7 & 0.7\\
CMa19 & 245 & 233 & 95 & 106.67$^{+0.16}_{-0.11}$  &  -10.03$^{+0.16}_{-0.13}$  &  -0.77$^{+0.14}_{-0.16}$  &  -0.66$^{+0.13}_{-0.16}$  &  0.58$^{+0.08}_{-0.10}$  &  1654$^{+213}_{-265}$  &  1.09$^{+0.32}_{-0.46}$  &  7.9 & 1.9\\
CMa23 & 239 & 217 & 91 & 106.66$^{+0.33}_{-0.23}$  &  -12.35$^{+0.31}_{-0.24}$  &  -0.96$^{+0.17}_{-0.19}$  &  -0.28$^{+0.21}_{-0.16}$  &  0.62$^{+0.12}_{-0.11}$ &  1583$^{+220}_{-384}$  &  0.81$^{+0.12}_{-0.26}$  &  7.7 & 2.0\\ \hline
\end{tabular}
}
\label{tab1}
\end{center}

NOTES:
(a) Number of stars in the group;
(b) Number of star with membership probability P~$\geq$~50\%;
(c) Percentage of star with membership probability P~$\geq$~50\%;
(d) Right ascension (ICRS) at Ep = 2015.5;
(e) Declination (ICRS) at Ep = 2015.5;
(f) Right ascension proper motion;
(g) Declination proper motion;
(h) Parallax;
(i) Astrometric distance from \citet{bailerjones2018}; 
(j) Visual extinction from three-dimensional maps of dust \citep{bayestar2019};
(k) Tangential velocity;
(l) Velocity dispersion in 2-dimension (R.A. and Dec.).

\end{table*}

In the first cross-matching, implemented in CMa06, we found  251 members of CMa~R1 in our group. We noticed that all these objects were classified as P $>$75\% by GH21, even considering 18 stars with membership probability P $<$ 50\% obtained by us,  showing that our method is more conservative than theirs. Actually, this is not surprising since, in addition to the proper motion, used by the authors to calculate membership probabilities, \texttt{HDBSCAN}  also takes into account parallax and spatial distribution for the selection of groups.
On the other hand, \texttt{HDBSCAN} selected, in our group CMa06,  about  72\%~(251/347) of the CMa~R1 members with P $>$ 75\% that follow our selection criteria, or more conservatively,  85\% (201/238) of P~$\geq$~90\% CMa~R1 members, proving that our method was effective in finding high probability CMa~R1 members.  Moreover, GH21 estimated, using the Cross Entropy technique, the  mean  proper motion of CMa~R1:  $\mu_{\alpha}* = -4.1 \pm 0.6$ mas yr$^{-1}$ and {$ \mu_{\delta} = 1.5 \pm 0.4$ mas~yr$^{-1}$  which is compatible with our group CMa06, with $\mu_{\alpha}* = -4.18 \pm 0.36$~mas~yr$^{-1}$ and $ \mu_{\delta} = 1.52 \pm 0.21$~mas~yr$^{-1}$ (see Table~\ref{tab1}).

By means of the proper motion distribution of the CMa~R1 members (see Fig. \ref{fig4} (b)), we also note that part of the CMa~R1 candidates, selected within a 3$\sigma$ distribution around the main cluster, coincides with our group CMa05. For this reason, we performed another cross-matching and found 6 CMa~R1 members candidates (P $<$ 50\%) in CMa05.  

In short, our methodology proved to be quite efficient to find clusters known in the literature, as well as detected new groups that could potentially be new  identified clusters.
In addition, comparing stars selected by \texttt{HDBSCAN} in our groups (specially, CMa05 and CMa06 with CMa~R1 members from GH21 and objects well known in the literature, including bright stars, H$\alpha$ emitters,  X-ray sources and YSOs - see discussion in Appendix \ref{sec:know_stars}), we verify that our method is also efficient to find  they fiducial members.  This is particularly true for objects with a high membership probability,  since we found 98\% of CMa06 members in  T80S FOV fulfilling the criteria selection of GH21, classified by the authors as P $>$ 75\%.  On the other hand, due to our conservative method, we were unable to find 28\% of members CMa~R1 with P $>$ 75\% according to GH21. 

\section{Groups Characterisation}
\label{sec:characterization}

Focusing on groups with high probability of being associated with CMa~OB1, with a distance of $\sim$ 1200  pc \citep{2019ApJ...879..125Z,2020A&A...633A..51Z}, and avoiding groups that may not have been well determined by our choice of \texttt{HDBSCAN} parameter (see Appendix \ref{sec:met_constraints}),  we select a sample of the 15 physical groups. These 15 groups are those which more than 75\% of their members have  membership probability P~$\geq$~50\% and  D$_{A}$~$<$~1700 pc (see Table~\ref{tab1}). In order to better understand these groups, we derived their astrometric parameters using a statistical approach and we applied two algorithms, which use different isochrones fitting methods, combined with the {\it Gaia} DR2 photometric data, to determine fundamental parameters: age, distance,  visual extinction and metallicity of our groups. All parameters are derived considering only members with P $\geq$ 50\%.

A more detailed discussion about unreliable groups are presented in Appendix \ref{subsec:unreliabe_cluster}. We also show a list of all possible physical groups that were not well determined in this work, but deserve to be studied in a future work, considering different ranges of parameters in the sample selection.

\subsection{Astrometric Parameters}
\label{sec:astrometric_par}
 
After determining all groups using \texttt{HDBSCAN}, we calculate their astrometric parameters, from the distribution of the individual parameters for the objects with membership probabilities greater than 50\% in each group. The position, proper motion and parallax attributed to each group are given by the 50\%  percentile (median), and for their respective errors, we assume the 16\% - 84\% percentile ranges. The astrometric distances and the visual extinction of our groups are derived in the same way using values described in Sect. (\ref{subsec:extinction}). Furthermore, we calculate the tangential velocity of each star from its individual proper motions and astrometric distances. We assigned the tangential velocity of each group also from the 50\% percentile, and the 2-dimensional velocity dispersion from the standard deviation of the tangential velocity distribution. All these parameters for each group are presented in Table~\ref{tab1}.

\begin{table*}
\caption{Fundamental parameters obtained from isochrones fitting.}
{\centering
{\small
\begin{tabular}{l|cccc|cccc|l}
\hline

Groups & Age$^{M}$ & D$_{P}^{M}$  & A$_{VP}^{M}$ & [Fe/H]$^{M}$ & Age$^{f}$   & D$_{P}^{f}$   & A$_{VP}^{f}$   & [Fe/H]$^{f}*$   & Cluster name \\     
 & (Myr) & (pc)  & (mag) &  & (Myr)   & (pc)   & (mag)   &    & \\ \hline \hline  
 
CMa00 & 85 $\pm$ 21 & 570.0 $\pm$ 1.0 & 0.34 $\pm$ 0.05 & 0.00 $\pm$ 0.07 & 70$^{+5}_{-20}$ & 549.3$^{+1.3}_{-6.3}$ & 0.31$^{+0.01}_{-0.05}$ & 0.07$^{+0.05}_{-0.05}$ & Alessi 21 $^{D \ K \ L  \ C}$ \\
CMa01 & 172 $\pm$ 36 & 1059 $\pm$ 19 & 0.42 $\pm$ 0.08 & 0.07 $\pm$ 0.09 & 200$^{+13}_{-13}$ & 1015$^{+16}_{-43}$ & 0.43$^{+0.10}_{-0.02}$ & 0.05$^{+0.03}_{-0.01}$ &  \\      
CMa02 & 137 $\pm$ 67 & 1071 $\pm$ 35 & 0.99 $\pm$ 0.12 & -0.15 $\pm$ 0.12 & 125$^{+13}_{-13}$ & 1020$^{+76}_{-113}$ & 1.25$^{+0.09}_{-0.05}$ & -0.48$^{+0.32}_{-0.04}$ &  \\      
CMa03 & 536 $\pm$ 419 & 1531 $\pm$ 35 & 1.71 $\pm$ 0.17 & -0.12 $\pm$ 0.11 & 650$^{+25}_{-25}$ & 1542$^{+58}_{-76}$ & 1.18$^{+0.19}_{-0.05}$ & 0.30$^{+0.01}_{-0.01}$ & Gulliver 13 $^{C}$ \\     
CMa05 & 17 $\pm$ 27 & 1229 $\pm$ 28 & 1.27 $\pm$ 0.16 & -0.14 $\pm$ 0.06 & 18.0$^{+7.0}_{-2.0}$ & 1130$^{+9}_{-33}$ & 1.11$^{+0.05}_{-0.03}$ & -0.10$^{+0.04}_{-0.02}$ &  \\      
CMa06 & 10.1 $\pm$ 1.0 & 1099 $\pm$ 29 & 0.96 $\pm$ 0.09 & -0.15 $\pm$ 0.05 & 9.0$^{+0.5}_{-0.5}$ & 1069$^{+25}_{-8}$ & 1.18$^{+0.04}_{-0.03}$ & 0.07$^{+0.02}_{-0.01}$ & VdB 92 $^{D \ K \ L  \ C}$ \\
CMa07 & 13.1 $\pm$ 1.5 & 1149 $\pm$ 45 & 0.67 $\pm$ 0.08 & -0.03 $\pm$ 0.11 & 14.0$^{+1.0}_{-1.0}$ & 1092$^{+4}_{-5}$ & 0.74$^{+0.02}_{-0.03}$ & 0.07$^{+0.02}_{-0.02}$ & FOF 2216 $^{L}$\\     
CMa08 & 18 $\pm$ 5 & 1138 $\pm$ 46 & 1.01 $\pm$ 0.07 & 0.05 $\pm$ 0.09 & 18.0$^{+1.0}_{-1.0}$ & 1005$^{+55}_{-12}$ & 0.99$^{+0.01}_{-0.05}$ & 0.05$^{+0.03}_{-0.01}$ &  \\      
CMa09 & 329 $\pm$ 186 & 1346 $\pm$ 9 & 1.08 $\pm$ 0.29 & 0.07 $\pm$ 0.21 & 325$^{+13}_{-13}$ & 1219$^{+109}_{-21}$ & 1.08$^{+0.13}_{-0.11}$ & 0.14$^{+0.02}_{-0.01}$ & NGC 2318 $^{D \ K \ L  \ C}$ \\
CMa13 & 208 $\pm$ 63 & 873 $\pm$ 4 & 0.27 $\pm$ 0.07 & 0.07 $\pm$ 0.04 & 225$^{+13}_{-13}$ & 855$^{+19}_{-61}$ & 0.37$^{+0.05}_{-0.03}$ & 0.02$^{+0.03}_{-0.05}$ & FOF 2302 $^{L}$ \\    
CMa15 & 114 $\pm$ 36 & 1074 $\pm$ 63 & 0.53 $\pm$ 0.06 & -0.13 $\pm$ 0.11 & 150$^{+13}_{-13}$ & 1251$^{+53}_{-62}$ & 0.45$^{+0.05}_{-0.02}$ & 0.28$^{+0.01}_{-0.01}$ & NGC 2353 $^{D \ K \ L  \ C}$ \\
CMa17 & 163 $\pm$ 81 & 986 $\pm$ 34 & 0.91 $\pm$ 0.11 & -0.15 $\pm$ 0.06 & 175$^{+13}_{-13}$ & 933$^{+19}_{-74}$ & 0.86$^{+0.10}_{-0.05}$ & 0.07$^{+0.07}_{-0.02}$ & NGC 2323 $^{D \ K \ L  \ C}$ \\
CMa18 & 181 $\pm$ 43 & 1024 $\pm$ 25 & 0.63 $\pm$ 0.10 & -0.03 $\pm$ 0.06 & 150$^{+13}_{-13}$ & 1050$^{+24}_{-77}$ & 0.59$^{+0.09}_{-0.05}$ & 0.14$^{+0.02}_{-0.01}$ & NGC 2343 $^{D \ K \ L  \ C}$ \\
CMa19 & 42 $\pm$ 30 & 1446 $\pm$ 99 & 1.54 $\pm$ 0.11 & -0.09 $\pm$ 0.11 & 60$^{+5}_{-15}$ & 1413$^{+66}_{-436}$ & 1.24$^{+0.01}_{-0.10}$ & 0.30$^{+0.01}_{-0.20}$ & NGC 2335 $^{D \ K \ L  \ C}$\\ 
CMa23 & 131 $\pm$ 65 & 1374 $\pm$ 99 & 1.20 $\pm$ 0.10 & -0.16 $\pm$ 0.13 & 150$^{+25}_{-13}$ & 1552$^{+39}_{-182}$ & 1.20$^{+0.09}_{-0.05}$ & 0.14$^{+0.02}_{-0.01}$ &  \\\hline   
\end{tabular}
}
}

NOTES:
 (M) Parameters; Age, photometric distance (D$_{P}$), visual extinction (A$_{VP}$) and metallicity, obtained using \textsc{M20 code}, see also Appendix. \ref{Hektor}; 
(f) Same parameters obtained using \textsc{fitCMD}, see Appendix. 
\ref{Charles};
(*) Metallicity - [Fe/H] = $log_{10}(Z/Z_{\odot})$.
Clusters found in: 
$^{(D)}$ \citet{2002A&A...389..871D};
$^{(K)}$ \citet{2016A&A...585A.101K};
$^{(L)}$ \citet{2019ApJS..245...32L};
$^{(C)}$ \citet{2020A&A...640A...1C},
\citet{2020A&A...633A..99C} and \citet{2021MNRAS.504..356D}.
\label{tab2}

\end{table*}


\subsection{Fundamental Parameters}
\label{sec:fund_param}

The first code that we adopted to obtain the fundamental parameters, developed by \citet{2020MNRAS.499.1874M} --  henceforth \textsc{M20 code},  is based on the cross-entropy continuous multi-extremal optimisation method (CE), which takes into account the astrometric membership of the stars obtained in Sect. \ref{subsec:seachrclusters}, as well as the nominal errors of the data. The \textsc{M20 code} uses theoretical isochrones from Padova (PARSEC version 1.2S) database of stellar evolutionary tracks and isochrones \citep{Bressan2012}   fitted to the {\it Gaia} DR2 $G_{BP}$ and $G_{RP}$ photometric data. A more detailed description of the code is provided in Appendix~\ref{Hektor}.

For the \textsc{M20 code}, we applied the priors for distance and visual extinction to the mean values present in Table \ref{tab1}.  For [Fe/H] we use the default prior of the code, estimated from the Galactic metallicity gradient published by \citet{OCCAMgradient20}. The age, distance, A$_{V}$ and [Fe/H] for the 15 groups are presented in the first 4 columns in Table \ref{tab2}. Additional parameters estimated by \textsc{M20 code} are given in the Table \ref{tabC1}.

The second code we use here, developed by \citet{2019MNRAS.483.2758B} - \textsc{fitCMD},  is a statistical approach to extract fundamental parameters of star clusters from the photometric information contained in observed  Colour-Magnitude Diagrams (CMDs). \textsc{fitCMD} searches for physical parameters able to build a synthetic CMD, based on properties of the Initial Mass Function (IMF) obtained from isochrones also from Padova database, which best reproduce the observed one. The detailed description of \textsc{fitCMD} are present in Appendix \ref{Charles}.

The \textsc{fitCMD} input parameters used by us here were:  ages from 1 Myr to 1 Gyr; apparent distance modulus (DM) from 8.5 mag to 12 mag; colour excess (CEx) from 0 mag to 9 mag; metallicities from 0.1 Z$_{\odot}$ to 2.5 Z$_{\odot}$ with [$\alpha/Fe$]~$=$~0.0; and mass cluster (M$_{cl}$) was default (from 1 M$_{\odot}$ to 10$^{5}$ M$_{\odot}$). The DM range corresponds to the distances adopted here (500 pc - 2500 pc) and metallicities are the values expected to open clusters in the Milky Way disk. The partial results of \textsc{fitCMD} are also given in Table \ref{tab2} (last four columns)  and Table \ref{tabC2} presents additional parameters estimated by this code.

\subsection{Overall comparison among groups}
\label{clusters}

The results in Table \ref{tab2}, evaluated by us with \textsc{M20 code} and \textsc{fitCMD}, show that the groups found in the neighbourhood of CMa~OB1 have ages ranging from 9~Myr to $\sim$ 650 Myr and the photometric distances (D$_{P}$) varying from $\sim$  550 pc to $\sim$ 1600 pc.  The interstellar visual extinction in the direction of these groups is less than 2~mag,  with photometric values (A$_{VP}$) ranging from 0.27~mag to 1.71~mag. Metallicity obtained from the \textsc{M20 code} varies between -0.16~$<$~[Fe/H]~$<$~0.07, while with \textsc{fitCMD} we obtain values in the range -0.48~$<$~[Fe/H]~$<$~0.28.

Most of the parameters obtained by the two methods are in good agreement and are consistent with each other (see Fig.~\ref{fig5}). The ages of all groups differ by less than 30\% from one method to another. Distances have values determined by both methods with a percentage difference smaller than the other three parameters, only for the groups CMa08, CMa15 and CMa23 this difference is greater than 10\%. However,  CMa00 and CMa05 have incompatible distances between the two methods, mostly because both codes determined small errors for these measurements. The visual extinction of all groups is also compatible between the methods, except for CMa03 which has A$_{V}$ determined by \textsc{M20 code} almost 45\% higher than the value estimated by \textsc{fitCMD}.

\begin{figure*}
\begin{center}

\includegraphics[width=2.0\columnwidth, angle=0]{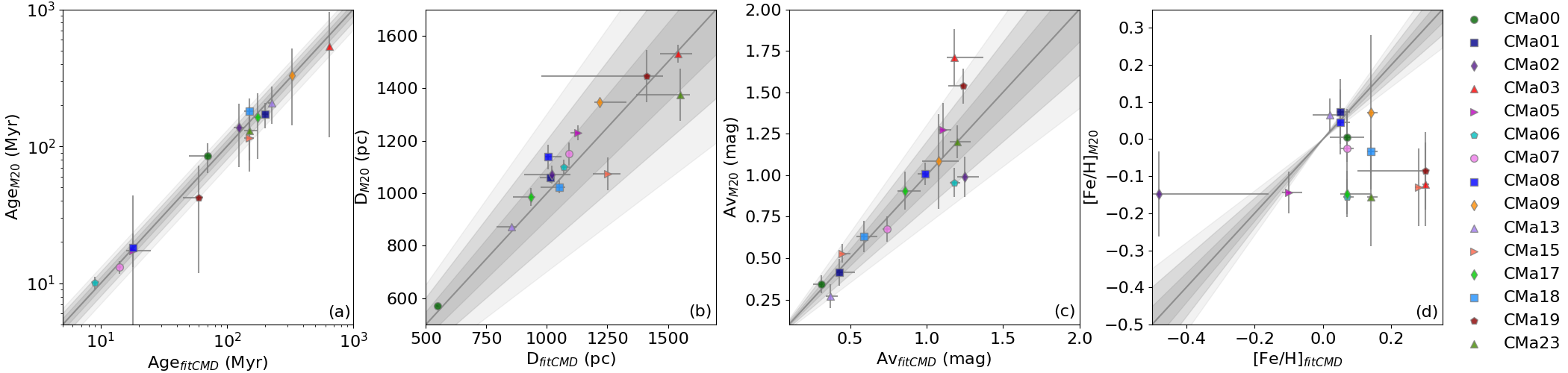}

\end{center}

\caption{Comparison between fundamental parameters obtained using both isochrones fitting methods, \textsc{M20 code} and \textsc{fitCMD}: (a) ages; (b) distances; (c) visual extinction and (d) metallicities  for 15 groups. The dark grey line represents equality between the two methods while the hatched areas show differences of 10\% (grey), 20\% (light grey) and 30\% (very light grey) from this line.}

\label{fig5}

\end{figure*}

Metallicity is the parameter that has the largest differences between values determined by the two methods. This can be explained in part by the use of a prior in metallicity in \textsc{M20~code}, which is based on the metallicity gradient of the Galaxy (Appendix \ref{Hektor} and \citealt{OCCAMgradient20,2020MNRAS.499.1874M}), while for for \textsc{fitCMD} no prior is used. The range used to fit this parameter is -0.9~$<$~[Fe/H]~$<$~0.7 for \textsc{M20~code} and of -1.0~$<$~[Fe/H]~$<$~0.4 for \textsc{fitCMD} (typical for open clusters, see Sects. \ref{sec:fund_param} and \ref{Charles}). It is important to note that due to the low sensitivity of the data to metallicity, the M20 code fits values distant from the gradient used as prior only if the weight of the evidence is significant. However, the values found by both methods are compatible for most groups, since the errors are significantly large in either case. Moreover, when looking at the other panels in Fig. \ref{fig5}, we notice that the different metallicity values do not significantly affect age and distance determinations and may only have a small effect on the determination of visual extinction of some groups such as CMa02, CMa03 and CMa019, which are still within 30\% differences when taking uncertainties into account.


\begin{figure}
\begin{center}

\includegraphics[width=1.0\columnwidth, angle=0]{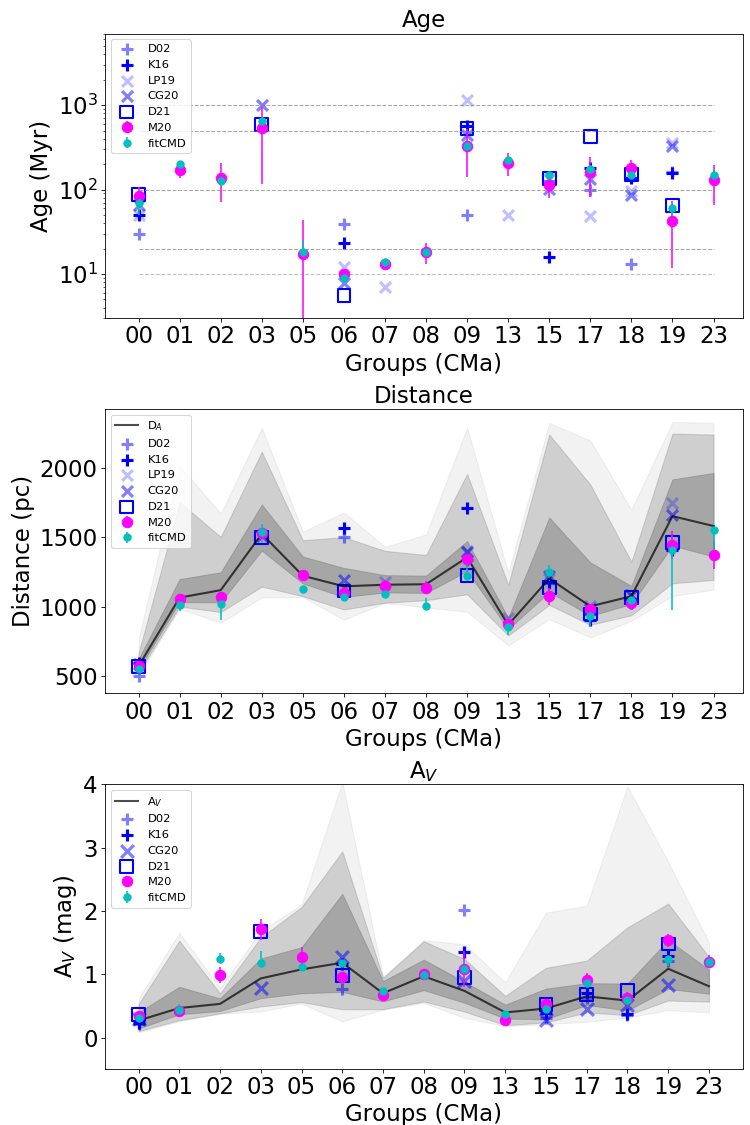}

\end{center}

\caption{Comparison of ages (top panel), distances (middle panel) and  visual extinction (bottom panel)  for 15 groups in the line of sight CMa~OB1 (D$_{A}$ $<$ 1 700 pc) derived using both: \textsc{M20 code} - magenta points - (Sect. \ref{Hektor}) and \textsc{fitCMD} - cyan points - (Sect. \ref{Charles}). Parameters from the literature: \citet[][D02]{2002A&A...389..871D}, dark, and \citet[][K16]{2016A&A...585A.101K}, light blue crosses; \citet[][LP19]{2019ApJS..245...32L}, light, and \citet[][C20]{2020A&A...640A...1C} blue~``X'' and \citet[][D21]{2021MNRAS.504..356D}, blue squares. Grey dashed lines  indicate ages of 10 Myr, 20 Myr, 100 Myr, 500 Myr and 1 Gyr.
Mean values of astrometric distances (D$_{A}$) and mean visual extinction from 3D maps (A$_V$) are presented by black lines and their errors (1$\sigma$, 2$\sigma$ and 3), see Sect. \ref{subsec:seachrclusters}) are presented by grey shaded areas.}

\label{fig6}

\end{figure}

To validate the ages, distances and visual extinction determined by the isochrones fittings, in Fig.~\ref{fig6} we compared them with parameters available in the literature \citep[][]{2002A&A...389..871D, 2016A&A...585A.101K, 2019ApJS..245...32L,2020A&A...640A...1C}, for 10 open clusters compatible with our groups (see Sect. \ref{subsec:knowclusters}), including parameters of 8 clusters determined by \citet{2021MNRAS.504..356D} using \textsc{M20 code} in objects with membership probabilities determined by \citet{2020A&A...640A...1C}. Astrometric distances and visual extinction from 3D maps of dust extinction for each group are also compared in middle and bottom panels, respectively. 

Four of our groups, CMa05, CMa06, CMa07 and CMa08, have young populations with ages under 20 Myr. Most groups have intermediate ages, older than 100 Myr and younger than 500 Myr. CMa03 is the older cluster with $\sim$ 600 Myr. This parameter is the one with the greatest divergence among values from the ones in the literature catalogues for most of the clusters compared to our groups. However, almost all groups have some ages similar to those found in the literature, except CMa07 and CMa13 having discrepant values provided for their corresponding clusters, which are present in only in \citet{2019ApJS..245...32L}.

The isochronal distances of our groups have similar values determined by the two methods and are also within the range of astrometric distances (Table \ref{tab1}). While only the A$_{V}$ determined for CMa02, by both methods and the A$_{V}$ from \textsc{M20 code} are outside the range of visual extinction obtained from the 3D maps. Both parameters of our groups are also in good agreement with the literature. Although VdB 92, comparable with CMa06, has distance overestimated in the two catalogues before {\it Gaia}. The cluster corresponding to CMa09, NGC~2318, also has overestimated distance by  \citet{2016A&A...585A.101K} and it is also the only one with the highest visual extinctions provided by the two catalogues before {\it Gaia}.

It is important to emphasise that all recent parameters estimated by \citet{2021MNRAS.504..356D}, using one of the same code than us (\textsc{M20 code}), are similar to those found in our groups, except for the ages of VdB 92 ($\sim$ 6 Myr) and NGC 2323 ($\sim$ 470 Myr), reinforcing that some groups were well detected by \texttt{HDBSCAN}, selecting members that preserve the same characteristic that cluster known in the literature with membership determined by other method.

These comparisons helped us better understand the relationship between our groups and some literature clusters, as discussed in the following section.

\section{Stellar content}
\label{sec:comparison}

We employ parameters evaluated using astrometric and photometric data (see Tables \ref{tab1} and \ref{tab2}) to  recognise and confirm that some groups found in this work are, in fact, clusters already known in the literature or new candidates.

\subsection{Known open clusters}
\label{subsec:knowclusters}

We consider that our groups are previously known clusters if they have the entire astrometric information compatible with objects present in at least one stellar cluster catalogue discussed in Sect.~\ref{subsec:lit_clusters}, preferably in the catalogues with parameters determined from {\it Gaia}~DR2 data \citep{2019ApJS..245...32L,2020A&A...640A...1C}. Among the 15 groups discussed in Sect.~\ref{clusters},  we confirm that 10 of them are associated with already known open clusters.

The  CMa00, CMa03, CMa09, CMa17, CMa18 and CMa19 groups were recognised to be the clusters Alessi~21, Gulliver 13, NGC~2318, NGC~2323, NGC~2343 and NGC~2335  respectively, having all 5 parameters similar to those provided in both catalogues based on {\it Gaia} DR2 data. In addition, similar ages were found in at least 2 catalogues, considering also \citet{2002A&A...389..871D}, \citet{2016A&A...585A.101K} and \citet{2021MNRAS.504..356D}, and helped us to confirm them, except for Gulliver 13 and NGC~2335 having similar ages estimated only by \citet{2021MNRAS.504..356D} than CMa03 ($\sim$ 600 Myr) and CMa19 ($\sim$ 50 Myr), respectively. Gulliver 13 has 1 Gyr according to \citet{2020A&A...640A...1C}, and NGC~2335 has about 160~Myr provided by both catalogues before {\it Gaia} and $\sim$ 350 Myr by \citet{2019ApJS..245...32L} and \citet{2020A&A...640A...1C}.

The CMa06 and CMa15 groups also have all 5 parameters compatible with the vdB 92 and NGC 2353 clusters, respectively. However, their number of objects and spatial distribution suggests that these clusters are only part of their corresponding group (see Fig.~\ref{fig3}). CMa06  has 377 members with P $\geq$ 50\% distributed across the molecular cloud at CMa~R1, in which at least almost 270 of them have P $>$ 75\% membership in the star-forming region determined by Bayesian method (see Sect. \ref{subsec:spluscount}), while vdB 92 has less than 200 members in all catalogues.  Thus, we consider CMa06 as CMa~R1 population, containing vdB 92. 
On the other hand,  CMa15 appears to have two different spatially distributed subgroups, which we call CMa15-A and CMa15-B, with the same proper motion and parallax showing a slightly elongated distribution (see Fig.  \ref{fig2}).  The most populated subgroup (CMa15-A) is clearly NGC~2353, following the same steps to recognise other known clusters and including ages found in two catalogues that are the same as we found for CMa15 ($\sim$~125~Myr). However, a more detailed study of this group is necessary to state that CMa15-B is, in fact, part or not of NGC~2353.

 Finally, CMa07 and CMa13 has astrometric parameters very similar to FOF~2216 and FOF~2302 from \citet{2019ApJS..245...32L}. Although this catalogue does not have photometric parameters for comparisons, CMa13 has almost the same amount of members of FOF~2302, so it was enough for us to consider that both are the same object. On the other hand,  CMa07 has less members (26 stars with P $\geq$ 50\%) than the cluster (54 stars) and it is spatially smaller (see Fig. \ref{fig3}). Furthermore, although the age of FOF~2216 derived by \citet{2019ApJS..245...32L} is younger than our estimates for CMa07,  the ages assigned to both are less than 20 Myr, contributing for us to consider that CMa07 should be part of FOF~2216.

\subsection{New cluster candidates}
\label{subsec:new}

In addition to the known clusters, another 5 groups with D$_{A}$~$<$~1700~pc (CMa01, CMa02, CMa05, CMa08 and CMa23) are not reported in any catalogue discussed here. Thus, we classified them as new cluster candidates. We suggest that CMa 08 and CMa 05 ($<$ 20 Myr) are young stellar clusters. CMa01, CMa02, and  CMa23  have ages varying between 125 Myr and 200 Myr and have been recognised by us as in the intermediate evolutionary stage. Therefore,  in a next work, we propose a more detailed characterisation of these groups,  based on multi-wavelengths data from the S-PLUS collaboration, to confirm that these are new open clusters.

Moreover, there are 4 distant groups (D$_{A}$ $>$ 1700 pc), which are not related to clusters known in the literature: CMa04, CMa11, CMa20, and CMa26 (see Appendix \ref{subsec:unreliabe_cluster}).  Most of them have a very asymmetric distribution of parallaxes, indicating an absence of objects with $\varpi$~$<$~0.4~mas (see  Fig. \ref{fig2} (c)). Although these groups can not be connected to CMa~OB1 association, some of them may be new open clusters. However, it is necessary to apply our tool considering other parallax and spatial ranges to confirm them. Astrometric parameters of these clusters are present in Table \ref{tabA1}.

\begin{figure*}
\begin{center}

\includegraphics[width=1.4\columnwidth, angle=0]{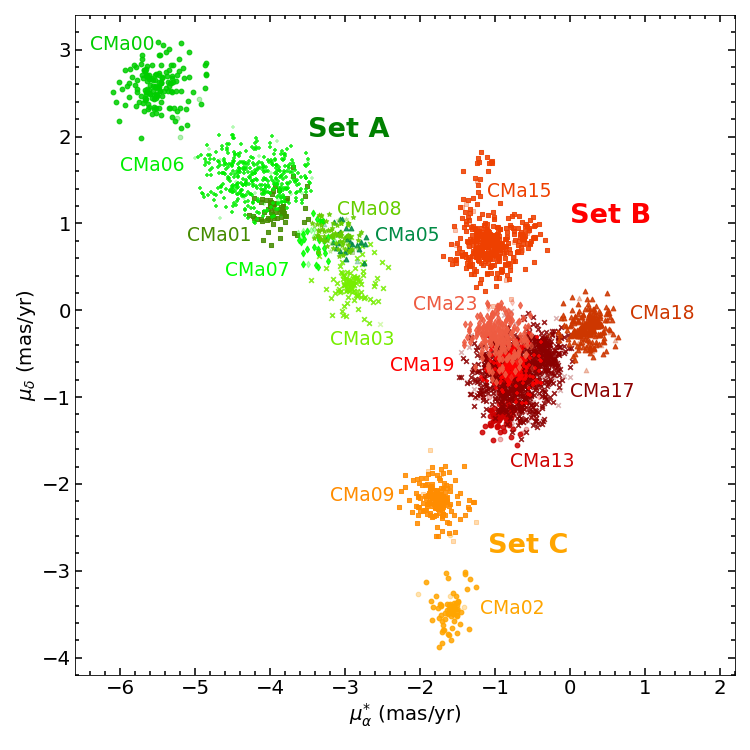}

\end{center}

\caption{Proper motion distributions of 15 groups with D$_{A}$ $<$ 1700 pc in CMa~OB1 association region. Green points represent groups in {\it Set A}, red points are {\it Set B} groups and orange points are {\it Set C}.  Light symbols present members P $<$ 50\%.}

\label{fig7}

\end{figure*}

\section{Discussion} 
\label{sec:discussion}

\citet{2019ApJ...879..125Z} estimated an astrometric distance, based on {\it Gaia} DR2 data,  of about 1200 pc for CMa~OB1 and \citet{2019A&A...628A..44F} showed that this association is formed by a  shell with a diameter of $\sim$ 60 pc, containing the CMa~R1 star-forming region, nested to a supershell with 140 pc diameter (see left panel in Fig. \ref{fig1}). Moreover,  \citet{2019A&A...628A..44F} suggested that three supernova explosions occurred in the region  $\sim$ 1 Myr, $\sim$ 2 Myr, and $\sim$ 6 Myr ago.  \citet{2018A&A...609A.127S} and \citet{2009A&A...506..711G} also noticed a mixture of populations younger than 5 Myr with one older than 10~Myr in CMa~R1, on the west side of CMa~OB1. 

Starting from the assumption that the groups composing the association must have similarities between them. Following the example of both families found by \citet{2020MNRAS.495.1349Y}  at the Sco OB1 association, we classed the 15 groups in the CMa~OB1  neighbourhood (D$_{A}$ < 1700~pc),  into three sets of groups, according to their proper motion (see Fig. \ref{fig7}). Then, we evaluate each set taking into account the distances, ages, spatial distribution and  tangential velocity of each group (see Fig. \ref{fig8}).  

Figure \ref{fig8} shows these sets separately.  {\it Set A}, {\it Set B} and {\it Set C} are shown respectively in the panels on the left, middle and right. The proper motion distribution of each one is presented considering the astrometric distances of the groups in the first row and the ages in the second row. The spatial distribution and the tangential velocity vectors of each group are presented in the third and fourth rows, respectively.  Each row follows a specific colour-map.

Bearing in mind that the  {\it CMa Supershell} can also be at least 140~pc deep, we consider as possible CMa~OB1 (D $\sim$ 1200 pc) contents, groups with astrometric distances in a range from 1000~pc to 1400~pc, in order to ensure that all groups having distances compatible with the association were not previously discarded. They are presented in the first row of Fig. \ref{fig8}, as blue groups. The foreground groups having D$_{A}$~$<$~1000 pc (cyan), and the background groups with D$_{A}$~$>$~1400 pc (magenta) were also analysed.
Taking into account different evolutionary stages among our groups, we also highlight, in different colours, four main ranges of ages in the second row of Fig. \ref{fig8}: very young groups, with less than 20 Myr (T0-20 - dark blue symbols); young groups, from 20~Myr to 100~Myr (T20-100 - blue); in an intermediate evolutionary stage, from 100~Myr to 500~Myr (T100-500 - cyan); and older groups, over 500 Myr (T500-1000 - magenta). Note that there are no objects older than 1 Gyr.  In the third row, the groups were coloured according to their location in relation to {\it CMa Shell} and {\it CMa Supershell} \citep{2019A&A...628A..44F}. And although we have no radial velocities for most stars (see Table \ref{tabC1}), to get some insight of the internal structure and internal kinematics of the groups we highlighted the tangential velocity vectors, in the last row, according to the 2-D velocity dispersion (see Table. \ref{tab1}). 

Hereafter, we evaluate each set, from Figs. \ref{fig7} and \ref{fig8}, until we find the CMa~OB1 groups contents and we discuss the formation history of the association based on these groups and previous works.

\subsection{Set of groups}
\label{subsec:families}

Considering the proper motion distribution in Figure \ref{fig7} it is possible to notice that {\it Set A}  forms an elongated structure composed of 7 groups with CMa00  at the top, CMa03  at the bottom and other groups, CMa01, CMa06, CMa05, CMa07 and CMa08, between the two (green symbols). {\it Set B} has 6 groups in which CMa13, CMa17, CMa19 and CMa23 are mixed, constituting the main structure with CMa15 above them and CMa18 on the right side (red symbols). The other two groups CMa02 and CMa09, having lower values of proper motion in declination ($\mu_{\delta}$ $<$ -1.5 mas yr$^{-1}$) are included in {\it Set C} (orange symbols).

\begin{figure*}
\begin{center}

\includegraphics[width=1.82\columnwidth, angle=0]{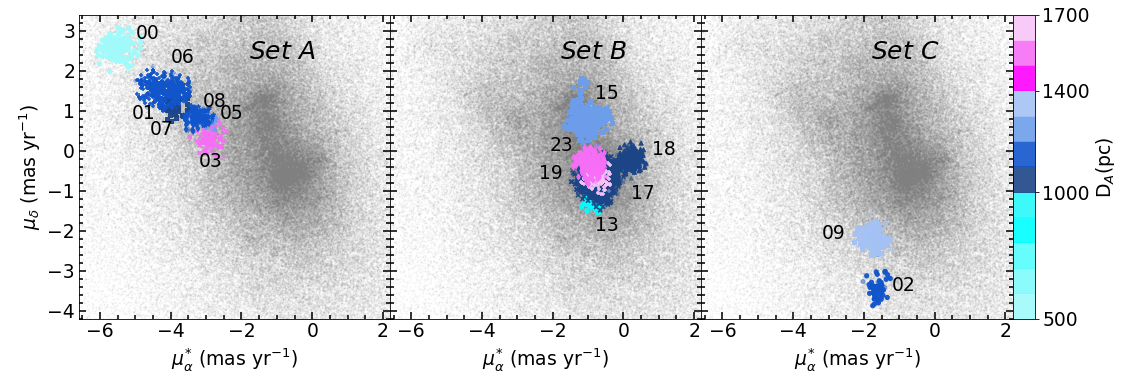}
\includegraphics[width=1.82\columnwidth, angle=0]{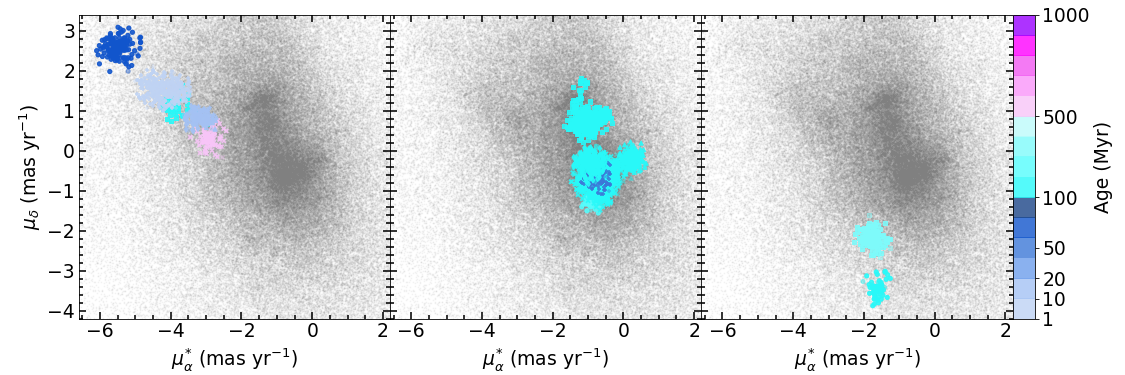}
\includegraphics[width=1.82\columnwidth, angle=0]{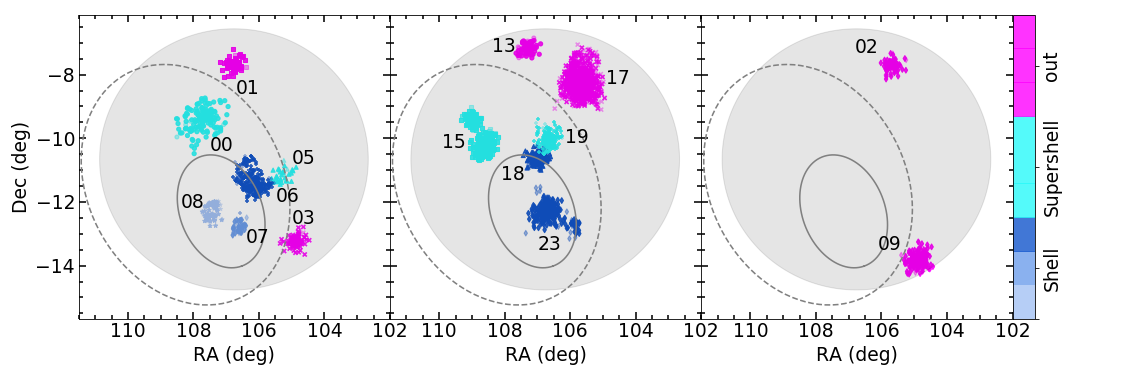}
\includegraphics[width=1.82\columnwidth, angle=0]{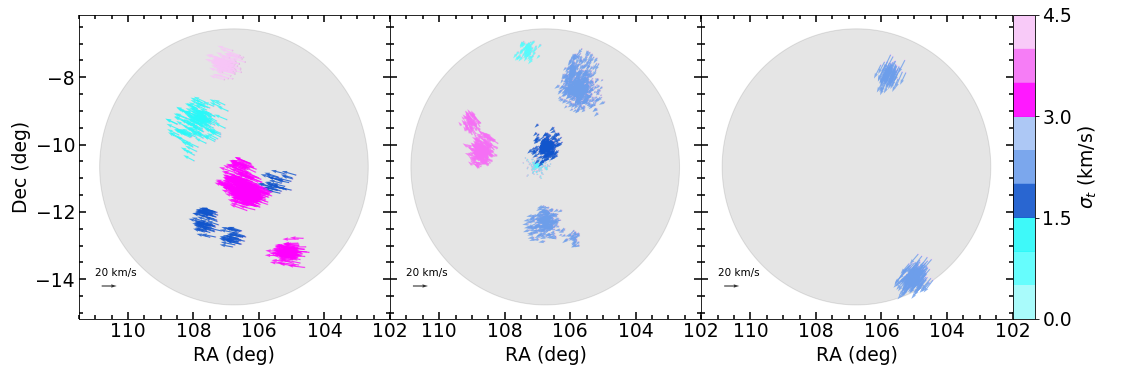}

\end{center}

\caption{Proper motion and spatial distributions of 15 groups with D$_{A}$ $<$ 1700 pc in each  set of groups: {\it Set A} (left column), {\it Set B} (centre column) and {\it Set C} (right column). First row: Proper motion  highlighted according the distances of each group. Cyan groups are foreground stars with  500 $<$ D$_{A}$ (pc) $<$ 1000,  blue points are at the same distance of CMa~OB1, 1000  $\leq$ D$_{A}$ (pc) $\leq$ 1400 and magenta points are background stars (1400 $<$ D$_{A}$ (pc) $<$ 1700). Second row: Proper motion distribution of these groups highlighting their ages.  Blue groups have ages younger 100 Myr, cyan groups have 100 $<$ Age (Myr) $<$ 500 and magenta groups between 500 $<$ Age (Myr) $<$ 1000. Third row: Spatial distribution are coloured according to the {\it CMa Shell} (grey line) and {\it CMa Supershell} (dashed grey line) position \citep{2019A&A...628A..44F}. Groups in the {\it CMa Shell} are blue, highlighting CMa08, almost in the centre, light-blue, and groups in edge, dark-blue. Groups inside {\it CMa Supershell} are cyan and out of both structure are magenta. Fourth row:  Projection of tangential velocities indicating their direction and magnitude according to the tangential velocity dispersion in cyan (0.0 $<$ $\sigma_{V2D}$ (km/s) $<$ 1.5), blue (1.5 $<$ $\sigma_{V2D}$ (km/s) $<$ 3.0) and magenta (3.0~$<$~$\sigma_{V2D}$~(km/s)~$<$~4.5).
}

\label{fig8}

\end{figure*}

{\it Set A} has 5 groups (CMa01, CMa05, CMa06, CMa07, and CMa08)  with distances comparable to the CMa~OB1 association (1000 $<$ D$_{A}$ pc $<$  1400), and there is one (CMa00) in the foreground and the other (CMa03) in the background. {\it Set A} is younger than the other two sets, having four groups (CMa05, CMa06, CMa07 and CMa08) younger than 20~Myr. Although it also has the older group, CMa03, the only one T500-1000. Other two groups, CMa00 and CMa01 are in distinct evolutionary stages: T20-100 and T100-500, respectively. In the projected spatial distribution, the CMa06 and CMa05  groups appear to be connected to each other at the edge of the {\it CMa Shell} (see Figs. \ref{fig8} and \ref{fig9}). The CMa07 and CMa08 groups are on the east side of them, within this structure  with CMa08 almost in the centre. CMa00 and CMa01 are in the north of them, out of the {\it CMa Shell}. CMa00, as well as CMa05, is inside the {\it CMa Supershell}, while CMa01 and CMa03 (in the south-west) are out of both structures. As expected by own set of groups selection criteria, based on their proper motion, we can see by the projection of the  tangential velocity vectors (last panel in the left side of Fig. \ref{fig8}), that almost all groups of this set are moving in the same direction and in a similar tangential velocity ($\sim$ 20 km/s), see Table \ref{tab1}. However, CMa00 has lower 2-D velocity dispersion ($\sigma_{V2D}$~$\sim$~1~km/s). CMa05, CMa07 and CMa08  have intermediate $\sigma_{V2D}$ (between 1.5 and 2.0 km/s), while CMa03 and CMa06 have 3.0~$<$~$\sigma_{V2D}$~(km/s)~$<$~3.5. CMa01 is the only group with $\sigma_{V2D}$ $>$ 4.0 km/s. 
Half of {\it Set B} (CMa15, CMa17 and CMa18) has similar distances to CMa~OB1. CMa19 and CMa23 are background groups,  while CMa13 is in the foreground. This set of groups is mainly composed of groups with intermediate ages between 100 and 500 Myr. Only CMa19 is T20-100. In concern to location, CMa23 is the only group whose most stars are inside the {\it CMa Shell} while CMa18 is on the edge, following CMa06, with some overlapping stars from CMa19. This group is inside {\it CMa Supershell} with CMa15 on the opposite side, while CMa13 and CMa17 are outside, in the north of the CMa~OB1 gas structures. The groups in this set have smaller tangential velocities than the other two sets, about 6~km/s, except for CMa18 having V$_{t}$ $<$ 2 km/s, which makes it difficult to identify which preferred direction it is moving.  On the other hand, while CMa15 seems to move in the north-east direction, as groups from {\it Set A}, other 4 groups are moving in the south-east direction. 
CMa18 and CMa13 are the only ones with 2-D velocity dispersion less than 1 km/s. CMa17, CMa 19 and CMa23 have $\sigma_{V2D}$ $\sim$ 2 km/s and CMa15 have higher $\sigma_{V2D}$, about 3.5 km/s.

\begin{table*}
\caption{Parameters of CMa~OB1 contents.}
\begin{center}
{
\small
\begin{tabular}{lcccccccccccc}

\hline 

Group   & N$_{50}^{(a)}$ &  R.A.$^{(b)}$  &    Dec.$^{(c)}$    &  Age$^{M}_{(d)}$    &  Age$^{f}_{(e)}$      &  D$_{A}^{(f)}$      &  A$_{V}^{(g)}$      &  $\mu_{\alpha}\cos\delta^{(h)}$       &   $\mu_{\delta}^{(i)}$      &    $\varpi^{(j)}$   & V$_{2}^{(k)}$ & $\sigma_{V2D}^{(l)}$ \\      
                                                    
   & (stars) &  (deg)  &   (deg)    &  (Myr)    &  (Myr)      &  (pc)      &  (mag)      &  (mas yr$^-1$)      &  (mas yr$^-1$)      &   (mas) & (km/s)  & (km/s) \\ \hline \hline      

CMa05  & 28 & 105.32 & -11.20 & 17 $\pm$ 27 & 16.0$^{+0.5}_{-0.5}$ &  1224$^{+55}_{-139}$ &  1.08$^{+0.38}_{-0.35}$ & -2.98$^{+0.11}_{-0.15}$ &  0.79$^{+0.08}_{-0.14}$ &  0.78$^{+0.10}_{-0.05}$ &  18.5 & 1.9\\
CMa06  &  377 & 106.11 & -11.45 & 10.1 $\pm$ 1.0 & 9.0$^{+0.5}_{-0.5}$ &  1147$^{+77}_{-133}$ &  1.18$^{+0.46}_{-1.09}$ & -4.20$^{+0.35}_{-0.38}$ &  1.52$^{+0.19}_{-0.21}$ &  0.85$^{+0.09}_{-0.07}$ &  24.3 & 3.4\\
CMa07  &  26 & 106.60 & -12.79 & 13.1 $\pm$ 1.5 & 14.0$^{+0.5}_{-0.5}$ &  1159$^{+55}_{-69}$ &  0.70$^{+0.13}_{-0.19}$ & -3.36$^{+0.20}_{-0.08}$ &  0.79$^{+0.13}_{-0.13}$ &  0.84$^{+0.04}_{-0.06}$ & 19.2 & 1.6\\
CMa08  & 64 & 107.43 & -12.30 & 18 $\pm$ 5 & 18.0$^{+1.0}_{-4.0}$ &  1162$^{+62}_{-62}$ &  0.97$^{+0.22}_{-0.26}$ & -3.13$^{+0.19}_{-0.17}$ &  0.87$^{+0.13}_{-0.13}$ &  0.84$^{+0.05}_{-0.06}$ & 18.2 & 1.6\\\hline

\end{tabular}
}
\label{tab3}
\end{center}

NOTES:
(a) Number of stars with P $\geq$ 50\% in the group;
(b) Right ascension (ICRS) at Ep = 2015.5;
(c) Declination (ICRS) at Ep = 2015.5;
(d) Age obtained using \textsc{M20 code} (M), see Sect. \ref{Hektor};
(e) Age obtained using \textsc{fitCMD} ({\it f}), see Sect. \ref{Charles};
(f) Astrometric distance from \citep{bailerjones2018}; 
(g) Visual extinction from three-dimensional maps of dust \citet{bayestar2019};
(h) Right ascension proper motion;
(i) Declination proper motion;
(j) Parallax;
(k) Tangential velocity;
(l) Velocity dispersion in 2-dimension (R.A. and Dec.).

\end{table*}

\begin{figure*}
\begin{center}

\includegraphics[width=0.72\columnwidth, angle=0]{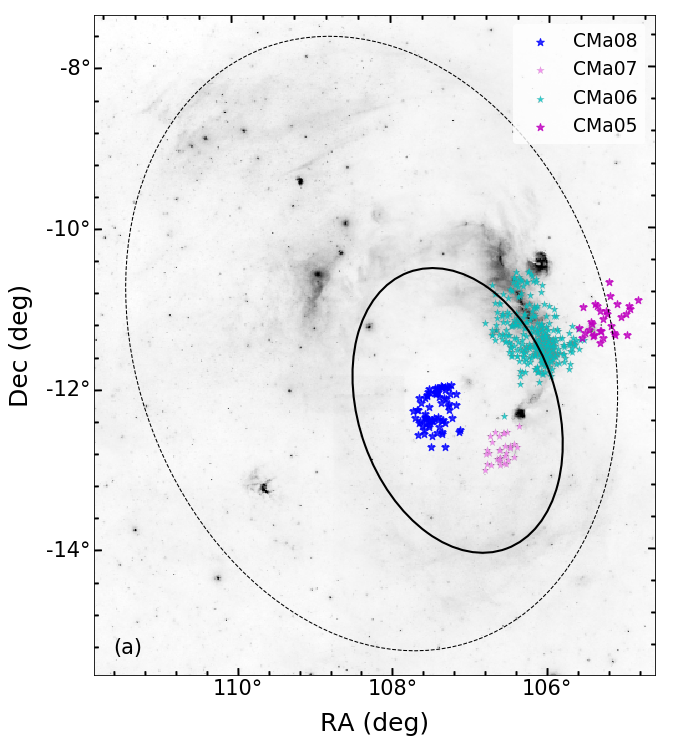}
\includegraphics[width=0.72\columnwidth, angle=0]{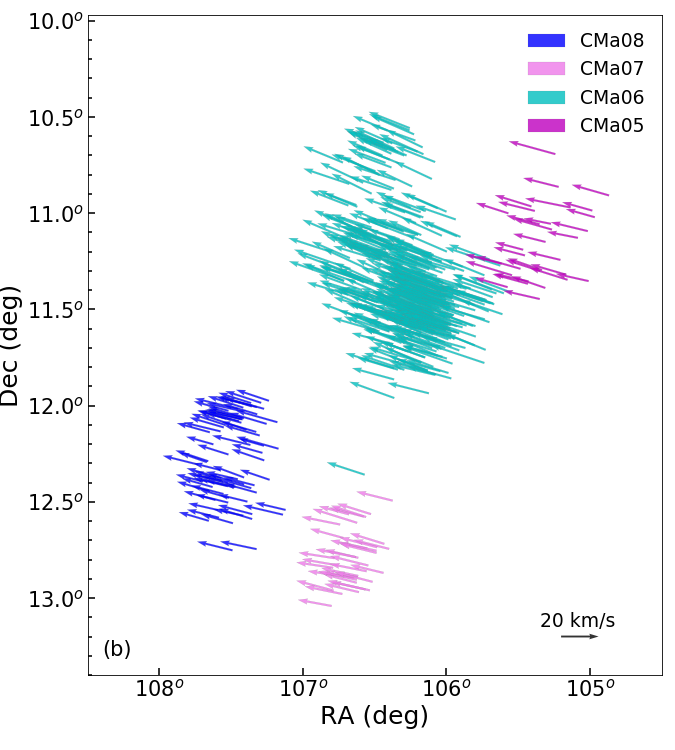}
\includegraphics[width=0.56\columnwidth, angle=0]{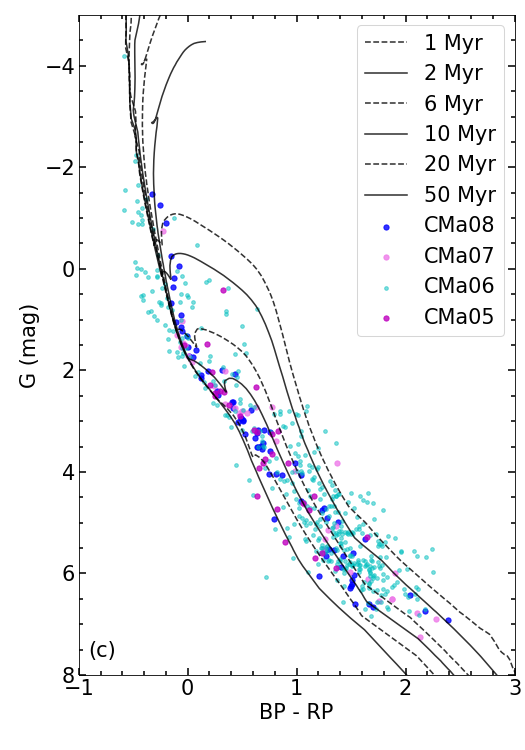}

\end{center}

\caption{ (a) Spatial distribution on  Digital Sky Survey image (640 nm); (b) Tangential velocity vectors; (c) Colour-magnitude diagram of the groups that compose CMa~OB1: CMa05 (magenta), CMa06 (cyan), CMa07 (light pink) and CMa08 (blue).}

\label{fig9}

\end{figure*}

The two groups of {\it Set C} have similar characteristics. Both are at the same range of distance of the Association and have intermediate age, being CMa02 about 200 Myr younger than CMa09. Moreover, both are outside {\it CMa Supershell},  moving in the same direction with tangential velocity about 20 km/s, as well as groups in {\it Set A}, and they have 2.0 $<$ $\sigma_{V2D}$ (km/s) $<$ 3. However, both are quite far from each other, while CMa02 is in the north-west side of the {\it CMa Supershell}, CMa09 is in the south-west. 

\subsection{CMa~OB1 stellar groups}
\label{subsec:CMacontent}

From Figs. \ref{fig7} and \ref{fig8},  we identified four groups (CMa05, CMa06, CMa07 and CMa08) in  {\it Set A} with similar characteristics. A summary of the main parameters of these groups is present in Table \ref{tab3}. These groups are spatially close to each other, at the same distance as the CMa~OB1 association and they seem to be moving together regardless of the other groups. 
Moreover, they are the youngest groups among all in the line of sight of CMa~OB1, with ages younger than 20 Myr. 

CMa06 is the youngest one,  with 9 Myr and 10 Myr, estimated by us using both methods. It is associated with the molecular dense gas present in the star formation region CMa~R1, since this group contains  many stars recognised in CMa~R1, according to the discussion in Sect.~\ref{subsec:spluscount} and Appendix~\ref{sec:know_stars}. The CMa07, CMa08 and CMa05 groups are older than 10 Myr and appear to be unrelated to any of the three small ($<$ 10$^{3}$ M$_{\odot}$) $^{13}$CO clouds surveyed by \citet{2004PASJ...56..313K}.
Moreover, although CMa08 and CMa07 are spatially on the opposite side to CMa05, in relation to CMa06, they are practically overlapping in the proper motion space (see Fig. \ref{fig7}), having very similar tangential velocities and 2-D velocity dispersion, followed by CMa06 with both parameters slightly larger (see Table \ref{tab3} and Fig.~\ref{fig9} (b)), suggesting that the internal structure and kinematics of this group are a little different from the others.

On the other hand, in the panel (c) of the same figure, it is possible to notice a large spread in the CMa06 CMD, mainly for low-mass stars. Most of the objects can be seen between isochrones from 1 Myr to 6 Myr, with part of them ranging up to about 20 Myr. Among the stars with G~$>$~2~mag there is also a spreading, indicating that some bright stars may have an overestimated visual extinction.  

Finally, CMa08 shows a very interesting projected spatial distribution with an almost circular area absent of stars in its inner region. Combined with its proximity to the crossing point predicted for the runaway stars  \citep[see Fig. 5 from][]{2019A&A...628A..44F} makes us suggest that this group may have been the progenitor cluster that expelled the runaway stars, and also any other O-type star once there are no other massive stars in the region, as well as cleared the low-mass stars from its central region.
It is possible that the absence of gas and dust in there and in the nearby group, CMa07, occurred also due to the exhaustion of the parental cloud caused by an older ($>$~10~Myr) generation of massive stars, that are no longer seen in this group now. 

\subsection{CMa~OB1 star forming-history}
\label{subsec:star_forming_region}

Although the monolithic, or multi-monolithic scenario \citep[e.g.][]{1991ASPC...13....3L, 1997MNRAS.285..479B, 2001MNRAS.321..699K} is not the most acceptable for the formation of OB associations today, all characteristics discussed in Sect. \ref{subsec:CMacontent} suggest that the four physical groups found within {\it CMa Shell} have formed in a small region and are expanding, as are Per OB1 and Car OB1  associations \citep{2017MNRAS.472.3887M}. During its expansion process, the most massive stars  appear to have been ejected from CMa08, at the same time that  its  gas is  being exhausted, as well as,  the gas from  other two groups (CMa07 and CMa05) around CMa06. The three supernova explosions reported by \citet{2019A&A...628A..44F}  are feeding back the star formation in CMa06, where gas and dust are still abundant, reinforcing the scenario of multiple star formation announced by \citet{2018A&A...609A.127S}.  
However the widely spaced stars of CMa05 can be interpreted as the older low-mass stars  (from the first episode of the star formation) that are dispersing from CMa06.  The CMa07 and CMa08 groups can also go through this process in a few  Myr, leaving to be a bound system and dissolving in field stars, since only 5\% of the clusters survive their first 20 Myr \citep{2003ARA&A..41...57L,2011A&A...536A..90P, 2009A&A...498L..37P}, remaining bound.  
Furthermore, a small group composed of stars known in the literature, that coincides with the BDSB~96 cluster (not found by {\tt HDBSCAN}, see Appendix A.2) and Sh 2-297 nebula \citep{2020A&A...633A..99C,2012ApJ...759...48M}, is found between CMa07 and CMa06 and it has astrometric characteristics and probably similar ages to these two groups, indicating that it is also part of the Association and is perhaps a remnant population from the distancing of these two groups. On the other hand, CMa06 must have a longer survival time as long as there is star-forming material,  but once finished it, the association reaches dissolution.

This process may have been essential for the maintenance of these physical groups until today. However, a hierarchical scenario can not be totally discarded in the star formation history of association CMa~OB1, since several small groups of bright stars, YSOs, H$\alpha$ emitters and X-ray sources are found mainly associated with edges of {\it CMa Shell} and {\it CMa Supershell} from \citet{2019A&A...628A..44F}, see Appendix~\ref{subsec:subgroups}. 

In particular, \citet{2019ApJS..240...26S} recently searched YSOs using infrared data, Hershel (HI-GAL survey), Spitzer, and 2MASS,  where they found nearly 300 YSOs in a region dubbed CMa-l224 (see Fig. \ref{fig1}~(a)), at the opposite side of CMa06, following the borders of {\it CMa Shell}. They report that the vast majority of these objects are associated with high H$_{2}$ column density regions and they argue that these stars have formed in filaments and become more dispersed over time, reinforcing a hierarchical star formation scenario.

These structures could not be found in this work for two main reasons, the first is the detection limit of {\it Gaia}, for which it is challenging to find embedded stars, and  the second is the configuration of {\tt HDBSCAN} chosen in this work (Sect. \ref{subsec:sampleselection}), which allows finding only groups with more than 30 stars. It is important to note that most of the objects reported by \citet{2019ApJS..240...26S} were detected by infrared surveys, and suffer from high extinction, being immersed in gas and dust and making it difficult their detection by optical surveys such as {\it Gaia}. For example, \citet{2019A&A...630A..90P} found only 98 stars with useful parallaxes determined by {\it Gaia}, among all the 334 H$\alpha$ star emitters found by them. On the other hand, most YSOs are found in small groups, such as \citet{2016ApJ...827...96F}, who found no groups containing more than 25 objects. A brief analysis of the objects studied by these authors in addition to bright sources from \citet{2008hsf2.book....1G}, YSOs from \citet{2015MNRAS.448..119F} and X-rays \citep{2009A&A...506..711G,2018A&A...609A.127S} shows that most of these objects are clustered on the edges of the {\it CMa Shell} but only part of them have good quality optical data from {\it Gaia} DR2. However, we found 4 subgroups containing between 6 and 14 known stars, having parameters similar to those of our CMa~OB1 groups (see Appendix \ref{subsec:know_stars_membership}), giving us a strong indication that although these structures are not large bound groups like clusters, they are also part of the Association.

\section{Summary and Conclusions}
\label{sec:conclusions}

In order to find physical groups or star clusters in the CMa~OB1 association, we use a clustering  algorithm --- {\tt HDBSCAN} \citet{hdbscan} --- in 5--dimensional parameter space: $\mu_{\alpha}\cos\delta$, $\mu_{\delta}$, $\varpi$, R.A. and Dec.  to search for different stellar groups in regions as large as stellar associations and  to provide their astrometric parameters  and characterise their properties. We applied this code in a sample of about 250 thousand stars with good astrometric and photometric quality from {\it Gaia} DR2, with parallaxes between 0.4 mas and 2.0 mas, covering an area of R $=$ 4.1$^{o}$ centred on R.A.~$=$~106.7$^{o}$ and Dec.~$=$~-10.6$^{o}$,  in which we found 29 groups containing from 31 to 1096 stars.  We also used the astrometric distances derived by \citet{bailerjones2018} and visual extinction computed from 3D maps to help us characterise the stellar groups identified in our study.

Fifteen groups were found by our method  having distances between 570 pc and 1650 pc around CMa~OB1 (1200 pc).  Seven of them have been recognised in the literature as open clusters. CMa06 and  CMa15 contain the known clusters VdB 92 and NGC 2353, respectively, and CMa07 is probably part of a new cluster, FOF 2216, identified by \citet{2019ApJS..245...32L}.  The other 5 groups are unpublished in the literature and they were considered by us as open cluster candidates.  In particular, CMa06 also contains a vast population of young objects associated with the CMa~R1 star formation region, including 166 objects know in the literature, of which 76\% are X-ray sources studied by \citet{2018A&A...609A.127S}, and many of them ($\sim$ 67\%) appear to be very-likely members as discussed by GH21 estimation of membership probabilities. Moreover, by comparing the members of  CMa05 and CMa06 with data from \citet{2018A&A...609A.127S}, we were able to corroborate that 55 X-ray sources confirmed by the authors are in fact members of CMa~R1 (CMa06). In addition, 57  objects classified by them as possible members and two as unknown origin can be reclassified as CMa~R1 members.

We used two different algorithms developed by \citet{2020MNRAS.499.1874M} and \citet{2019MNRAS.483.2758B}, both based on PARSEC isochrones fitting for the  {\it Gaia} magnitudes, to determine the ages of the physical groups found here. These fittings also provided distances, visual extinction and metallicities. The ages determined by the two algorithms were compatible between them for  all 15 groups around CMa~OB1. The photometric distances and visual extinction also  well agree between both estimations and with astrometric distances and mean visual extinction from 3D maps, respectively.  Only CMa02 and CMa03 have overestimated photometric A$_{V}$, obtained  by at least one method.

In order to identify groups membership of the association, we segregate the  15 groups in three distinct set of groups, according  with their proper motion distribution. {\it Set A}  has 7 groups (CMa00, CMa01, CMa03, CMa05, CMa06, CMa07 and CMa08). {\it Set B} is composed by CMa13, CMa15, CMa17, CMa18, CMa19 and CMa23 and {\it Set C} has two groups, CMa02 and CMa09. 

The younger groups in  {\it Set A} (CMa05, CMa06, CMa07 and CMa08)  are considered by us to be associated with CMa~OB1, mainly because they are located within {\it CMa Shell} reported by \citet{2019A&A...628A..44F}. Moreover, CMa06 ($\sim$ 10 Myr) contains the youngest stellar population associated with CMa~R1 \citep[][GH21]{2018A&A...609A.127S}. CMa07 has a smaller (26 stars) and more intermediary-aged population ($\sim$ 14 Myr) in the south of CMa~OB1. The CMa05 (28~stars) and CMa08 (64 stars) are aged about 18 Myr and are on the west and east side of CMa~R1, respectively.

Astrometric and photometric analysis of these 4 physical stellar groups in the association CMa~OB1 helped us to better clarify their star formation history. These relatively close physical groups within the 60 pc shell are consistent with the monolithic model of association formation. However, it does not explain the embedded stellar content associated with gas and dust filaments found by other authors \cite[see][]{2015MNRAS.448..119F, 2016ApJ...827...96F,2019A&A...630A..90P,2019ApJS..240...26S}, mainly at the edges of {\it CMa Shell} on the opposite side of CMa~R1 star formation, where there are no larger stellar physical groups probably due to our selection criteria that are constrained to visible stars.

We suggest that these groups were all born together from a smaller space in the centre of the CMa~OB1 association, such as Per OB1 and Car OB1 associations \citep{2017MNRAS.472.3887M}, and are expanding. A first generation of stars older than 10 Myr gave rise to CMa06, CMa07 and CMa08. The morphology of CMa08 suggests that it is probably the progenitor cluster of massive runaway stars expelled during the following episodes of star formation ($\sim$ 6 Myr , $\sim$ 2 Myr and $\sim$ 1 Myr) in CMa~OB1  \citep{2019A&A...628A..44F}, { phenomena also responsible for expelling the low-mass stars from  CMaO8 centre. CMa06 is immersed in a dense molecular cloud in which both, the first generation stars ($>$ 10 Myr) and the young population ($<$ 5 Myr) born from these recent episodes, coexist according to \citet{2018A&A...609A.127S} and \citet{2009A&A...506..711G} findings based on X-ray data.
On the other hand, both CMa07 and CMa08 are not having their populations renewed, as their gas and dust have already been exhausted during the first star-formation, and they may lose their members in the coming Myr,
since 95\% of the clusters are expected to not survive their first 20 Myr  \citep{2003ARA&A..41...57L,2011A&A...536A..90P, 2009A&A...498L..37P}. As it is happening on the west edge of CMa06, giving rise to CMa05, which are probably older low-mass stars leaving the larger group. CMa~R1 must still form stars for as long as its interstellar material lasts.

In the near future, we intend to find different characteristics in the existing population in the CMa~OB1 association, such as mass segregation or multiple ages from the characterisation of the new cluster candidates and those already studied, using multiband photometry from the S-PLUS collaboration. On the other hand,  astrometric data from next {\it Gaia} data releases combined with radial velocities will be essential to confirm the expansion of younger groups in CMa~OB1, as well as to unravel the mystery of the older population ($> $ 100 Myr) found in the same region as CMa~OB1.

\section*{Acknowledgements}

We thank our anonymous referee for a useful and constructive
comments and suggestions that improved our work.
TSS thanks FAPESP proc. 2018/06822-6,
HDP thanks FAPESP proc. 2018/21250-9,
FAF thanks FAPESP proc. 2018/20977-2,
JGH thanks FAPESP Proc. 2014/18100-4,
VJP thanks FAPESP Proc. 2015/24946-6,
We thank Friedrich Anders, R. Herpich, Luis A. G. Soto \& Ángeles Pérez-Villegasábio for their suggestions and comments on the manuscript.
This research was performed in part using the facilities of the Laborat\'orio de Astrof\'isica Computacional da Universidade Federal de Itajub\'a (LAC-UNIFEI).
This work has made use of data from the European Space Agency (ESA) mission {\it Gaia} (\url{https://www.cosmos.esa.int/gaia}), processed by the {\it Gaia} Data Processing and Analysis Consortium (DPAC,
\url{https://www.cosmos.esa.int/web/gaia/dpac/consortium}). Funding for the DPAC has been provided by national institutions, in particular, the institutions participating in the {\it Gaia} Multilateral Agreement.

\section*{Data Availability}

Most of the underlying data this article is available in the main part of the paper. The tables containing individual stars information (IDs and astrometric parameters from {\it Gaia} and probability membership estimated by us) of the 29 groups that we found using {\tt HDBSCAN} are available in machine-readable form at the CDS and in the online supplementary material, as well as their analysis figures, in PDF format, as presented in Appendix \ref{sec:figures}.



\bibliographystyle{mnras}
\bibliography{bibliography} 




\appendix

\section{\texttt{HDBSCAN} Caveats}
\label{sec:met_constraints}

It can be noted in Fig. \ref{fig2} (a) that groups CMa04, CMa22, CMa24, CMa27 and CMa28 are close to the edge of the sample on the celestial sphere, and in Fig. \ref{fig2} (c) several groups seem to be at the lower limit of parallax adopted in this work ($\varpi$ $>$ 0.4 mas). In order to avoid the bias, introduced by these constraints, in the inferred characteristics of 14 unreliable groups, they were excluded from our analysis (see Sect. \ref{sec:characterization}). 

\begin{table*}
\caption{Astrometric median values of {\it Gaia} 5-dimension parameters, distance and visual extinction of 7 distant possible groups not well determined by \texttt{HDBSCAN} in CMa~OB1 region.}
\begin{center}
{
\begin{tabular}{lcccccccccccc}
\hline 

Group	&	N $^{(a)}$	& N$_{50}$ $^{(b)}$ & P$_{N50}$ $^{(c)}$	&	R.A. $^{(d)}$	&	 Dec. $^{(e)}$ 	&	$\mu_{\alpha}\cos\delta$ $^{(f)}$  &		$\mu_{\delta}$ $^{(g)}$	&   $\varpi$ $^{(h)}$	&	D$_{A}$ $^{(i)}$ &	A$_{V}$ $^{(j)}$	\\

	&	(stars)	& (stars)	&  (\%)	& (deg)	&	(deg) 	&	(mas yr$^{-1}$)	&	(mas yr$^{-1}$)	&  (mas)   &	(pc)	& (mag)	\\ \hline \hline

CMa04 & 132 & 123 & 93 & 102.98 $^{+0.25}_{-0.29}$ & -9.96 $^{+0.17}_{-0.41}$ & -1.25 $^{+0.12}_{-0.21}$ & 1.26 $^{+0.17}_{-0.16}$ & 0.50 $^{+0.06}_{-0.09}$ &  1921 $^{+252}_{-233}$ & 0.96 $^{+0.14}_{-0.23}$  \\
CMa10* & 37 & 35 & 95 & 105.40 $^{+0.06}_{-0.12}$ & -13.56 $^{+0.08}_{-0.18}$ & -0.99 $^{+0.06}_{-0.09}$ & -1.43 $^{+0.08}_{-0.08}$ & 0.47 $^{+0.05}_{-0.05}$ &  2050 $^{+200}_{-185}$ & 0.84 $^{+0.06}_{-0.18}$ \\
CMa11 & 31 & 28 & 90 & 106.53 $^{+0.07}_{-0.11}$ & -7.41 $^{+0.11}_{-0.17}$ & -2.31 $^{+0.12}_{-0.08}$ & 0.02 $^{+0.09}_{-0.14}$ & 0.48 $^{+0.04}_{-0.08}$ &  1976 $^{+212}_{-164}$ & 0.91 $^{+0.18}_{-0.30}$ \\
CMa20 & 258 & 229 & 89 & 106.56 $^{+0.70}_{-0.51}$ & -9.37 $^{+0.26}_{-0.23}$ & -1.26 $^{+0.19}_{-0.15}$ & 0.78 $^{+0.19}_{-0.16}$ & 0.46 $^{+0.04}_{-0.09}$ &  2063 $^{+258}_{-183}$ & 1.15 $^{+0.27}_{-0.48}$ \\
CMa25** & 96 & 88 & 92 & 107.06 $^{+0.12}_{-0.12}$ & -13.22 $^{+0.17}_{-0.12}$ & -1.42 $^{+0.30}_{-0.17}$ & 1.26 $^{+0.18}_{-0.14}$ & 0.46 $^{+0.04}_{-0.06}$ &  2091 $^{+194}_{-173}$ & 1.18 $^{+0.20}_{-0.32}$ \\
CMa26 & 70 & 61 & 87 & 106.45 $^{+0.08}_{-0.16}$ & -12.27 $^{+0.17}_{-0.16}$ & -1.59 $^{+0.12}_{-0.13}$ & 1.17 $^{+0.12}_{-0.10}$ & 0.47 $^{+0.05}_{-0.07}$ &  2030 $^{+241}_{-206}$ & 0.85 $^{+0.08}_{-0.16}$ \\\hline

\end{tabular}
}
\label{tabA1}
\end{center}

NOTES:
(a) Number of stars in the group;
(b) Number of star with membership probability P $\geq$ 50\%;
(c) Percentage of star with membership probability P $\geq$ 50\%
(d) Right ascension (ICRS) at Ep = 2015.5;
(e) Declination (ICRS) at Ep = 2015.5;
(f) Right ascension proper motion;
(g) Declination proper motion;
(h) Parallax;
(i) Astrometric distance from \citet{bailerjones2018}. 
(j) Visual extinction from three-dimensional maps of dust \citep{bayestar2019}.
Probably clusters *Ruprecht 8 
and **NGC 2345.

\end{table*}

This appendix is dedicated to summarising the characteristics of some of the discarded groups that deserve to be studied in a forthcoming work (Sect. \ref{subsec:unreliabe_cluster}). On the other hand, in Sect. \ref{sec:cluster_not_found} we address the issue of some previously known stellar clusters that were missed, probably due to the conservative criteria adopted in our methodology.

\subsection{Unreliable groups detected by \texttt{HDBSCAN}}
\label{subsec:unreliabe_cluster}

 Checking the diagnostic figures (See Appendix D and online supplementary material) we realised that \texttt{HDBSCAN} may fail to characterise groups that have parameters near to the edges of our selection criteria (see Sect. \ref{subsec:sampleselection}). For instance, CMa04 has more than 100 stars and seems to be part of a population that could not be fully identified since some members may be outside the spatial distribution of the sample. Despite the lack of detected objects, this group has more than 85\% of the population showing P $\geq$ 50\%. This high membership probability for a large part of objects in CMa04, as well as for CMa10, CMa11, CMa20, CMa25 and CMa26, indicates they probably are physical groups that deserve to be better studied, considering a different range of the parameters. 

The median values of astrometric parameters from {\it Gaia} DR2; distance from \citet{bailerjones2018}; visual extinction from 3D-maps, and the number of members (total and P $\geq$ 50\%) of these six possible groups are present in Table \ref{tabA1}. However, they cannot be considered the actual parameters of these groups, but only a first guess to help search for the entire group.

Comparing these parameters with previously known clusters (see Sect. \ref{subsec:lit_clusters}) we found that Ruprecht 8 ($\varpi$ = 0.44 mas, \citealp{2020A&A...640A...1C})  and NGC 2345 ($\varpi$ = 0.35 mas) can be related to CMa10 and CMa25, respectively, once our groups have positions and proper motions similar to these clusters. These coinciding results reinforce our hypothesis that these are actual groups. The other candidates: CMa04, CMa11 and CMa26 probably are new stellar clusters, while CMa20 is suggested to be a large moving group, due to its extensive spatial dispersion.

\subsection{Clusters not found by \texttt{HDBSCAN}}
\label{sec:cluster_not_found}

As discussed in Sect. \ref{subsec:lit_clusters}, there are 30 clusters in the literature suggested being related to the CMa~OB1 region. However, \texttt{HDBSCAN} did not find physical groups associated with 19 of these previously known clusters. In fact, most of them (74\%)\footnote[11]{Collinder 466, FSR 1158, FSR 1159, FSR 1160, FSR 1164, FSR 1169, FSR 1178, FSR 1194, FSR 1199, FSR 1200, FSR 1202, FSR 1204, NGC 2349 and NGC 2351.} are not confirmed as cluster by any of the recent catalogues that are based on {\it Gaia} data \citep[e.g.][]{2019ApJS..245...32L,2020A&A...633A..99C}.

Our conclusion is that only 5 previously known clusters were missed by \texttt{HDBSCAN}: FOF 2304 \citep{2019ApJS..245...32L}, and BDSB~93, BDSB~96, FSR~1183 and UPK~452 \citep{2020A&A...633A..99C}. It is possible the some of these clusters were not detected due to our choice of the minimum number of objects constraining the identification as a group ({\tt min\_cluster} $=$ 30). \citet{2020A&A...633A..99C} list 28 and 36 members for UPK~452 and BDSB~93, respectively.  However, only 25 objects in each cluster would be selected according to our criteria (Sect \ref{subsec:sampleselection}). On the other hand, FSR 1183 has 42 members following these criteria, but several of them have low membership probability\footnote[12]{Among the 42 stars in FSR 1183 cluster following our selection criteria, 10 have probability membership P $\leq $ 10\%, 9 are 10\% $<$ P $< $ 50\% and only 23 have P $\geq$ 50\% according to \citet{2020A&A...633A..99C}.}, remaining less than 30 candidates to be selected and identified as a group by the code. Still, its parallax ($\varpi$~=~0.428 mas) is in the limit adopted by us, which is an additional reason of \texttt{HDBSCAN} has missed this cluster. Finally, despite FOF 2304 has 50 members, no data is available for comparison. It is also located on the edge of our sample spatial distribution seeming that less than 30 of its stars coincide with the area studied by us.


\begin{figure*}
\begin{center}

\includegraphics[width=0.36\columnwidth, angle=0]{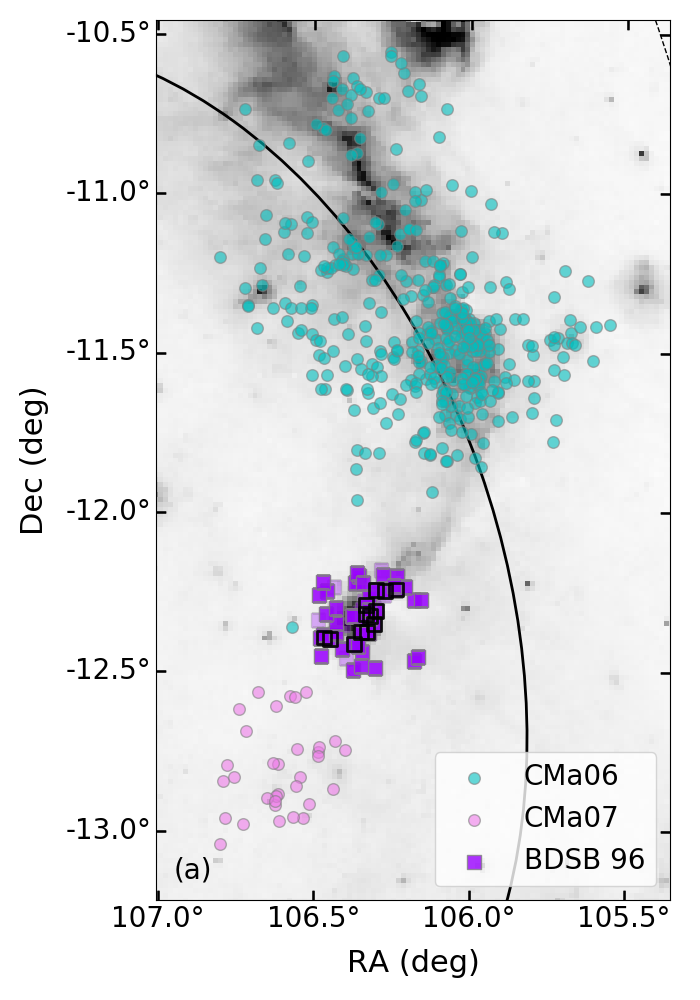}
\includegraphics[width=0.515\columnwidth, angle=0]{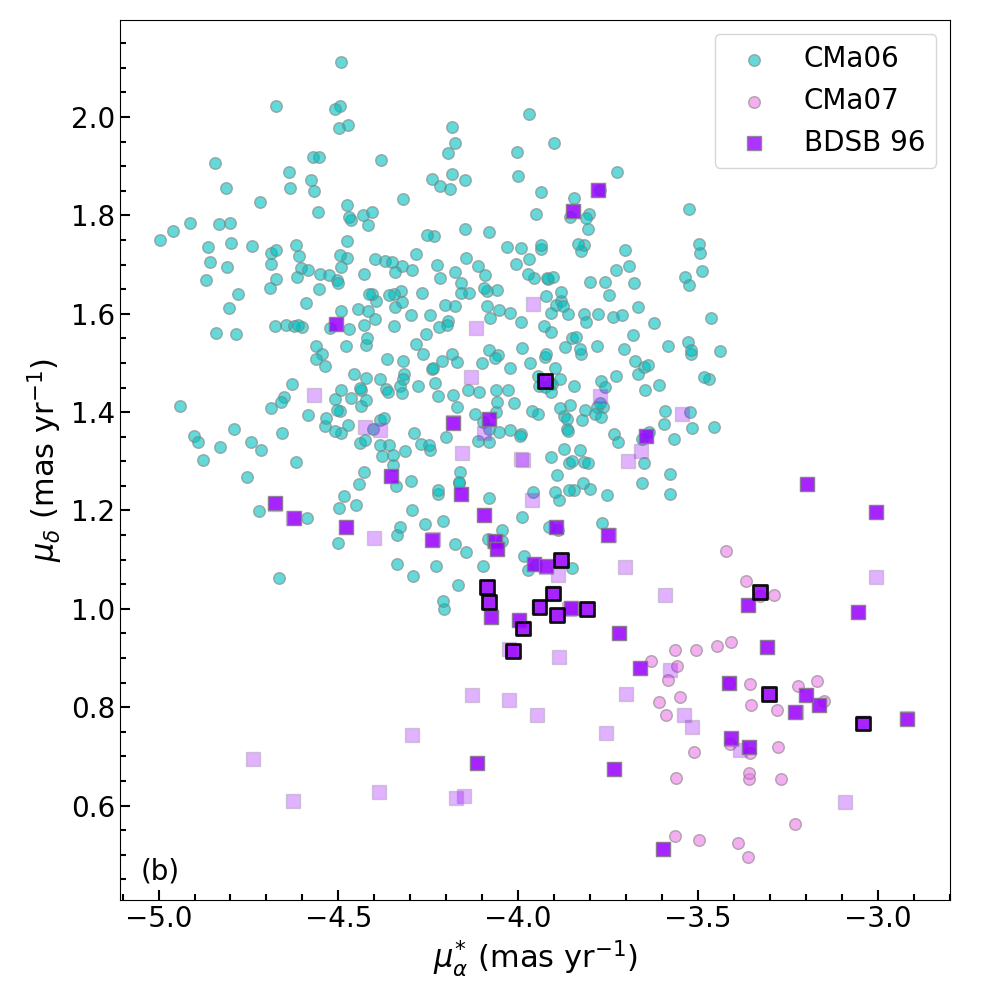}
\includegraphics[width=0.36\columnwidth, angle=0]{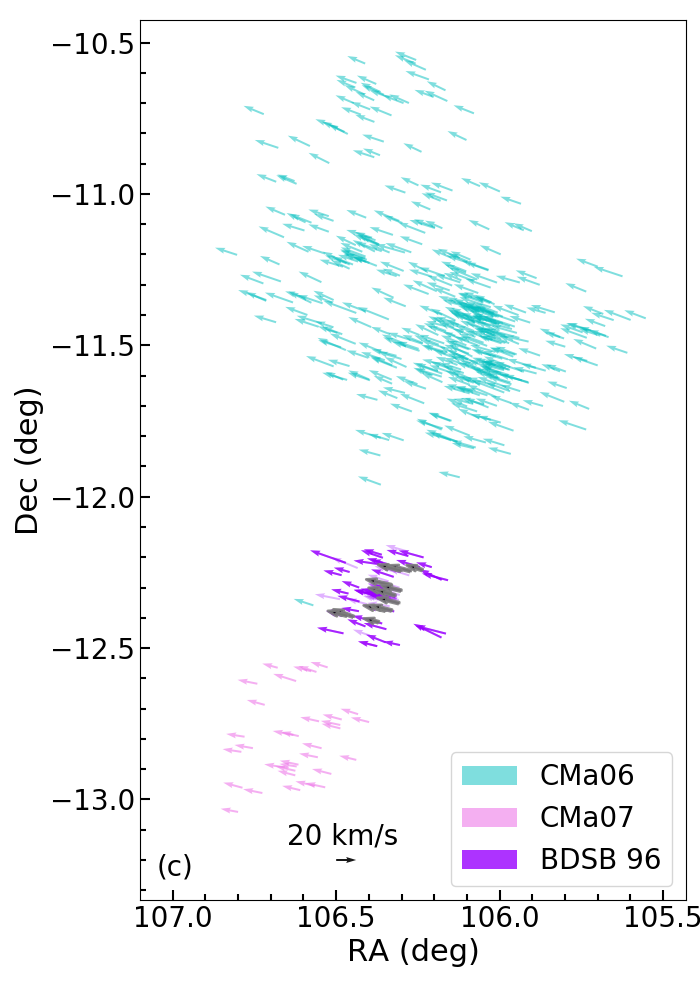}
\includegraphics[width=0.41\columnwidth, angle=0]{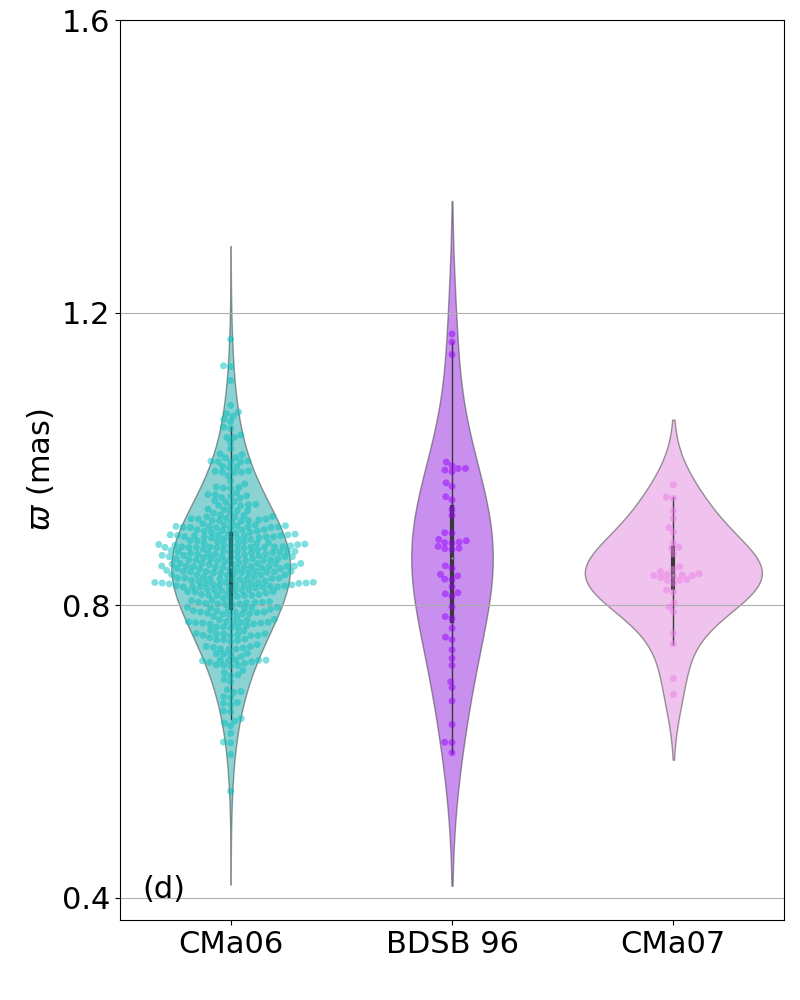}
\includegraphics[width=0.36\columnwidth, angle=0]{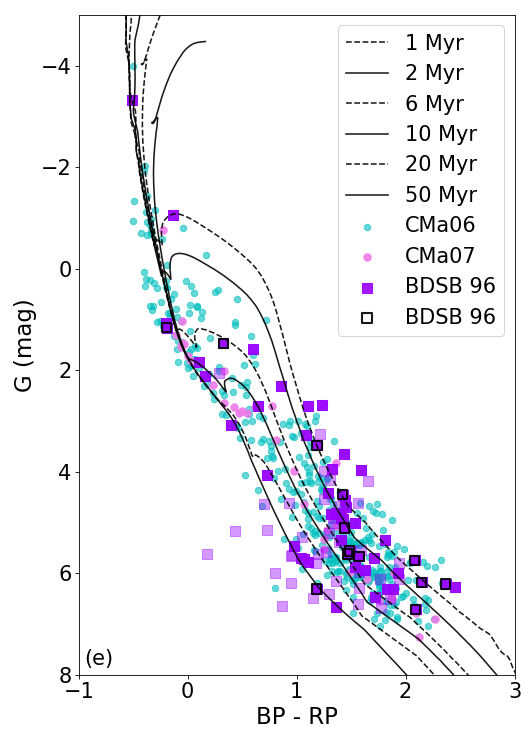}

\end{center}

\caption{(a) Spatial distribution on Digital Sky Survey image  (640 nm) of CMa06 (cyan) and CMa07 (light pink) compared with BDSB~96 cluster \citep[purple squares][]{2020A&A...633A..99C}. Its light points present members with P $<$ 50\% according to these authors and black contours highlights CMaS1 (see Appendix \ref{subsec:subgroups}). Black line show the {\it CMa Shell} edge from \citep{2019A&A...628A..44F}; (b) proper motion distribution; (c) tangential velocity vectors; (d) violin histograms; and (e) colour-magnitude diagram of both groups and BDSB~96.}

\label{figA1}

\end{figure*}


Another way to ensure the detection of these groups by the code is to decrease the values of {\tt min\_cluster}, however, groups with less than 30 objects are difficult to characterise because they are scarce and less cohesive. These kinds of groups deserve more detailed studies that are not the purpose of this work.  Moreover, we are looking for stellar groups at $\sim$ 1200 pc, related to the CMa~OB1 association, which does not seem to be the case of FOF 2304, BDSB 93 and FSR 1183 and UPK 452.

BDSB~96, present in 4 catalogues discussed in Sect. \ref{subsec:lit_clusters}, is the only important cluster, for our analyses,  that should be detected by \texttt{HDBSCAN} but did not. According with \citet{2020A&A...633A..99C} and \citet{2021MNRAS.504..356D} this cluster contains 91 objects, but only 52 of them follow our selection criteria. An intriguing issue is that BDSB~96 (d $\sim$ 1145 pc) shows the spatial distribution between our groups CMa06 and CMa07, and has characteristics similar to both. This led us to question why this cluster was not found by the code. Could it be an important part of the CMa~OB1 star formation scenario that is not being taken into account?

For this reason, we performed a comparison among data of BDSB~96 members with P $\geq$ 50 \% \citep{2020A&A...633A..99C} and our groups CMa06 and CMa07. Although spatially BDSB~96 looks like a physical group, the members have an elongated spread distribution of proper motion, partially covering the parameters of both, CMa06 and CMa07. Less than 20 objects are concentrated around $\mu_{\alpha}\cos\delta$~$\sim$~-4~mas~yr$^{-1}$ and $\mu_{\delta}$~$\sim$~1~mas~yr$^{-1}$, meaning that it is not as cohesive as the groups found by \texttt{HDBSCAN} (see Fig.~\ref{tabA1}~(b)).  The tangential velocity vectors (see Fig.~\ref{tabA1}~(c)) show that the members of BDSB~96 are moving in the same direction and having similar tangential velocity as the groups CMa06 and CMa07. The CMD shown in Fig.~\ref{tabA1}~(e) was obtained using the same visual extinction adopted for our groups. The ages distribution indicates that the BDSB~96 population is also similar to CMa06 and CMa07. Moreover, about 13 members of BDSB~96 are known in the literature as bright stars, H$\alpha$ emitters or YSOs (see discussion at Sect. \ref{subsec:know_stars_membership}). Six of these objects are also related to the HII region Sh2-297 \citep{2012ApJ...759...48M}.

Even relaxing the minimal number of members (e.g. {\tt min\_cluster} $=$ 10) to be searched with \texttt{HDBSCAN}, which would allow identifying less cohesive physical groups, BDSB~96 is not detected. This fact leads us to suggest it is not formed by a single physical group but may be the remaining population from the rupture of the two groups during the expansion process of the entire CMa~OB1 association. Besides it is associated to this complex region, BDSB~96 may have a particular and located star formation history that could be supported by the interaction between the ionising star HD 53623 (spectral type B0 V, see Fig. \ref{fig1}(c)), and the HII region Sh2-297 for which a dynamical age of 1.07~Myr was estimated by \citet{2012ApJ...759...48M}. However, to confirm this hypothesis, a more detailed multiwavelength study is needed, including the embedded objects reported by \citet{2012ApJ...759...48M} and data from {\it Gaia} EDR3 and S-PLUS, obtained for the region between CMa06 and CMa07.

\section{Previously known stellar contents}
\label{sec:know_stars}


\begin{figure*}
\begin{center}

\includegraphics[width=2.0
\columnwidth, angle=0]{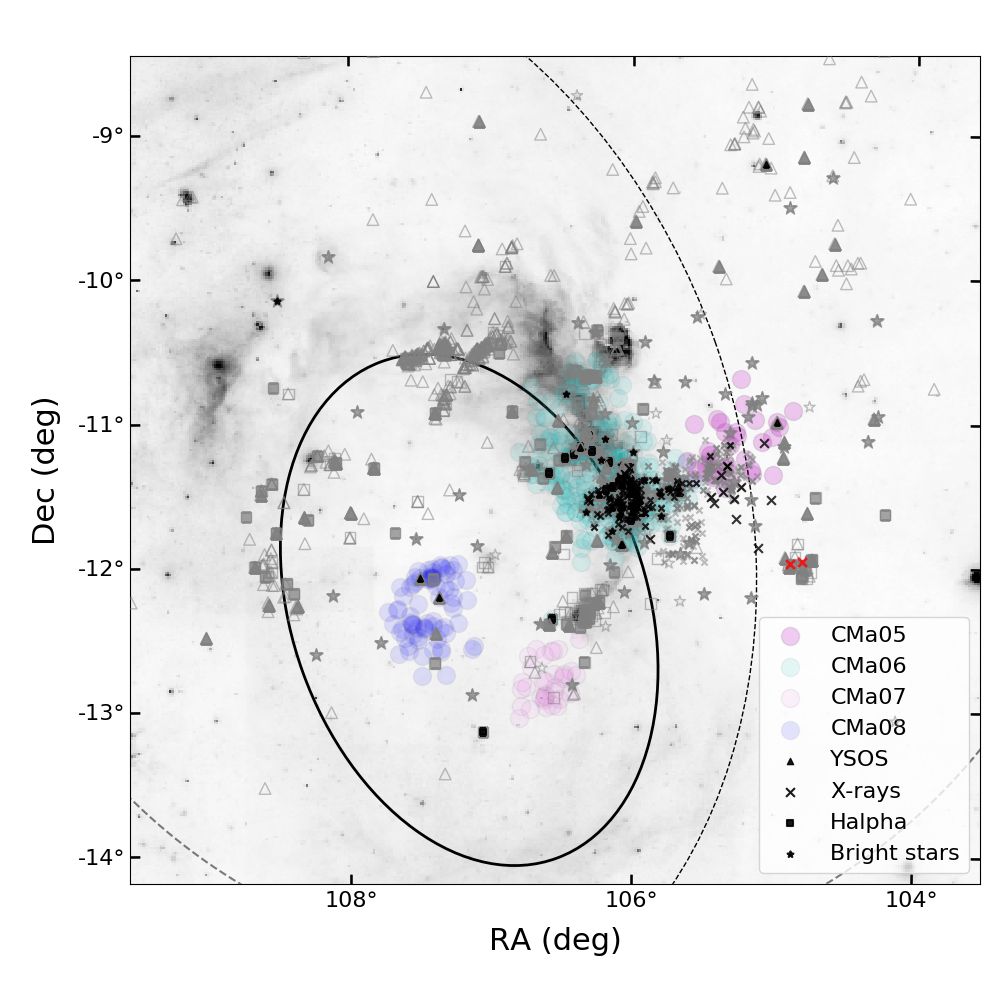}

\end{center}

\caption{Spatial distribution of 4 groups of CMa~OB1 and stars known in the literature on Digital Sky Survey image  (640 nm) of CMa~OB1: Bright stars from \citet{2008hsf2.book....1G} (stars); H$\alpha$ emitters from \citet{2019A&A...630A..90P} (squares); X-ray sources from \citet{2009A&A...506..711G} and \citet{2018A&A...609A.127S} (X); and YSOs from \citet{2015MNRAS.448..119F} and \citep{2016ApJ...827...96F} (triangle)). Grey open and filled symbols represent objects without and with {\it Gaia} counterparts, respectively, while black symbols highlight objects known in the literature present in our groups. Red ``X'' are X-ray sources CMaX06 and CMaX07 from \citet{2009A&A...506..711G}.
}

\label{figB0}

\end{figure*}


Still trying to better understand the stellar population of CMa~OB1, we performed a search in the literature for objects identified in association with this region (see Fig. \ref{figB0}). As mentioned in Sect.~\ref{sec:intro} and Sect.~\ref{subsec:star_forming_region}, part of the CMa~OB1, mainly its famous star formation region CMa~R1, has been widely studied at different wavelengths. In this Appendix we describe the comparison of members of our groups with published catalogues related to CMa~OB1.

\subsection{Known stellar groups membership}
\label{subsec:know_stars_membership}

As compiled by \citet{2008hsf2.book....1G}, there are 114 bright stars in the CMa region, most of them having spectral type B, including objects  from catalogues published by \citet{1974AJ.....79.1022C} and \citet{1999MNRAS.310..210S}. About 334 H$\alpha$ emitters were detected by \citet{2019A&A...630A..90P} using the ESO 1m Schmidt telescope at the La Silla Observatory, covering a field of $\sim$~5$^{o}$~$\times$~5$^{o}$ on the sky in CMa~OB1. \citet{2009A&A...506..711G} found 98 X-ray sources using ROSAT observations and \citet{2018A&A...609A.127S} detected most of these sources and almost 3 hundred new ones using sensitive observations from {\it XMM-Newton} satellite, which gives a total of 387 X-ray sources found in CMa~R1. Additionally, \citet{2015MNRAS.448..119F} characterised 56 young stars in CMa~R1, using multi-objects spectroscopy performed by the Gemini South telescope, based on the optical spectral features, mainly H$\alpha$ and lithium lines. The authors confirmed 41 T~Tauri stars: 7 Classical T Tauri (CTT), 34 Weak~T~Tauri~(WTT) and 15 very likely pre-main sequence stars. \citet{2016ApJ...827...96F} also searched for young stellar objects (YSOs) using the Wide-field Infrared Survey Explorer (WISE). They found 144 Class~I  and 335 Class~II candidates in a FOV of about 10$^{o}$ $\times$ 10$^{o}$, covering the entire region of CMa~OB1. Finally, \citet{2019ApJS..240...26S} identified 293 YSO candidates in the  opposite side of CMa06 ($\sim$~l~=~224$^{o}$) in relation to the {\it CMa Shell} (magenta rectangle in Fig. \ref{fig1} (a)), using data from far-, mid- and near-infrared surveys from {\it Hershell Space Observatory}, {\it Spitzer Space Telescope}, 2MASS and WISE. The authors characterised 210 of these YSOs and classified them into 3 groups: 16 objects with ``envelope-only''; 21 with ``envelope and disk'' and 173 ``disk-only''.

In Fig. \ref{figB0} we present all these objects highlighting with filled grey symbols the stars with {\it Gaia} counterparts that satisfy our selection criteria (Sect. \ref{subsec:sampleselection}) and we use black symbols to represent objects found in our groups. We consider 4 sets of objects based on their characteristics in each survey: bright stars \citep{2008hsf2.book....1G}, H$\alpha$ emitters \citep{2019A&A...630A..90P}, X-rays sources \citep{2009A&A...506..711G,2018A&A...609A.127S}  and YSOs \citep{2015MNRAS.448..119F, 2016ApJ...827...96F,2019ApJS..240...26S}.

The first step in identifying known stars in association with our physical groups was to perform a cross-matching with objects detected in the surveys mentioned above.  We found: 26 bright stars, 34 H$\alpha$ emitters, 126 X-ray sources and 41 YSOs in our physical groups. Taking into account the overlap among the catalogues that occurs for some of the objects, there are in our groups a total of  166 objects already known in the literature, 29\% (48/166) of them appear in two or more of the inspected catalogues. Among the YSOs that were previously classified, we found: 1 Class I, 17 Class II (CTT),  18 Class III (WTT), and 5 very-likely pre-main sequence stars.
Most of the known population is associated to CMa06. Only 10 objects are found in the other groups: CMa05 contains 4 known stars, CMa08 has 2 stars, while CMa 15, CMa25, CMa 14, and CMa20 have only 1 known object each.

As can be seen in Fig. \ref{figB0}, there is a large number of objects that were previously identified in the direction of the CMa~OB1 region, mainly YSOs and H$\alpha$ emitters. However, most of them do not have {\it Gaia} data or they are out of the astrometric and kinematic criteria adopted by us to identify reliable members of the Association.

\subsection{X-ray sources}
\label{subsec:xraycount}
 
X-ray sources correspond to 76\% (126/166) of the objects from the literature that coincide with our groups, for this reason, we dedicate this subsection to confirm the efficiency of our method in finding objects previously known and to discuss the nature of some of these objects.


\begin{figure}
\begin{center}

\includegraphics[width=1.0\columnwidth, angle=0]{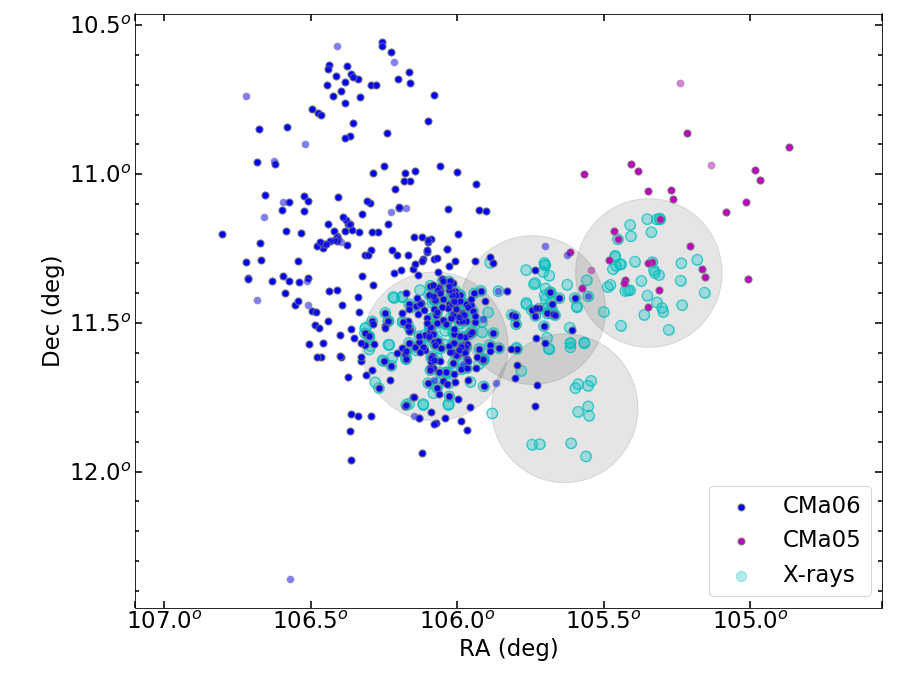}

\end{center}

\caption{Spatial distributions of CMa06 and CMa05 (blue and magenta circles, respectively)  compared to  X-ray sources (cyan circles) from \citet{2018A&A...609A.127S}. Groups members P $<$ 50\% are light points. The hatched area represents the fields observed by the {\it XMM-Newton} satellite.}

\label{figB1}

\end{figure}


Our analysis started with a sample of 387 X-ray sources studied by \citet{2018A&A...609A.127S} in CMa~R1 in which 118 were suggested to be very likely members, 185 are possible members and 84 remained without classification. The spatial distribution of these sources is constrained to the same region as two groups found by us: CMa05 and CMa06 (see Fig. \ref{figB1}).

 We performed a cross-match of the X-ray sources  with {\it Gaia} objects that fulfil our selection criteria (see Sect. \ref{subsec:sampleselection})\footnote[13]{Among 387 X-rays sources from \citet{2018A&A...609A.127S}: 222 have reliable optical counterparts, i. e. stars from {\it Gaia} DR2 following our selection criteria (see Sect.~\ref{subsec:sampleselection});  157 have counterparts out of these criteria;  and 8 have no {\it Gaia}  counterpart.}. We apply the same methodology used by \citet{2018A&A...609A.127S} to search for infrared counterparts and we find 270 stars from {\it Gaia} DR2 associated with 222 X-ray sources. Note that some X-ray sources have more than one optical counterparts.

Three {\it Gaia} counterparts are in CMa05 and 123 are in CMa06 (9 X-ray sources from this group have two optical counterparts),  see grey ``X'' in Fig. \ref{figB0} and cyan circles in Fig. \ref{figB1}.  Other 144 {\it Gaia} counterparts  related to 105 X-ray sources were not found associated with any of our groups. There are different reasons for these sources have not been selected with \texttt{HDBSCAN}: 42 have parallax incompatible with CMa05 or CMa06; 77 do not have proper motion compatible with these groups, 12  have parallax and proper motion similar to one group, but the spatial distribution does not agree with that group. Finally, there are  13 objects having all the {\it Gaia} parameters compatible with CMa05 or CMa06, but they were not detected by our \texttt{HDBSCAN} methodology. This means that, among the analysed sources,  we missed $\sim$ 3\%  (13/387) that could be associated with CMa~OB1. We also conclude that a significant part of the sources that were previously considered possibly related to CMa~OB1, are in fact foreground or background objects.

It is interesting to note that, among 117 X-ray sources identified in association with our groups, only 57 were previously classified as very likely members of CMa~R1, while the other sources were suggested to be candidates or had an unknown origin \citep{2018A&A...609A.127S}.  In other words, thanks to the present study, it was possible to confirm that the other 60 X-ray sources are indeed members of CMa~OB1 association.


\begin{figure}
\begin{center}

\includegraphics[width=1.0\columnwidth, angle=0]{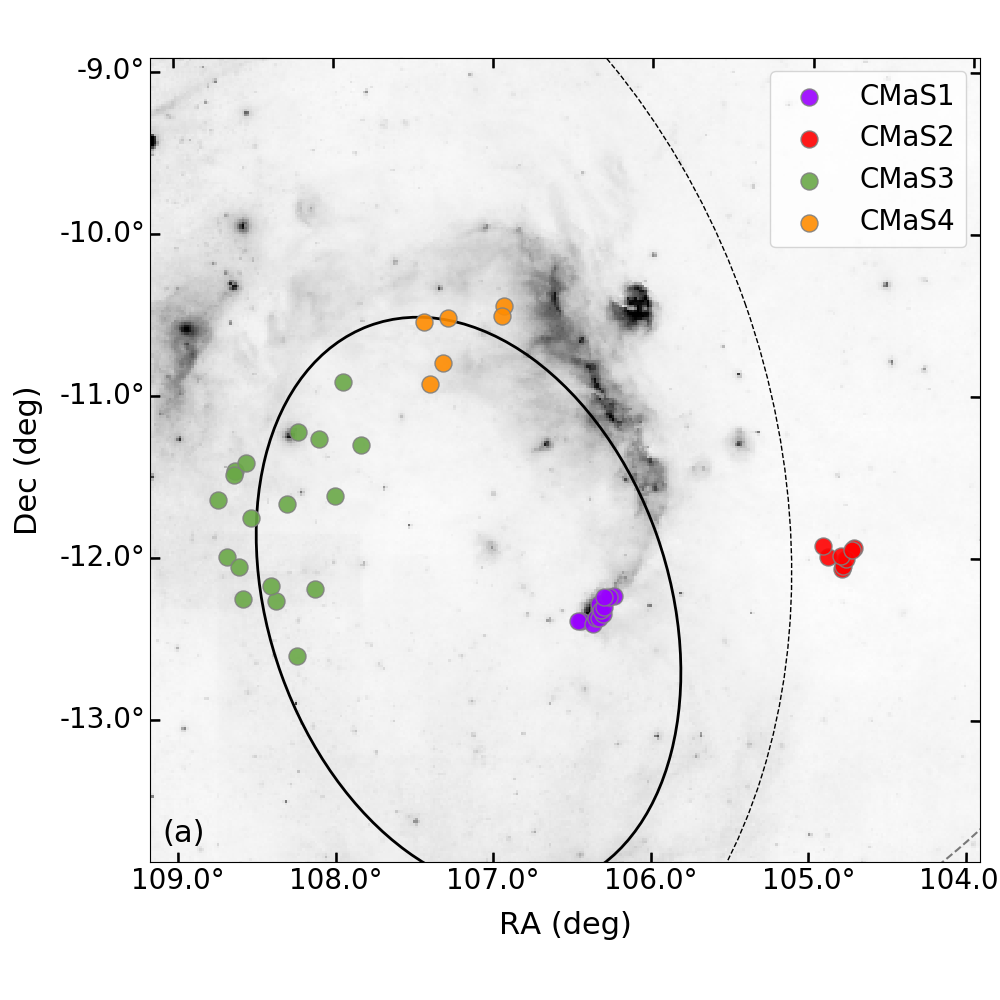}
\includegraphics[width=1.0\columnwidth, angle=0]{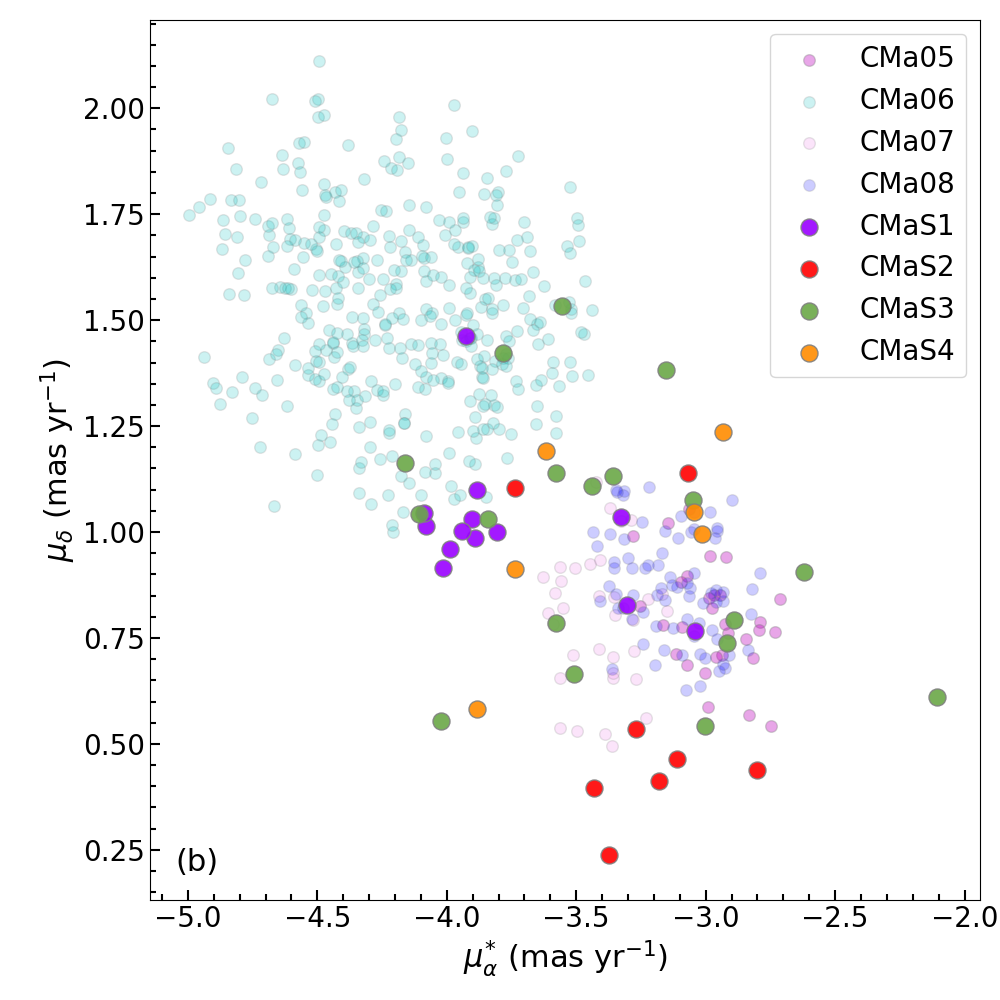}
\includegraphics[width=1.0\columnwidth, angle=0]{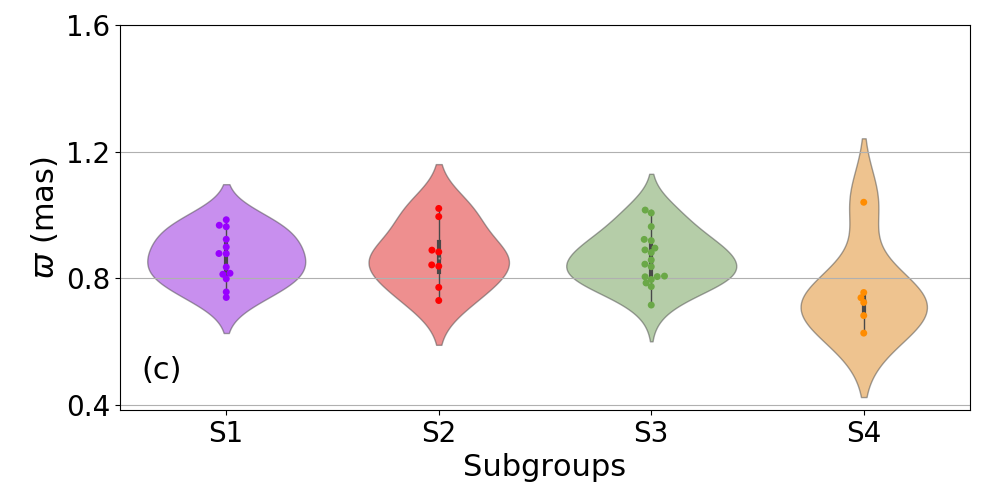}

\end{center}

\caption{(a) Spatial distribution of 4 subgroups of stars known in the literature, potentially CMa~OB1 membership on Digital Sky Survey image  (640 nm) of CMa~OB1. The black and dashed line show the {\it CMa Shell} and {\it CMa Supershell} from \citep{2019A&A...628A..44F}. (b) Proper Motion distribution of subgroups, including 4 groups of CMa~OB1 for comparisons; (c) Violin histograms of parallaxes for each subgroup.
}

\label{figB2}

\end{figure}


\subsection{Subgroups in CMa~OB1}
\label{subsec:subgroups}

In order to investigate if there are other known objects that should be detected as members of our groups, we inspected the spatial distribution shown in Fig. \ref{figB0} searching for possible additional sub-groups. By selecting objects with distances $<$ 1700~pc that form sub-groups with at least 6 objects, we detected 4 small clusters, which are called CMaS1, CMaS2, CMaS3 and CMaS4 in Fig. \ref{figB2}.

The sub-group CMaS1 coincides with the BDSB~96 cluster (Sh~2-297~nebula), discussed in Sect. \ref{sec:cluster_not_found}, while CMaS3 and CMaS4 are scattered around the borders of the {\it CMa Shell} \citep{2019A&A...628A..44F}, and CMaS2 appears a few arcminutes to the west side of the supershell structure. It is interesting to note that despite CMaS2 seems to be outside the considered structures, this sub-group coincides with two unresolved  (positional error $\sim$30'')  X-ray sources detected by ROSAT, called CMaX06 and CMaX07. For both objects, \citet{2009A&A...506..711G} identified infrared counterparts with age in the range 2 to 5 Myr and mass 2 to 3~M$_{\odot}$.

By analysing the proper motion distribution of these sub-groups (see middle panel of Fig. \ref{figB2}), we note that all the sub-groups show similarities with  results found for our groups CMa05, CMa07,  CMa08 and marginally coinciding with CMa06. CMaS2, CMaS3 and CMaS4 are scattered in the same space of proper motion  ($\mu_{\alpha}\cos\delta$ $\sim$~-3.3,  $\mu_{\delta}$~$\sim$~+0.75)~mas~yr$^{-1}$,  while CMaS1 is mainly concentrated around ($\mu_{\alpha}\cos\delta$ $\sim$ -4,  $\mu_{\delta}$ $\sim$ +1) mas yr$^{-1}$ (excepting 4 dispersed objects). Considering the parallax values  (see Fig B2 bottom panel), we note that CMaS1, CMaS2 and CMaS3 have $\varpi$~$\sim$~0.9~mas, corresponding to distances compatible with our groups, while CMaS4 seems to be more distant ($\varpi$ $\sim$ 0.7 mas), but still within the range of distances expected for the overall structure of the Association.

In a recent study based on CO observations of a molecular cloud found in the East side of the {\it CMa Shell}, \citep{GH2021sub} investigated the stellar population associated with the dense cores revealed by the emission peaks in the CO map. Nine dense cores were found, distributed in filamentary structures that follow the same spatial distribution of dust filaments observed in the infrared.  They found 44 cloud members in the region $\alpha$ = 108.51~$\pm$~0.08, $\delta$ = -12.1~$\pm$~0.14, showing {\it Gaia} parameters ($\mu_{\alpha}\cos\delta$~=~-3.3~$\pm$~0.3, $\mu_{\delta}$~=~0.97~$\pm$~0.2)~mas~yr$^{-1}$ and $\varpi$~$\sim$ 0.86 $\pm$ 0.3 mas. These results are similar to those found here for CMaS3,  which can be considered an independent confirmation of the existence of this sub-group belonging to the Association. In addition, several of the objects in CMaS3 studied by us were detected in the YSO survey by \citet{2019ApJS..240...26S}.

All these previous results confirming the young nature of objects appearing in sub-groups, as well as the astrometric and kinematic similarities with our physical groups lead us to suggest that they are actually associated with CMa~OB1. As discussed in Sect. \ref{subsec:knowclusters}, suggesting that CMa15 has two sub-groups, it is possible that some other large groups have complex distributions because they are composed of several sub-groups more or less scattered in the 5-parameters space, as also CMa23. This seems to be the case of CMa06, which shows possible small groups in the spatial distribution (see Fig. \ref{figA06}), which could not be resolved by the \texttt{HDBSCAN} method because the sub-groups are too scarce. These small groups need to be analysed in more detail, using additional observational information. An example is CMaS1 that coincides with BDSB~96, which we suggest to be related to  CMa06 and CMa07 (see Sect.~\ref{sec:cluster_not_found}).

Despite the evidence are less strong to the other two sub-groups, there are similarities indicating they may remain as possible members of this complex Association. CMaS2 is located at the same distance as the other groups and coincides in position with X-ray (ROSAT) sources. Its proper motion is comparable with group CMa03, and at the lower limit of CMa05, but the declination has an offset of $\sim$~1$^{o}$ from each of these main groups. Finally, CMaS4 has parallax compatible to CM05, as well as the proper motion, but the position is more than 1$^{o}$ separate in the sky.

\section{Isochrone fitting codes}
\label{Isos_fitting}

Here we present the two codes, \textsc{M20 code} and \textsc{fitCMD} used by us to estimate fundamental parameters of the groups found in this work. 

\subsection{Cross-entropy method}
\label{Hektor}

The Cross-entropy method, we called \textsc{M20 code} was developed by \citet{2020MNRAS.499.1874M}. The code uses theoretical isochrones that are fitted to the {\it Gaia} DR2 $G_{BP}$ and $G_{RP}$ photometric data and is based on the cross-entropy continuous multi-extremal optimisation method (CE), which takes into account the astrometric membership of the star, as well the nominal errors of the data. The method has been explained in detail in \citet{2020MNRAS.499.1874M} and references therein. 
In summary, the CE method involves an iterative statistical procedure, where the following operations are performed in each iteration:

\begin{enumerate}
\item random generation of the initial sample of  fitting parameters, respecting predefined criteria based on an assumed distribution;
\item selection of the $10\%$ best candidates based on weighted likelihood values;
\item generation of a new random sample of  fitting parameters derived from the distribution obtained based on the $10\%$ best candidates of the previous step;
\item repeats until convergence or stopping criteria is reached.
\end{enumerate}

\begin{table*}
\caption{Groups parameters obtained by \textsc{M20 code}.}

{\centering
\begin{tabular}{lcccccccc}
\hline
Group  &        r50 $^{(a)}$  &        r50 $^{(a)}$  &     r$_{MAX}$ $^{(b)}$  &      log(Age) $^{(c)}$    &       A$_{G}$ $^{(d)}$   &           RV $^{(e)}$   &     NRV $^{(f)}$ \\ 
   &   (pc)  &  (arcmin)  &  (arcmin)  &  (dex)    &  (mag)    &  km/s    &    \\ \hline \hline
                     

CMa00 & 0.363 & 22 & 67 & 7.93 $\pm$ 0.11 & 0.38 $\pm$ 0.36 & 39 $\pm$ 7 & 9  \\
CMa01 & 0.163 & 10 & 28 & 8.24 $\pm$ 0.09 & 0.52 $\pm$ 0.41 &  --  & 0  \\
CMa02 & 0.162 & 10 & 27 & 8.14 $\pm$ 0.21 & 0.65 $\pm$ 0.23 &  --  & 0  \\
CMa03 & 0.136 & 8 & 30 & 8.73 $\pm$ 0.34 & 1.07 $\pm$ 0.53 & 56.9 $\pm$ 0.7 & 4  \\
CMa05 & 0.217 & 13 & 29 & 7.24 $\pm$ 0.67 & 1.17 $\pm$ 0.59 &  --  & 0  \\
CMa06 & 0.278 & 17 & 50 & 7.00 $\pm$ 0.05 & 1.54 $\pm$ 0.90 & 30 $\pm$ 4 & 2  \\
CMa07 & 0.178 & 11 & 19 & 7.12 $\pm$ 0.05 & 0.87 $\pm$ 0.52 &  --  & 0  \\
CMa08 & 0.219 & 13 & 29 & 7.26 $\pm$ 0.12 & 1.05 $\pm$ 0.55 &  --  & 0  \\
CMa09 & 0.185 & 11 & 33 & 8.52 $\pm$ 0.25 & 0.96 $\pm$ 0.43 & 8.7 $\pm$ 0.3 & 2  \\
CMa13 & 0.148 & 9 & 25 & 8.32 $\pm$ 0.13 & 0.31 $\pm$ 0.22 & 3 $\pm$ 10 & 4  \\
CMa15 & 0.219 & 13 & 63 & 8.06 $\pm$ 0.14 & 0.48 $\pm$ 0.37 & 67.1 $\pm$ 0.3 & 2  \\
CMa17 & 0.245 & 15 & 68 & 8.21 $\pm$ 0.22 & 0.67 $\pm$ 0.30 & 15.7 $\pm$ 0.7 & 13  \\
CMa18 & 0.132 & 8 & 27 & 8.26 $\pm$ 0.10 & 0.57 $\pm$ 0.30 & 15.7 $\pm$ 0.3 & 2  \\
CMa19 & 0.146 & 9 & 42 & 7.62 $\pm$ 0.31 & 1.27 $\pm$ 0.55 &  --  & 0  \\
CMa23 & 0.315 & 19 & 63 & 8.12 $\pm$ 0.22 & 0.97 $\pm$ 0.32 &  --  & 0  \\\hline

\end{tabular}
}

NOTES:
(a)  Radius containing 50\% of groups members;
(b) Maximum radius;
(c) Age logarithm;
(d) Photometric extinction in G band;
(e) Radial velocity;
(f) Number of star used to calculate group radial velocity. 

\label{tabC1}

\end{table*}

The code interpolates on the Padova (PARSEC version 1.2S) database of stellar evolutionary tracks and isochrones \citep{Bressan2012}, generated for the {\it Gaia} filter passbands of \cite{Maiz18}, scaled to solar metal content with $Z_{\odot} = 0.0152$. The grid is constructed from isochrones with steps of 0.05 in $log(age)$ and 0.002 in metallicity. The search for the best solution can be performed in the following parameter space:

\begin{itemize}
\item age: from log(age) = 6.60 to log(age) =10.15;
\item distance: from 1 to 25000 parsec;
\item $A_V$: from 0.0 to 5.0 mag; 
\item $[Fe/H]$: from -0.90 to +0.70 dex.
\end{itemize}

Given the large passbands of {\it Gaia} filters, the extinction coefficients depend on the colour and the extinction itself. To account for this effect, we use an updated extinction polynomial for the {\it Gaia} DR2 photometric band-passes, as presented in detail by \citet{2020MNRAS.499.1874M}. 

The code also allows the user to specify priors in the parameters distance, $A_V$ and [Fe/H]. In general, when not specified in the text, we adopt distance $\mathcal{N}(\mu,\,\sigma^{2})$ obtained with Bayesian inference from the parallax ($\varpi$) and its uncertainty ($\sigma_{\varpi}$). The variance ($\sigma^2$) is obtained from the distance interval calculated from the inference using the uncertainty as $1\sigma_{\varpi}$. The prior in $A_V$ is also adopted as a normal distribution with $\mu$ and variance ($\sigma^{2}$) for each cluster taken from the 3D extinction map produced by \citet{3Debv}\footnote[14]{The 3D extinction map is available online at \url{https://stilism.obspm.fr/}}. The prior for [Fe/H] was estimated from the Galactic metallicity gradient published by \citet{OCCAMgradient20}. For the age we adopt a flat prior so that $P(X_n) = 1$.

To estimate uncertainties on the fundamental parameters, we used a Monte-Carlo technique, re-sampling in each run with a replacement in the original dataset, to perform a bootstrap procedure. The isochrones are also re-generated in each run from the adopted initial mass function (IMF).
The final fundamental parameters and respective errors were estimated by the mean and one standard deviation of ten runs.  The code also provides radial velocities, radius containing 50\% of members (r50) and maximum radius (r$_{max}$), containing all the stars in each group. The radial velocities (RV) are estimated from {\it Gaia} DR2 data of the individual stars. However, there are no groups with more than 13 stars with RV values, most of them have only one or two objects having this parameter. So that,  the estimates of RV for our groups are not reliable and  they were not used in this work.  The additional parameters estimated  using \textsc{M20 code}, are present in Table~\ref{tabC1}.

\subsection{\textsc{fitCMD}}
\label{Charles}


\begin{table*}
\caption{Groups fundamental parameters  evaluated by \textsc{fitCMD}.}

{\centering
\begin{tabular}{lccccc}
\hline
Group  & M$_{g}$ $^{(a)}$        & (m-M)o $^{(b)}$            & E(B-V) $^{(c)}$            & Z$^{(d)}$     & Z/Z$_{\odot}$ $^{(e)}$      \\       
   &  (M$_{\odot}$)     &    (mag)          &    (mag)          &     &               \\ \hline \hline
CMa00 & 317 $^{+146}_{-116}$ & 8.699 $^{+0.004}_{-0.025}$ & 0.100 $^{+0.003}_{-0.020}$ & 0.0180 $^{+0.0020}_{-0.0020}$ & 1.18 $^{+0.13}_{-0.13}$  \\
CMa01 & 213 $^{+70}_{-62}$ & 10.03 $^{+0.03}_{-0.09}$ & 0.14 $^{+0.03}_{-0.01}$ & 0.0170 $^{+0.0010}_{-0.0005}$ & 1.12 $^{+0.06}_{-0.03}$  \\
CMa02 & 310 $^{+147}_{-117}$ & 10.04 $^{+0.16}_{-0.24}$ & 0.40 $^{+0.03}_{-0.02}$ & 0.0050 $^{+0.0037}_{-0.0005}$ & 0.33 $^{+0.24}_{-0.03}$  \\
CMa03 & 507 $^{+199}_{-169}$ & 10.94 $^{+0.08}_{-0.11}$ & 0.38 $^{+0.06}_{-0.02}$ & 0.0300 $^{+0.0005}_{-0.0005}$ & 1.97 $^{+0.03}_{-0.03}$  \\
CMa05 & 265 $^{+173}_{-120}$ & 10.26 $^{+0.02}_{-0.06}$ & 0.37 $^{+0.02}_{-0.01}$ & 0.0120 $^{+0.0010}_{-0.0005}$ & 0.79 $^{+0.07}_{-0.03}$  \\
CMa06 & 1990 $^{+643}_{-573}$ & 10.15 $^{+0.05}_{-0.02}$ & 0.38 $^{+0.01}_{-0.01}$ & 0.0180 $^{+0.0010}_{-0.0005}$ & 1.18 $^{+0.07}_{-0.03}$  \\
CMa07 & 75 $^{+45}_{-32}$ & 10.19 $^{+0.01}_{-0.01}$ & 0.210 $^{+0.003}_{-0.010}$ & 0.0180 $^{+0.0010}_{-0.0010}$ & 1.18 $^{+0.07}_{-0.07}$  \\
CMa08 & 195 $^{+92}_{-74}$ & 10.01 $^{+0.12}_{-0.03}$ & 0.41 $^{+0.03}_{-0.02}$ & 0.0170 $^{+0.0010}_{-0.0005}$ & 1.12 $^{+0.07}_{-0.03}$  \\
CMa09 & 646 $^{+217}_{-192}$ & 10.43 $^{+0.19}_{-0.04}$ & 0.35 $^{+0.04}_{-0.03}$ & 0.0210 $^{+0.0010}_{-0.0005}$ & 1.38 $^{+0.06}_{-0.03}$  \\
CMa13 & 202 $^{+68}_{-59}$ & 9.66 $^{+0.05}_{-0.16}$ & 0.12 $^{+0.02}_{-0.01}$ & 0.0160 $^{+0.0010}_{-0.0020}$ & 1.05 $^{+0.06}_{-0.13}$  \\
CMa15 & 1510 $^{+390}_{-358}$ & 10.49 $^{+0.09}_{-0.11}$ & 0.15 $^{+0.02}_{-0.01}$ & 0.0290 $^{+0.0010}_{-0.0005}$ & 1.91 $^{+0.06}_{-0.03}$  \\
CMa17 & 3130 $^{+1360}_{-1110}$ & 10.57 $^{+0.03}_{-0.17}$ & 0.28 $^{+0.03}_{-0.02}$ & 0.0180 $^{+0.0030}_{-0.0010}$ & 1.18 $^{+0.20}_{-0.07}$  \\
CMa18 & 699 $^{+265}_{-224}$ & 10.11 $^{+0.05}_{-0.16}$ & 0.19 $^{+0.03}_{-0.02}$ & 0.0210 $^{+0.0010}_{-0.0005}$ & 1.38 $^{+0.06}_{-0.03}$  \\
CMa19 & 1600 $^{+1340}_{-808}$ & 10.75 $^{+0.10}_{-0.67}$ & 0.400 $^{+0.002}_{-0.030}$ & 0.0300 $^{+0.0005}_{-0.0140}$ & 1.97 $^{+0.03}_{-0.92}$  \\
CMa23 & 1400 $^{+583}_{-480}$ & 10.95 $^{+0.05}_{-0.26}$ & 0.39 $^{+0.03}_{-0.02}$ & 0.0210 $^{+0.0010}_{-0.0005}$ & 1.38 $^{+0.06}_{-0.03}$ \\\hline

\end{tabular}
}

NOTES:
(a) Total mass cluster;
(b) Apparent distance modulus;
(c) Colour excess;
(d) Metallicity;
(e) Metallicity in terms of Solar values, assuming $ Z_{\odot}$~=~0.0152;

\label{tabC2}

\end{table*}


The \textsc{fitCMD} code was developed by \citet{2019MNRAS.483.2758B}. In general lines, the code is a statistical approach intended to extract fundamental parameters of star clusters by means of the photometric information contained in the observed  Colour-Magnitude Diagram (CMDs). Based on properties of the  IMF obtained from isochrones, \textsc{fitCMD} searches for values of the total mass stored in stars (or cluster mass, M$_{cl}$), age (t$_{age}$), global metallicity (Z), foreground extinction (or colour excess, CEx) and the apparent distance modulus (DM), able to build a synthetic CMD that best reproduces the observed one. Magnitude-dependent photometric completeness and photometric scatter are also taken into account. Finally, the best-fitting parameters are found by minimising the residual differences between the synthetic and observed CMDs by means of the global optimisation algorithm Simulated Annealing \citep{GOFFE199465}. Other applications of \textsc{fitCMD} can be found in \citet{2019MNRAS.490.2414P,2020MNRAS.493.2688B}.

The \textsc{fitCMD} version used here (Apr, 14, 2020) works  on several photometric systems.  In particular, for this work we use  {\it Gaia} [VegaMags] (G, G$_{BP}$, G$_{RP}$) with PARSEC v1.2S + COLIBRI PR16/NBC models \citep{Bressan2012,2017ApJ...835...77M}.  The parameter space allowed by the code is: 

\begin{itemize}
\item masses: starting in 0.1 M$_{\odot}$;
\item DM: free; 
\item CEx: free;
\item ages: from 1 Myr and 13.5  Gyr;
\item Z: from 0.0001 to 0.03, assuming solar metallicity Z$_{\sun}$ $=$ 0.0152.

\end{itemize}

For adequate coverage ages vary in steps of $\Delta$t $=$ 1 Myr for 1-10 Myr, $\Delta$t $=$ 2Myr for 10-20 Myr, $\Delta$t $=$ 5Myr for 20-50 Myr, $\Delta$t $=$ 10 Myr for 50-100 Myr, $\Delta$t $=$ 25 Myr for 100-500 Myr, $\Delta$t $=$ 50 Myr for 500-1000 Myr and  $\Delta$t $=$ 250 Myr for 1000-13500 Myr, and the metallicites vary in steps of $\Delta$Z $=$ 10$^{-4}$ from 10$^{-4}$ to 10$^{-3}$ and $\Delta$Z $=$ 10$^{-3}$ from 10$^{-3}$ to 3 $\times$ 10$^{-2}$.

To derive the visual extinction, \textsc{fitCMD} uses the relation A$_{V}$~=~3.1 $\times$ E(B-V) and for the metallicities it uses [Fe/H]~=~$log_{10}(Z/Z_{\odot})~-~0.8[\alpha/Fe]~-~0.05[\alpha/Fe]^{2}$. For consistence with \textsc{M20 code} we adopt $[\alpha/Fe]$ = 0.0.

Additional  parameters computed \textsc{fitCMD} for the 15 clusters studied in this work are present in Table \ref{tabC2}.

\section{Tables and Analysis figures}
\label{sec:figures}

The tables containing the properties of the members of the 29 groups as Table \ref{tab_CDS} is available in its entirely in machine-readable form at the CDS. 

\begin{table*}
\caption{Parameters of the members of group CMa00 selected by HDBSCAN.}
\begin{center}
{
\begin{tabular}{lccccccccc}
\hline 

ID$_{Gaia}$ & P & R.A.   & Dec.   & $\mu_{\alpha}\cos\delta$   & $\mu_{\delta}$   & $\varpi$   & D$_{A}$ & A$_{V}$ & V$_{t}$\\
                             
 & (\%) & (deg)   & (deg)   & (mas yr$^{-1}$)   & (mas yr$^{-1}$)   & (mas)   & (pc) & (mag) & (km/s) \\ \hline\hline

3051226788077089664 & 100 & 107.36 $\pm$ 0.04 & -9.15 $\pm$ 0.05 & -5.24 $\pm$ 0.08 & 2.29 $\pm$ 0.08 & 1.74 $\pm$ 0.06 & 565 & 0.37 & 11.74 \\
3051233312124183168 & 100 & 107.46 $\pm$ 0.07 & -9.09 $\pm$ 0.07 & -5.56 $\pm$ 0.14 & 2.65 $\pm$ 0.15 & 1.64 $\pm$ 0.10 & 604 & 0.45 & 14.66 \\
3051226375760232960 & 99 & 107.34 $\pm$ 0.06 & -9.17 $\pm$ 0.07 & -5.88 $\pm$ 0.12 & 2.62 $\pm$ 0.10 & 1.55 $\pm$ 0.09 & 637 & 0.38 & 17.80 \\
3051254894342334592 & 90 & 107.17 $\pm$ 0.05 & -8.88 $\pm$ 0.05 & -5.61 $\pm$ 0.11 & 2.34 $\pm$ 0.09 & 1.81 $\pm$ 0.07 & 545 & 0.50 & 12.40 \\
3051255787695465344 & 72 & 107.19 $\pm$ 0.04 & -8.81 $\pm$ 0.05 & -5.15 $\pm$ 0.09 & 2.67 $\pm$ 0.08 & 1.75 $\pm$ 0.06 & 562 & 0.38 & 11.29 \\
3051247163401380992 & 66 & 106.94 $\pm$ 0.04 & -9.00 $\pm$ 0.04 & -5.66 $\pm$ 0.08 & 2.44 $\pm$ 0.09 & 1.79 $\pm$ 0.05 & 550 & 0.32 & 12.79 \\
3051203728391103232 & 56 & 106.98 $\pm$ 0.12 & -9.26 $\pm$ 0.13 & -5.45 $\pm$ 0.27 & 2.24 $\pm$ 0.28 & 1.52 $\pm$ 0.15 & 655 & 0.41 & 16.31 \\
3051241356605265408 & 99 & 107.38 $\pm$ 0.04 & -8.86 $\pm$ 0.05 & -5.34 $\pm$ 0.09 & 2.32 $\pm$ 0.09 & 1.81 $\pm$ 0.05 & 543 & 0.30 & 11.32 \\
3051338349846802560 & 80 & 107.74 $\pm$ 0.11 & -8.82 $\pm$ 0.12 & -5.57 $\pm$ 0.24 & 2.26 $\pm$ 0.27 & 1.73 $\pm$ 0.15 & 574 & 0.35 & 13.41 \\
3051330932445906560 & 100 & 107.50 $\pm$ 0.03 & -8.91 $\pm$ 0.04 & -5.58 $\pm$ 0.07 & 2.45 $\pm$ 0.07 & 1.86 $\pm$ 0.04 & 530 & 0.24 & 11.69 \\ \hline
\end{tabular}
}
\label{tab_CDS}
\end{center}

NOTES: We provide for each star:  the {\it Gaia} DR2 identifier, Membership probability (P), position, proper motion and parallax (without zero-point offset), distance obtained by \citet{bailerjones2018}, visual extinction from 3D-maps \citep{bayestar2019} and tangential velocities in the equatorial system. This entirety table containing 152 CMa00 members is available  
in machine-readable form at the CDS.

\end{table*}

Also, when running \texttt{HDBSCAN} our code also makes analysis images containing histograms and distributions of the parameter members  for each group: in the first row are presented R.A. and Dec.; in second row are proper motion. The last panel of theses rows the group members are contrasted with {\it Gaia} total sample (grey points); two first panels in third row are proper motion (pm) and tangential velocity (V$_{t}$) vectors, respectively. Third and forth panels are histograms of the visual extinction from 3D-maps \citep{bayestar2019} and the membership probability of the groups contents; and forth row we present the histograms of parallax, distances obtained from it (1/$\varpi$) and distances from \citet{bailerjones2018}. In the last panel are a comparison between both distances. The members with P~$\geq$~50\% are presented in bright colours, while P $<$ 50\% are in light colours. In the histograms, the median of each distribution, considering only stars P~$\geq$~50\%, which were used to determine the parameter of each group presented in Table \ref{tab1} (see Sect. \ref{sec:astrometric_par}) is indicated by a solid line, the 16\% and 84\% percentiles (used for the uncertainties) are the dashed lines. Moreover, the percentiles of 2.5\% and 97.5\% are represented by dot-dashed lines and 0.15\% and 99.85\% by dotted lines.
In this Appendix we present the diagnostic figures for 15 groups discussed in this work and the figures of other 14 groups are present in the online supplementary material.

In addition to the analysis figures we also present for the 15 groups, in the last line, the colour-magnitude diagrams provided by each fitting code: on the left is the CMD from \textsc{M20 code} and on the right is the CMD from \textsc{fitCMD}. In both panels the line represents the best-fit isochrone by the codes. The \textsc{M20 code} CMD shows the distribution of stars according to membership probability while the \textsc{fitCMD} diagram also presents the synthetic Hess diagram generated by this code (coloured hatched area).


\begin{figure*}
\begin{center}

\includegraphics[width=1.8\columnwidth, angle=0]{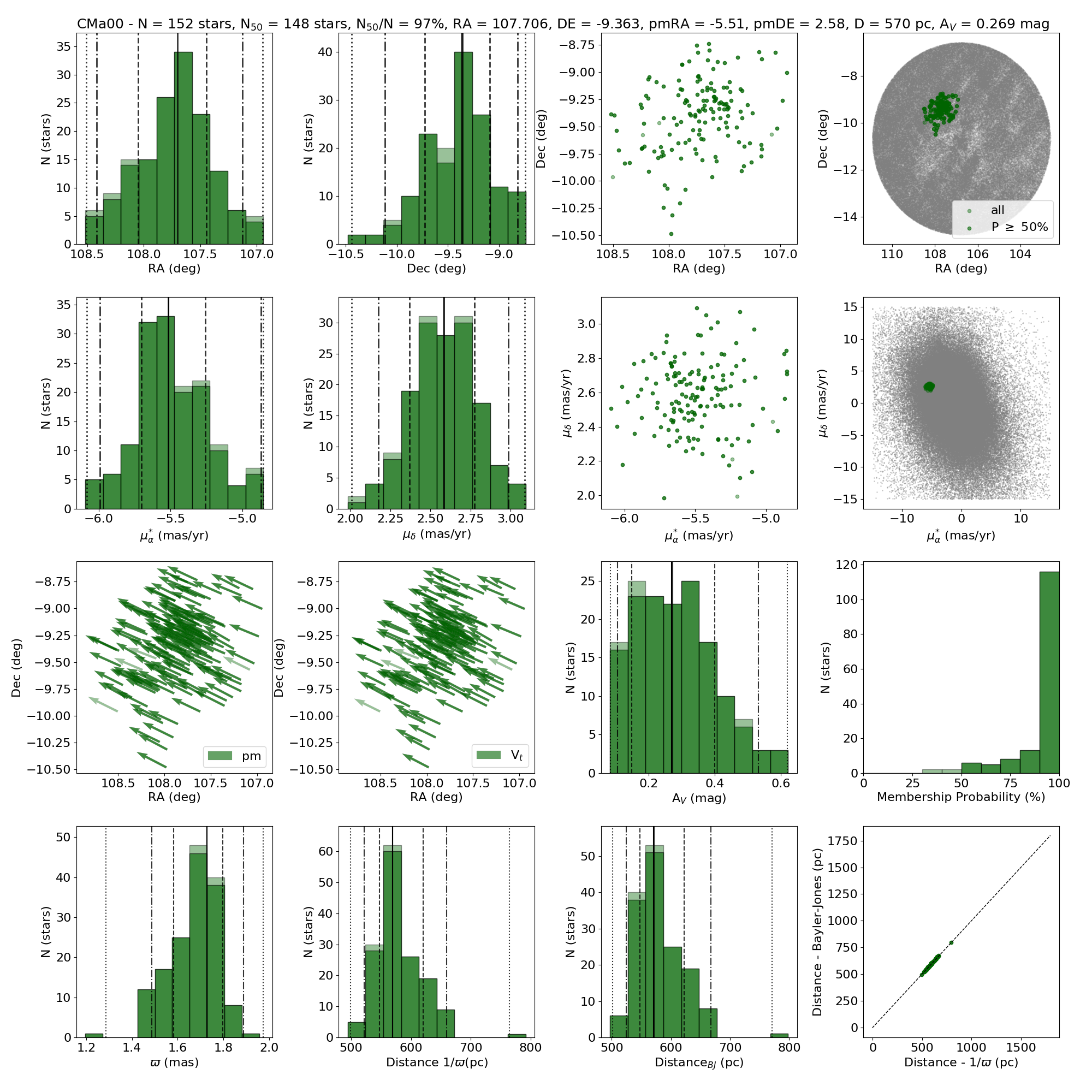}
\includegraphics[width=0.6\columnwidth, angle=0]{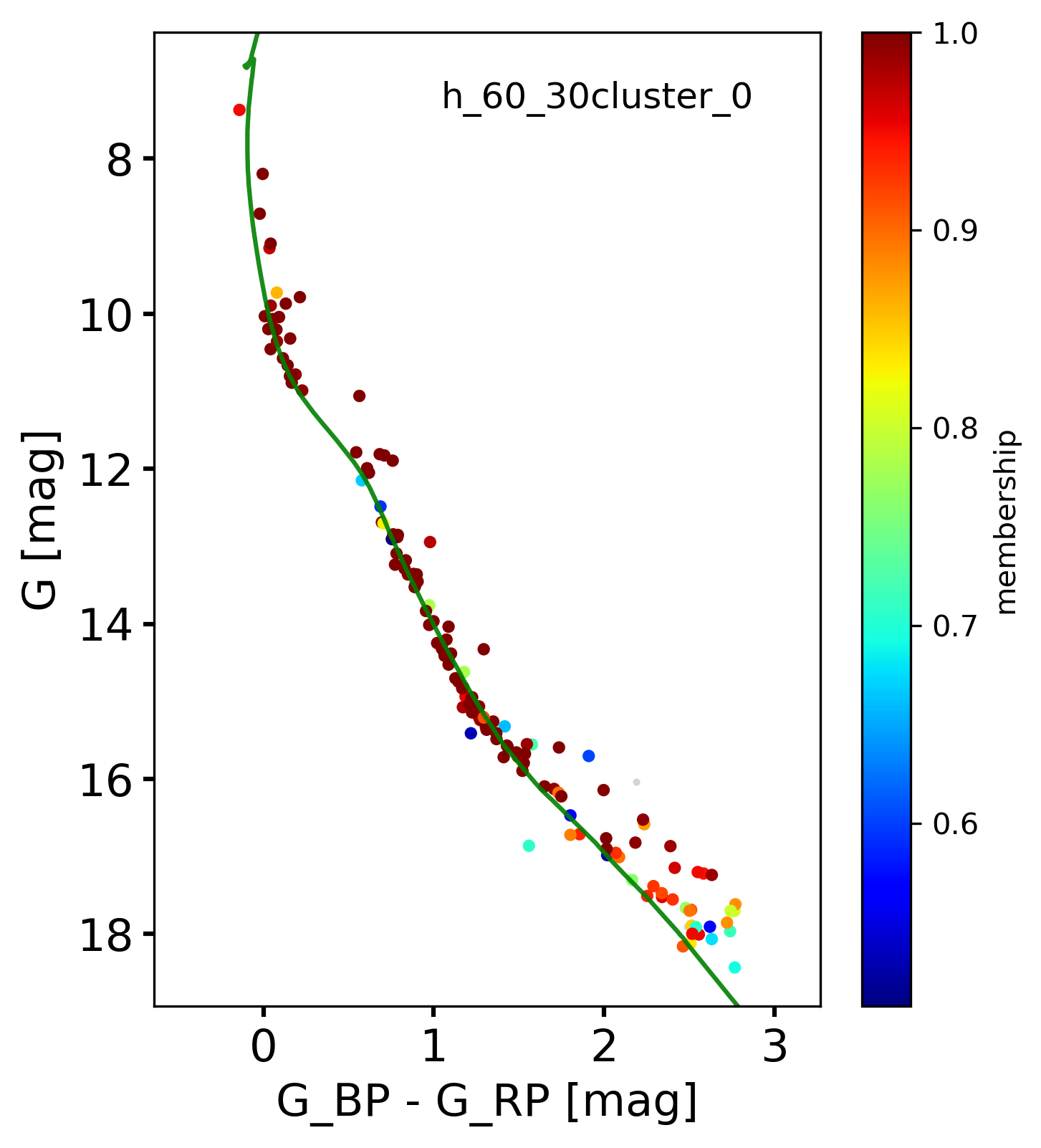}
\includegraphics[trim={0 1.4cm 0 0},clip, width=0.6\columnwidth, angle=0]{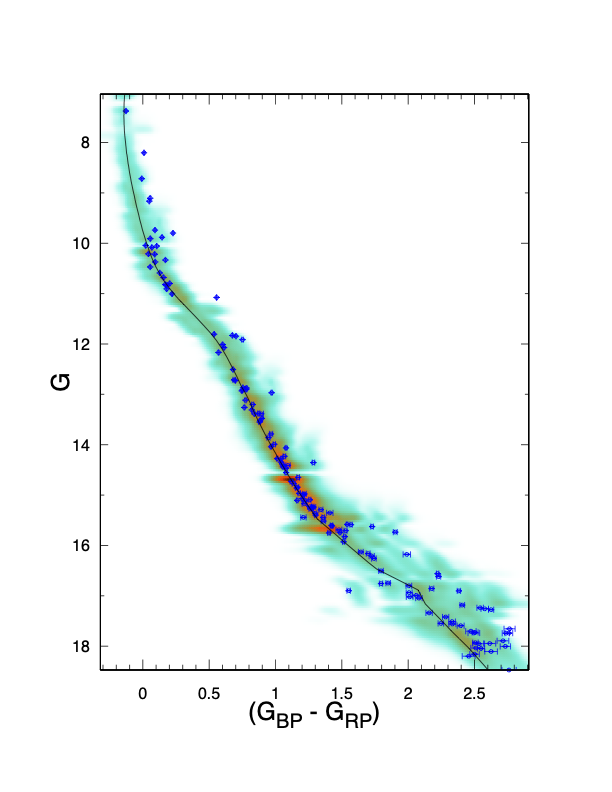}

\end{center}

\caption{Top - Analysis image; Bottom left - Colour-magnitude diagram (CMD) from \textsc{M20 code}; and Bottom right - CMD from \textsc{fitCMD} of the group CMa00.}

\label{figA00}

\end{figure*}



\begin{figure*}
\begin{center}

\includegraphics[width=1.8\columnwidth, angle=0]{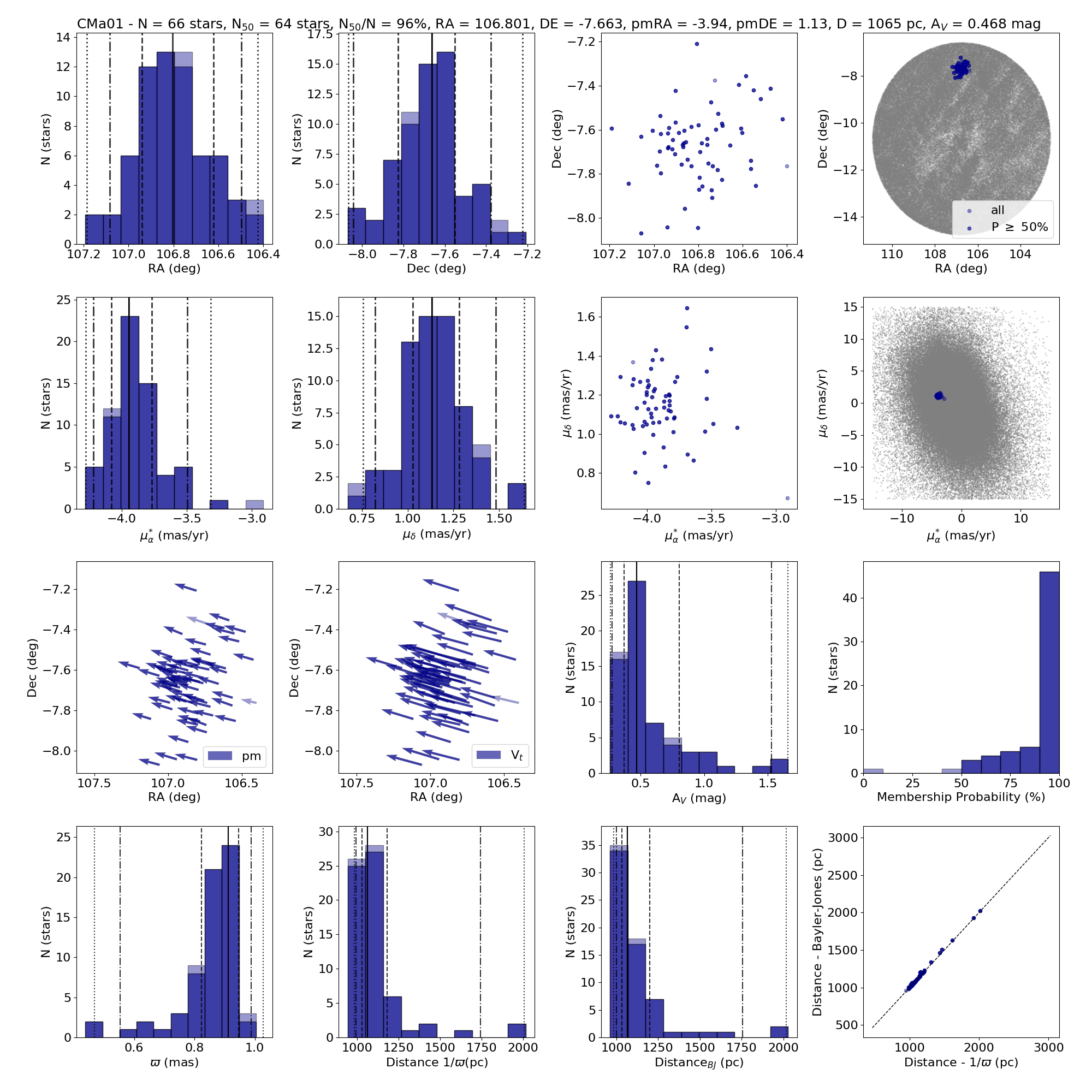}
\includegraphics[width=0.6\columnwidth, angle=0]{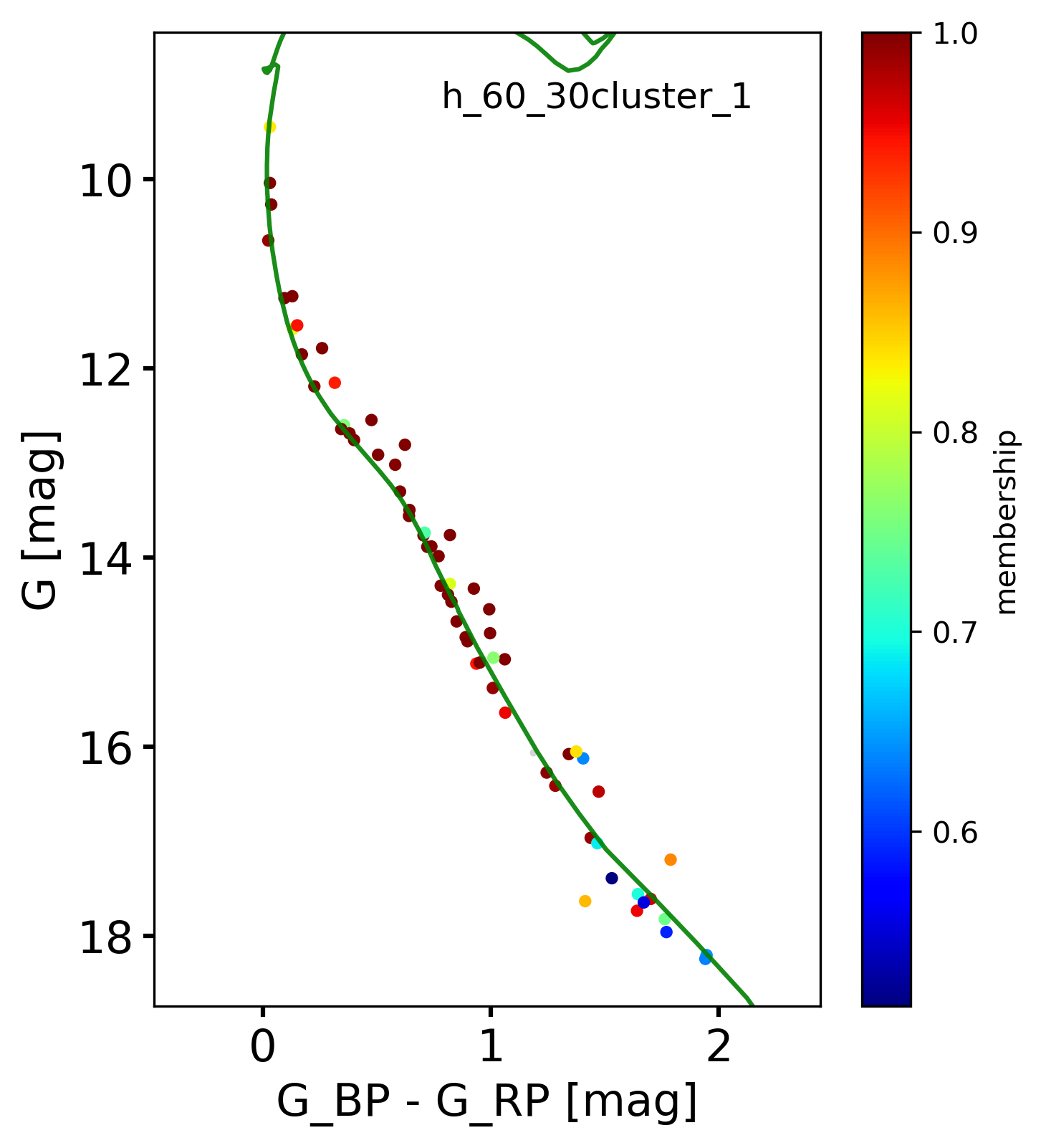}
\includegraphics[trim={0 1.4cm 0 0},clip, width=0.6\columnwidth, angle=0]{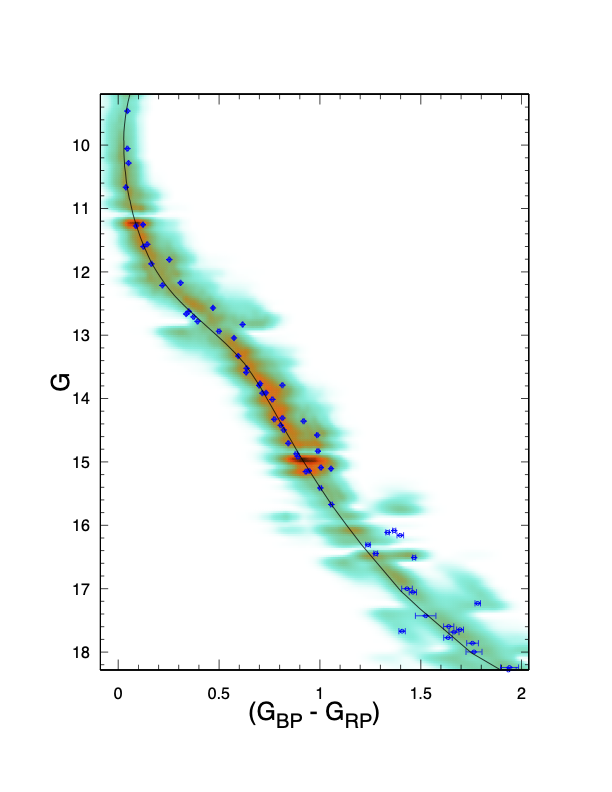}

\end{center}

\caption{The same of Fig. \ref{figA00} for CMa01.}

\label{figA01}

\end{figure*}



\begin{figure*}
\begin{center}

\includegraphics[width=1.8\columnwidth, angle=0]{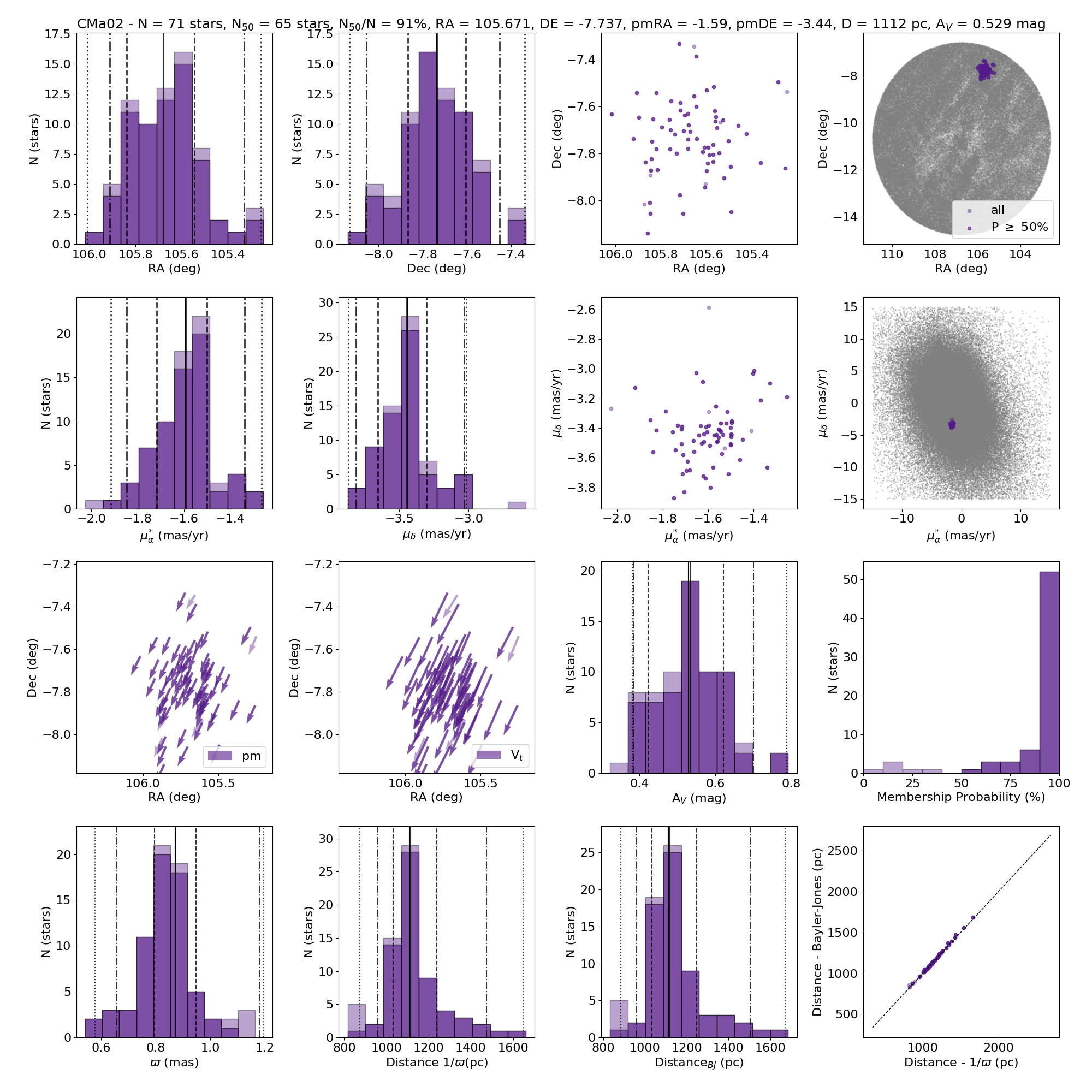}
\includegraphics[width=0.6\columnwidth, angle=0]{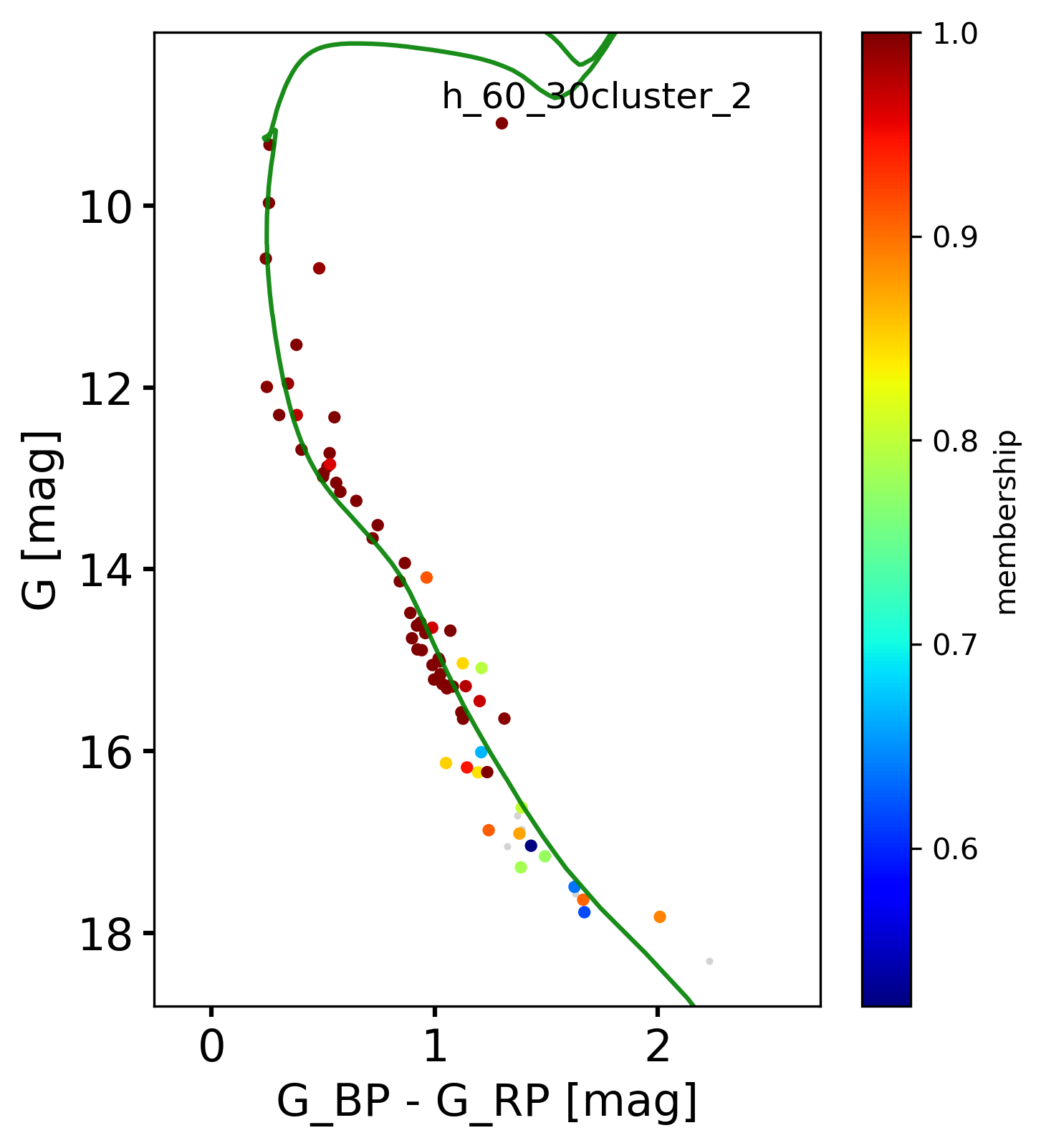}
\includegraphics[trim={0 1.4cm 0 0},clip, width=0.6\columnwidth, angle=0]{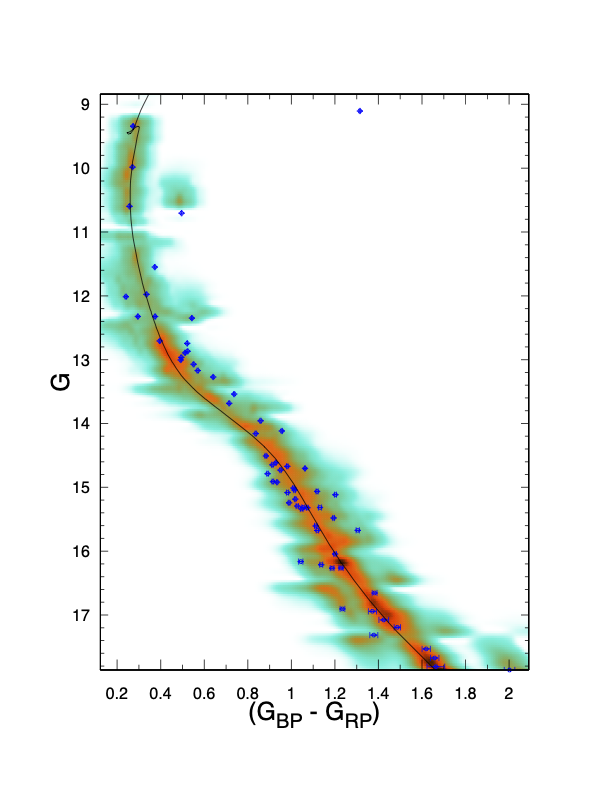}

\end{center}

\caption{The same of Fig. \ref{figA00} for CMa02.}

\label{figA02}

\end{figure*}



\begin{figure*}
\begin{center}

\includegraphics[width=1.8\columnwidth, angle=0]{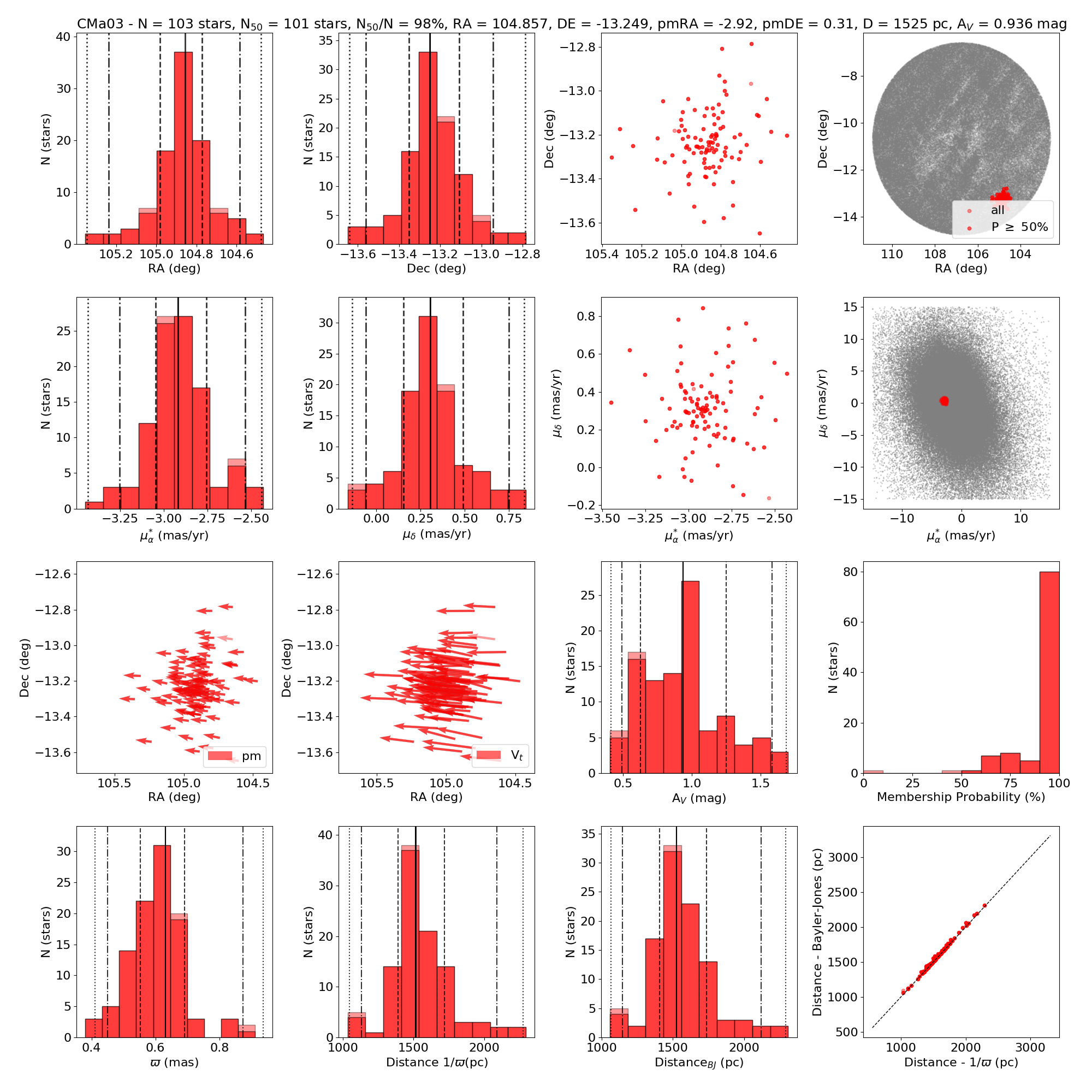}
\includegraphics[width=0.6\columnwidth, angle=0]{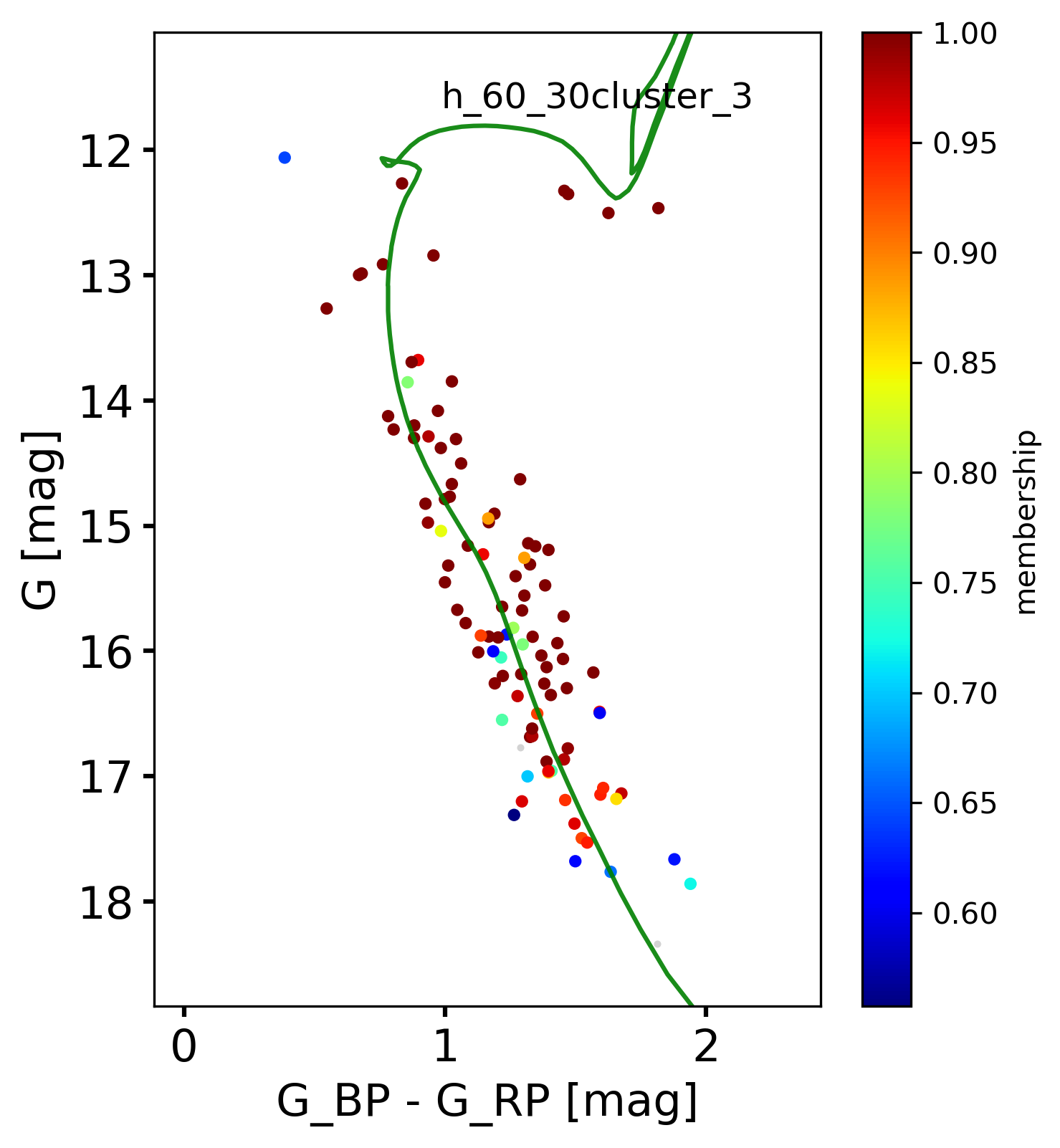}
\includegraphics[trim={0 1.4cm 0 0},clip, width=0.6\columnwidth, angle=0]{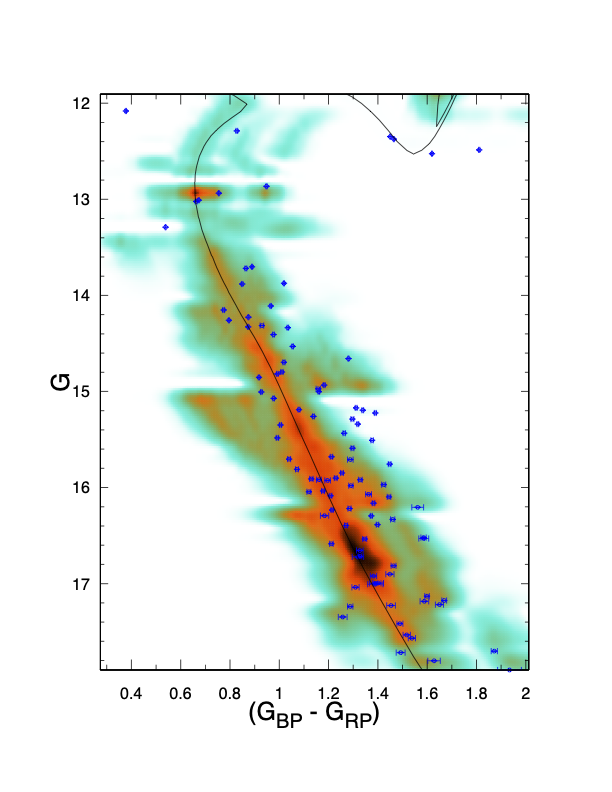}

\end{center}

\caption{The same of Fig. \ref{figA00} for CMa03.}

\label{figA03}

\end{figure*}



\begin{figure*}
\begin{center}

\includegraphics[width=1.8\columnwidth, angle=0]{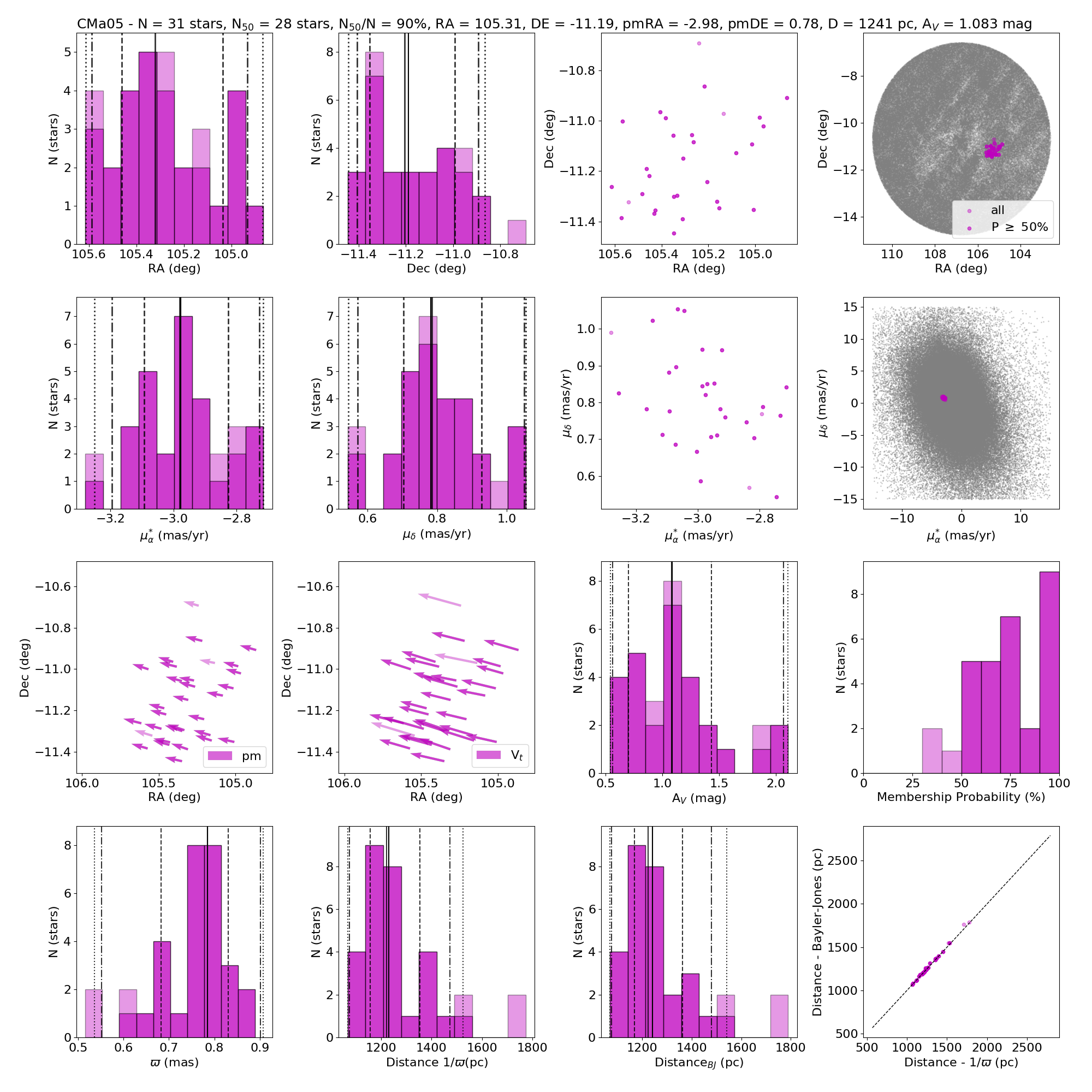}
\includegraphics[width=0.6\columnwidth, angle=0]{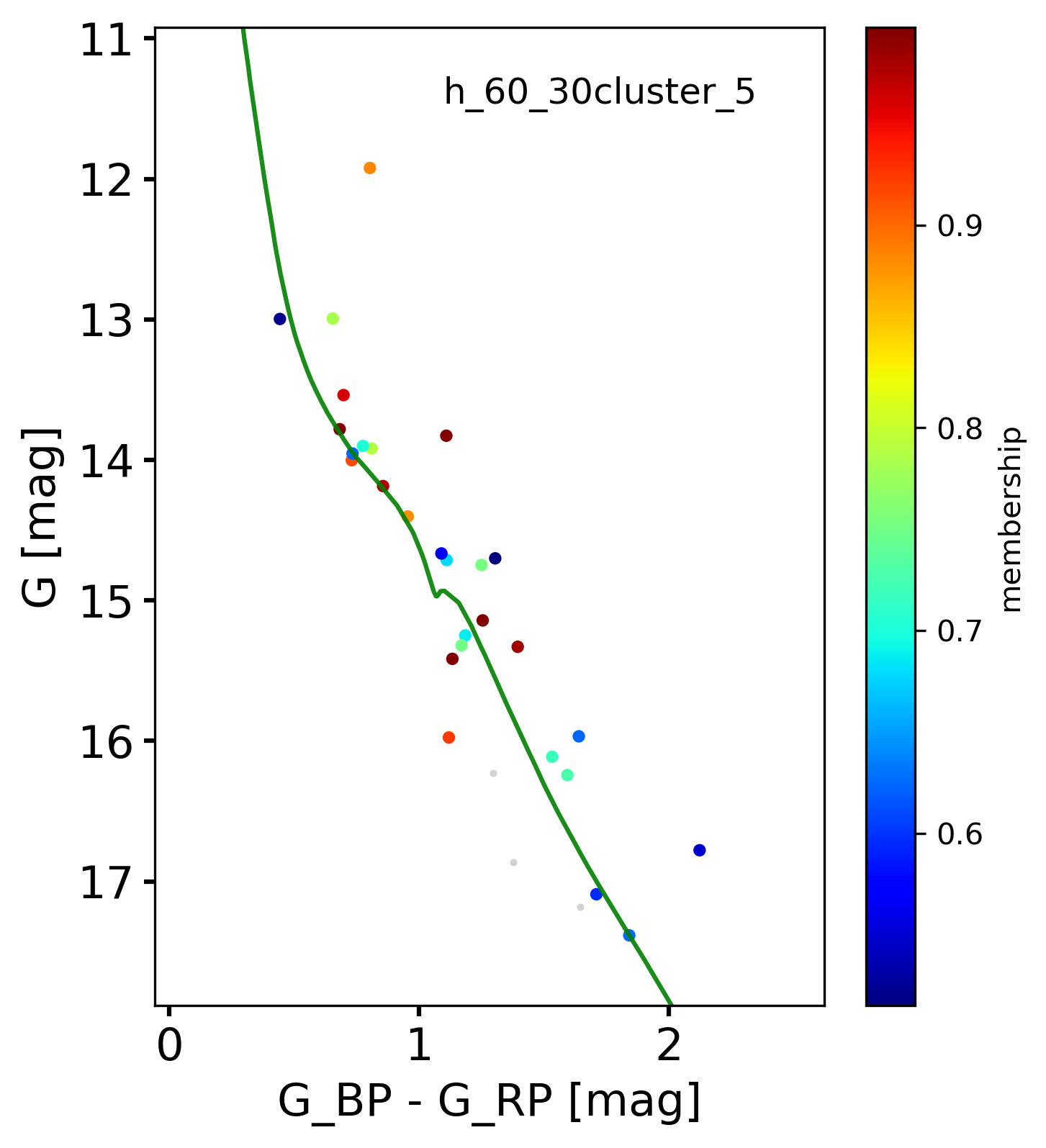}
\includegraphics[trim={0 1.4cm 0 0},clip, width=0.6\columnwidth, angle=0]{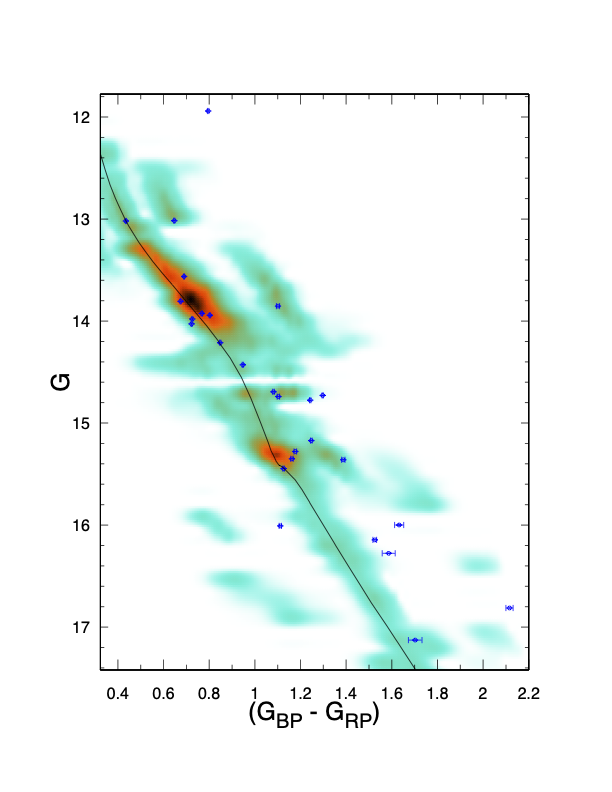}

\end{center}

\caption{The same of Fig. \ref{figA00} for CMa05.}

\label{figA05}

\end{figure*}



\begin{figure*}
\begin{center}

\includegraphics[width=1.8\columnwidth, angle=0]{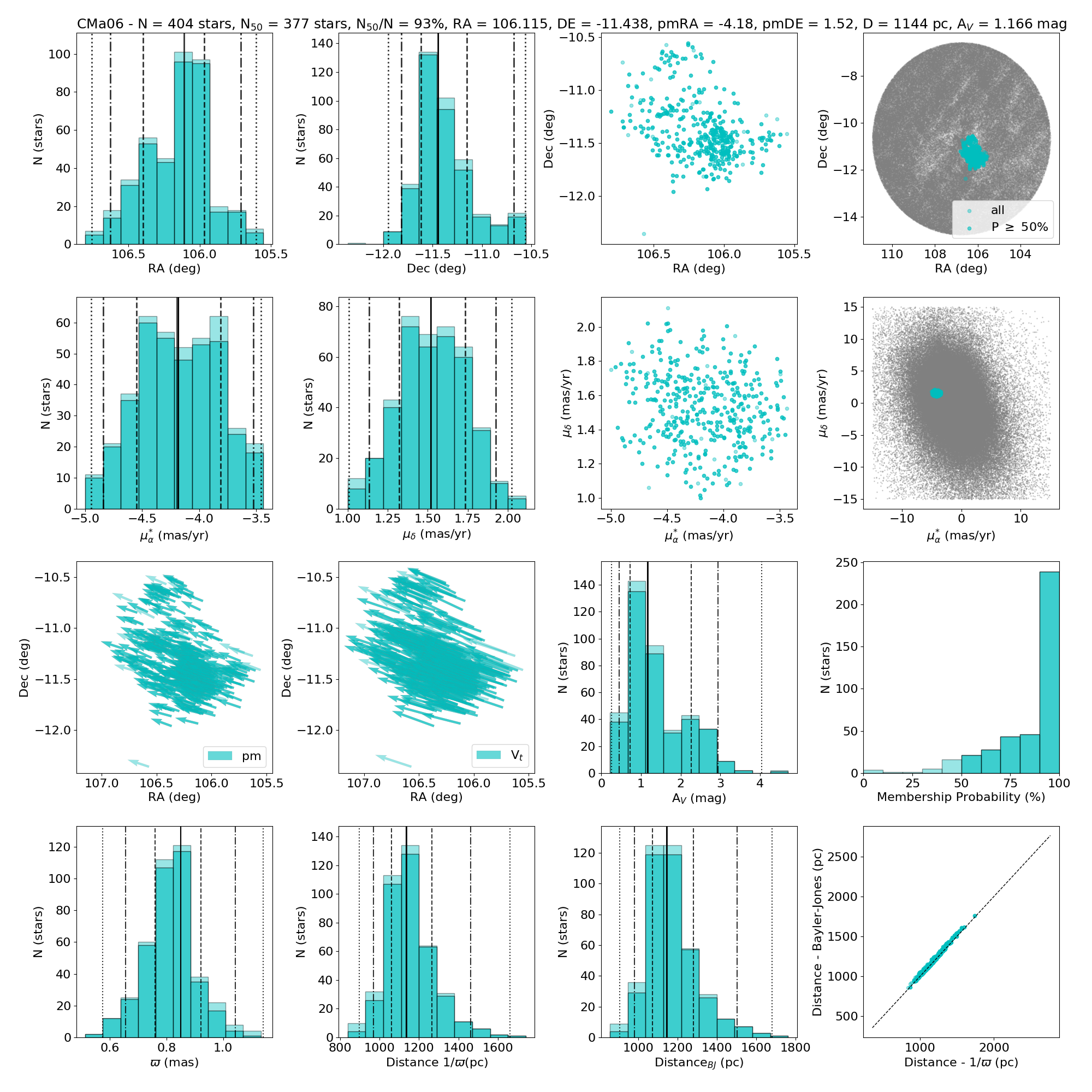}
\includegraphics[width=0.6\columnwidth, angle=0]{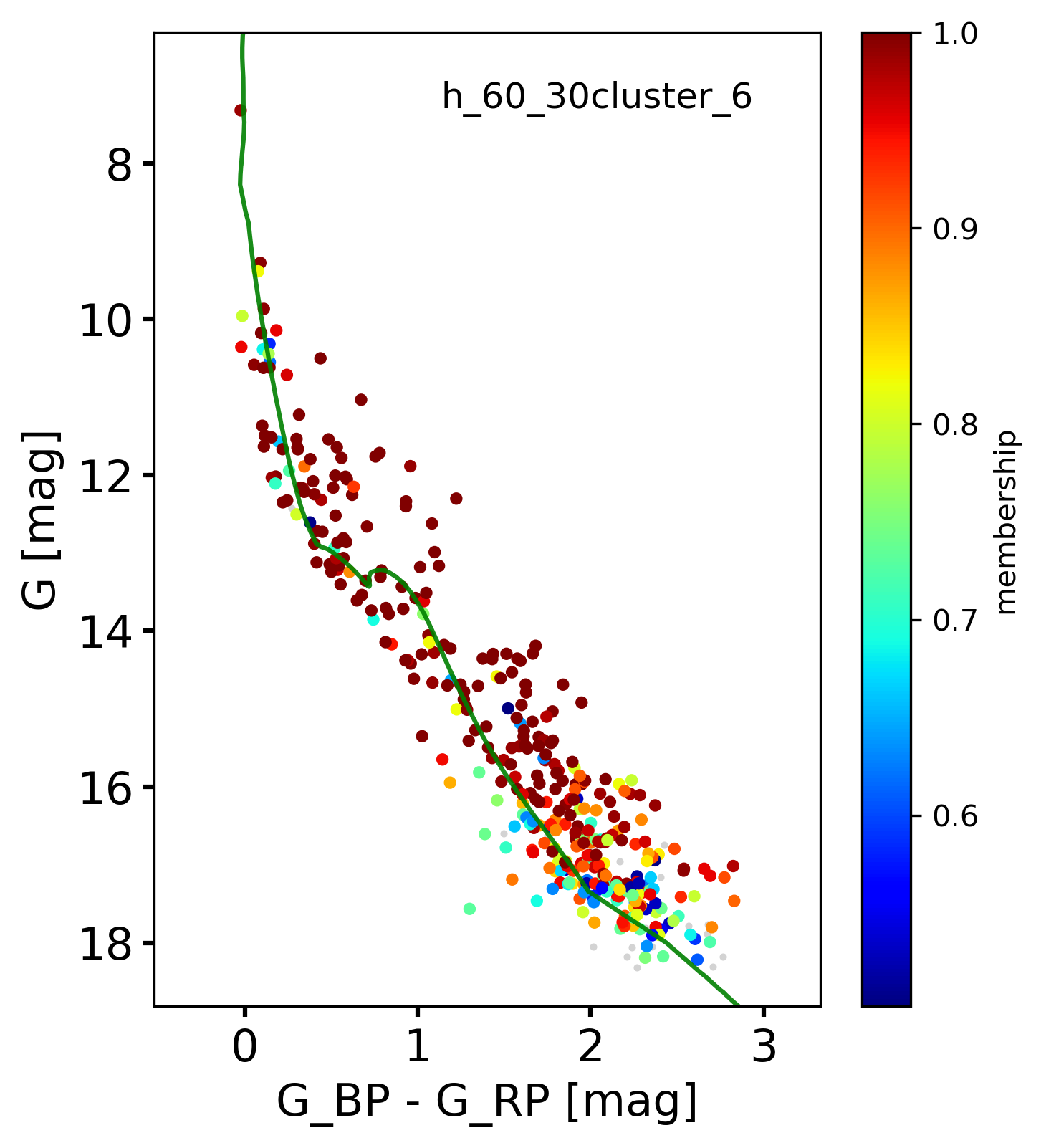}
\includegraphics[trim={0 1.4cm 0 0},clip, width=0.6\columnwidth, angle=0]{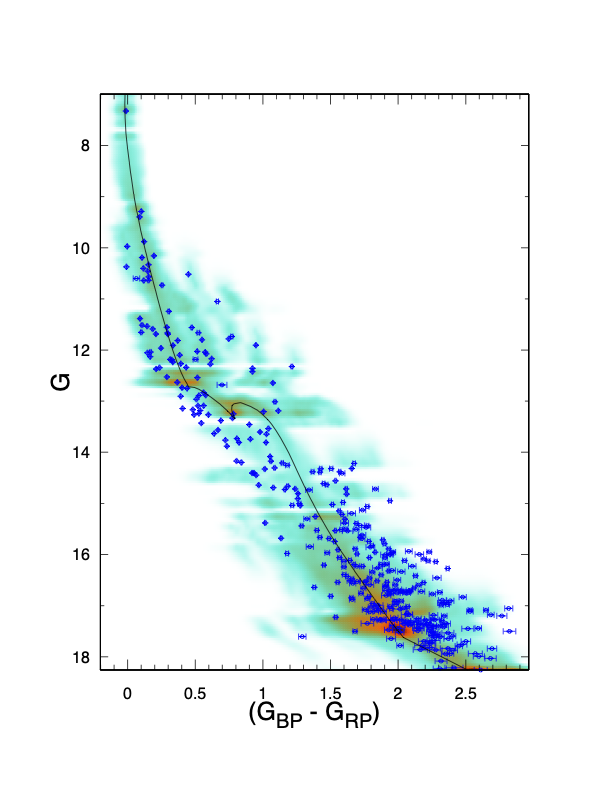}
\end{center}

\caption{The same of Fig. \ref{figA00} for CMa06.}

\label{figA06}

\end{figure*}



\begin{figure*}
\begin{center}

\includegraphics[width=1.8\columnwidth, angle=0]{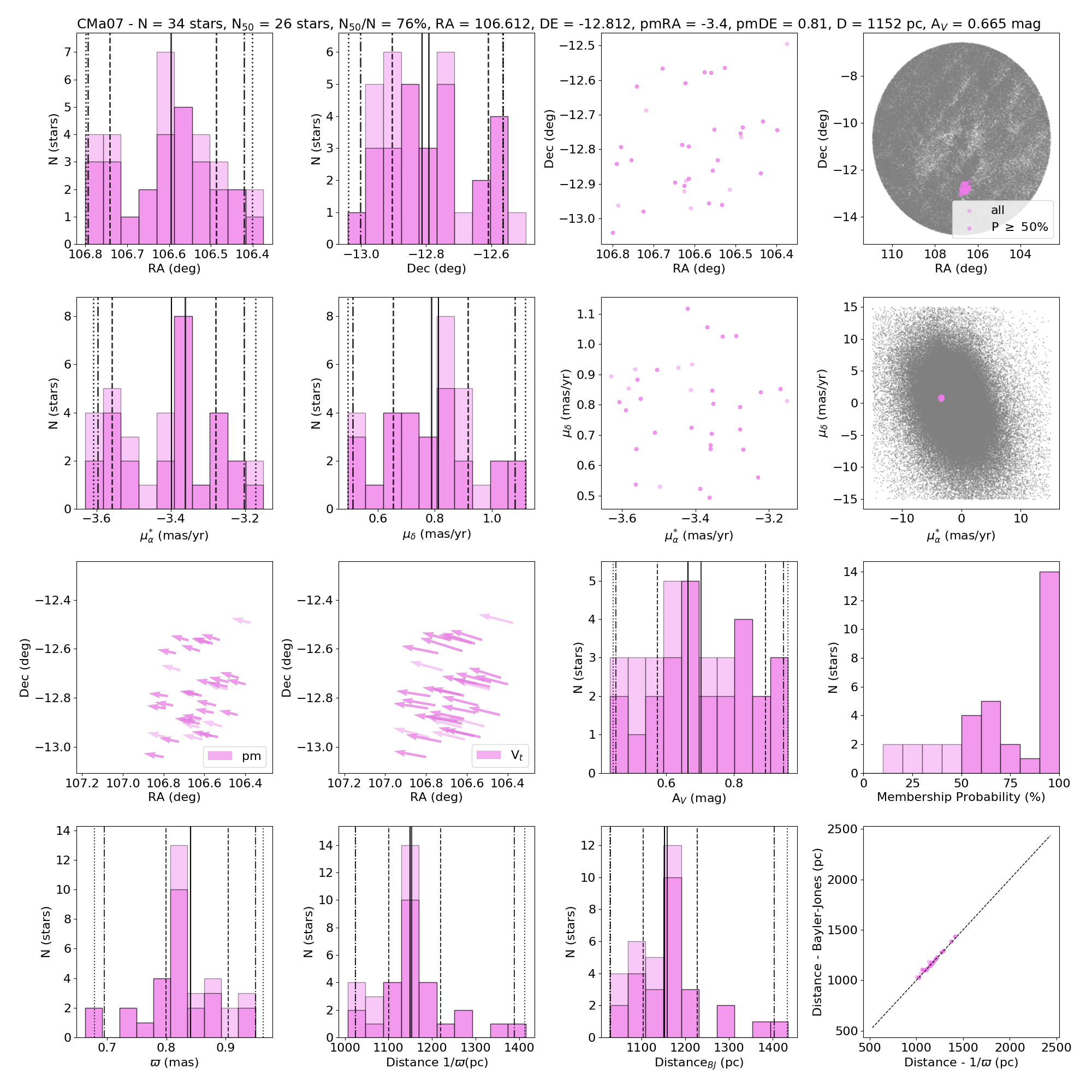}
\includegraphics[width=0.6\columnwidth, angle=0]{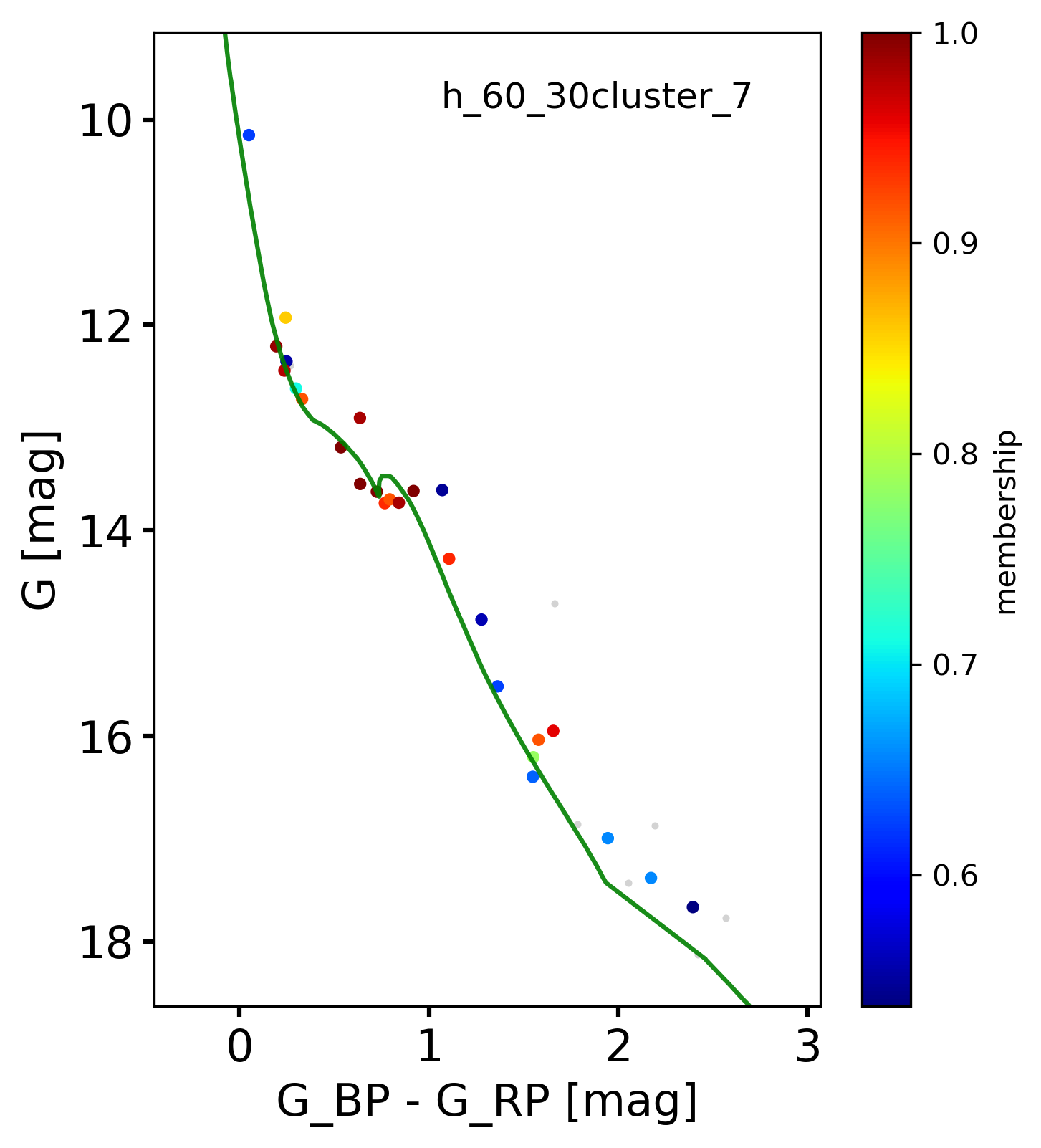}
\includegraphics[trim={0 1.4cm 0 0},clip, width=0.6\columnwidth, angle=0]{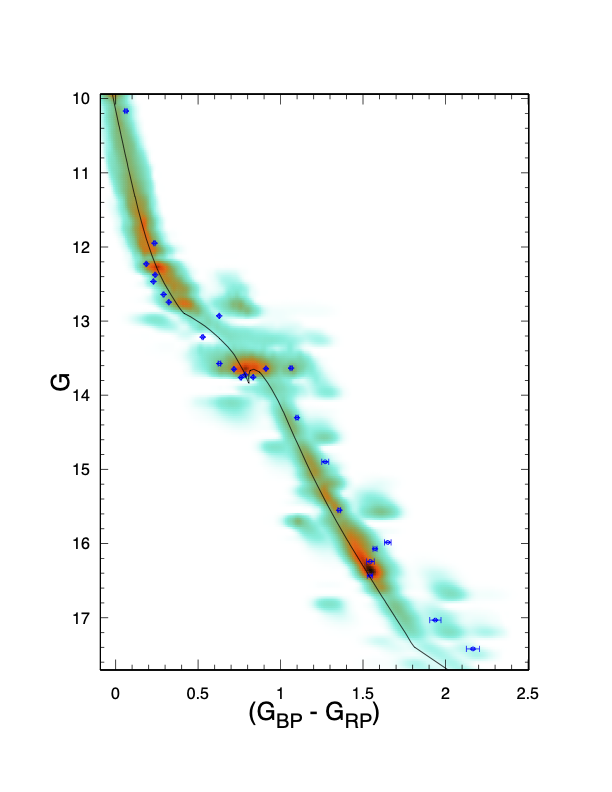}
\end{center}

\caption{The same of Fig. \ref{figA00} for CMa07.}

\label{figA07}

\end{figure*}



\begin{figure*}
\begin{center}

\includegraphics[width=1.8\columnwidth, angle=0]{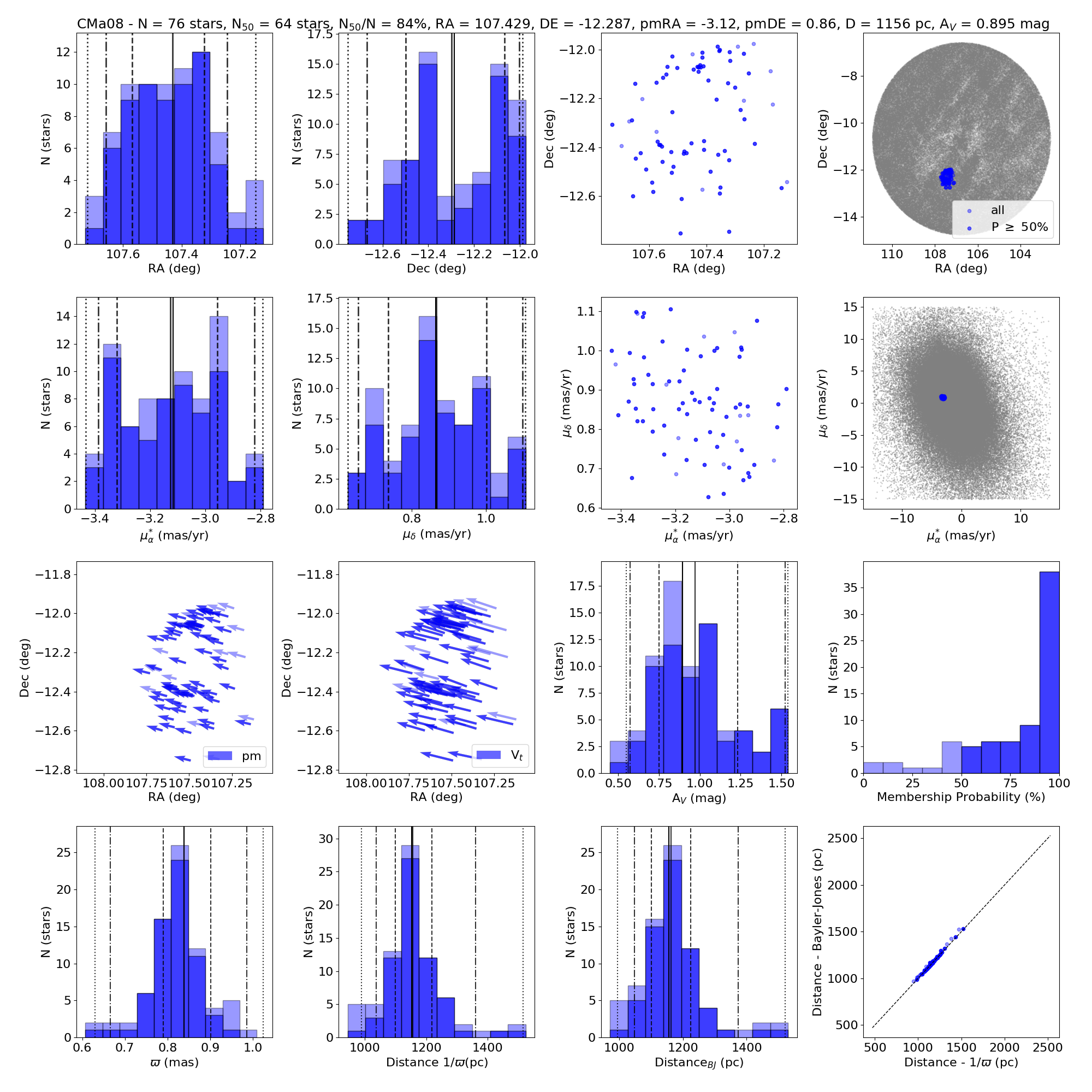}
\includegraphics[width=0.6\columnwidth, angle=0]{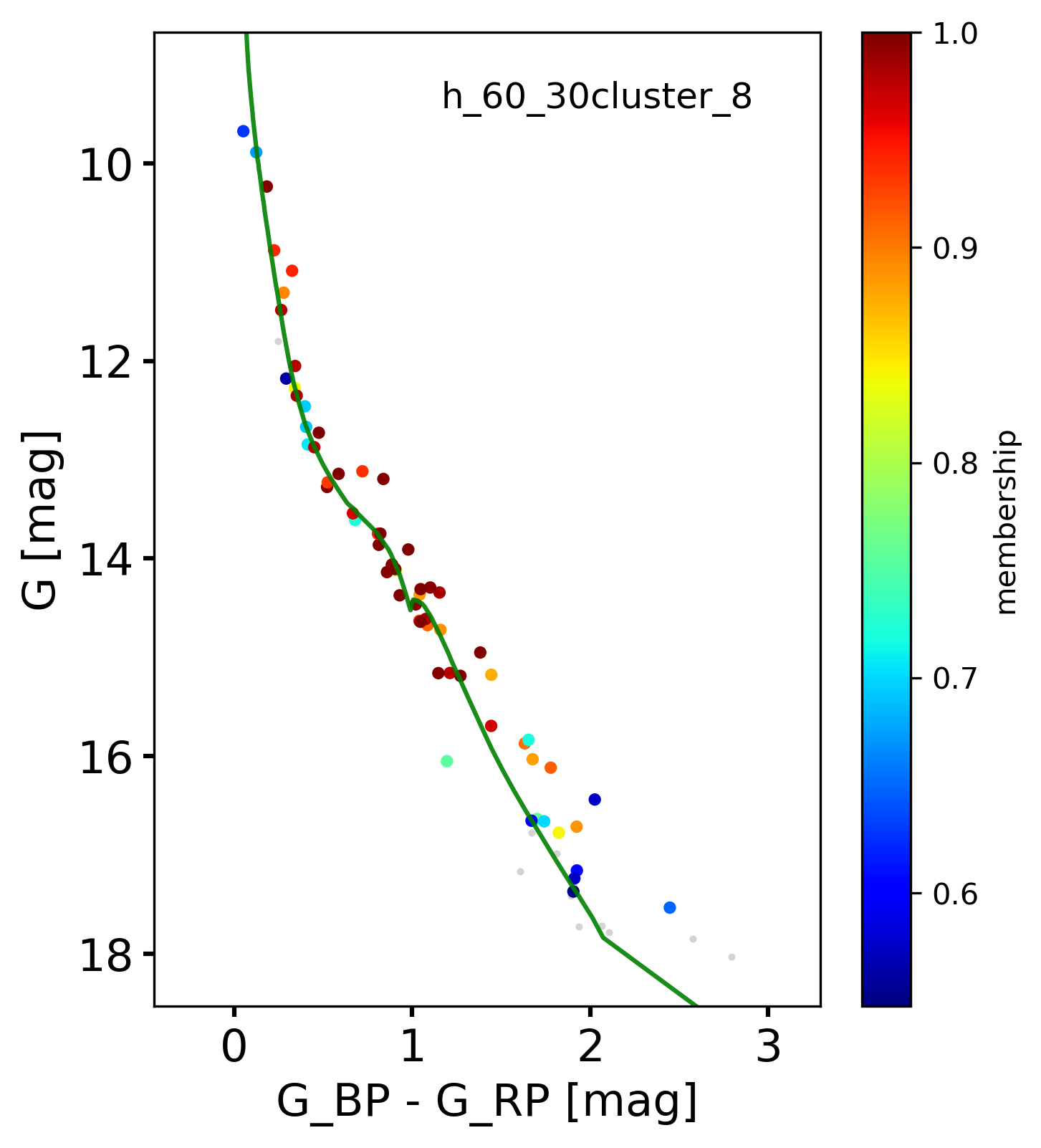}
\includegraphics[trim={0 1.4cm 0 0},clip, width=0.6\columnwidth, angle=0]{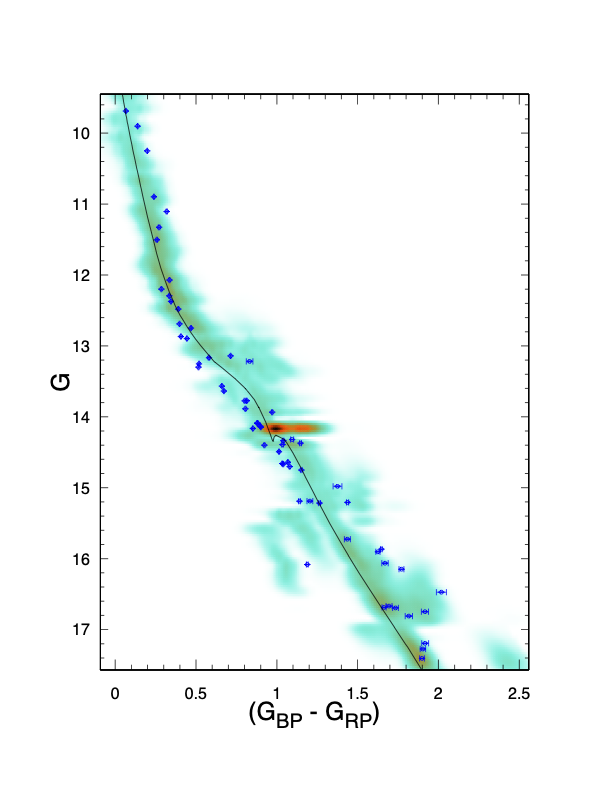}

\end{center}

\caption{The same of Fig. \ref{figA00} for CMa08.}

\label{figA08}

\end{figure*}



\begin{figure*}
\begin{center}

\includegraphics[width=1.8\columnwidth, angle=0]{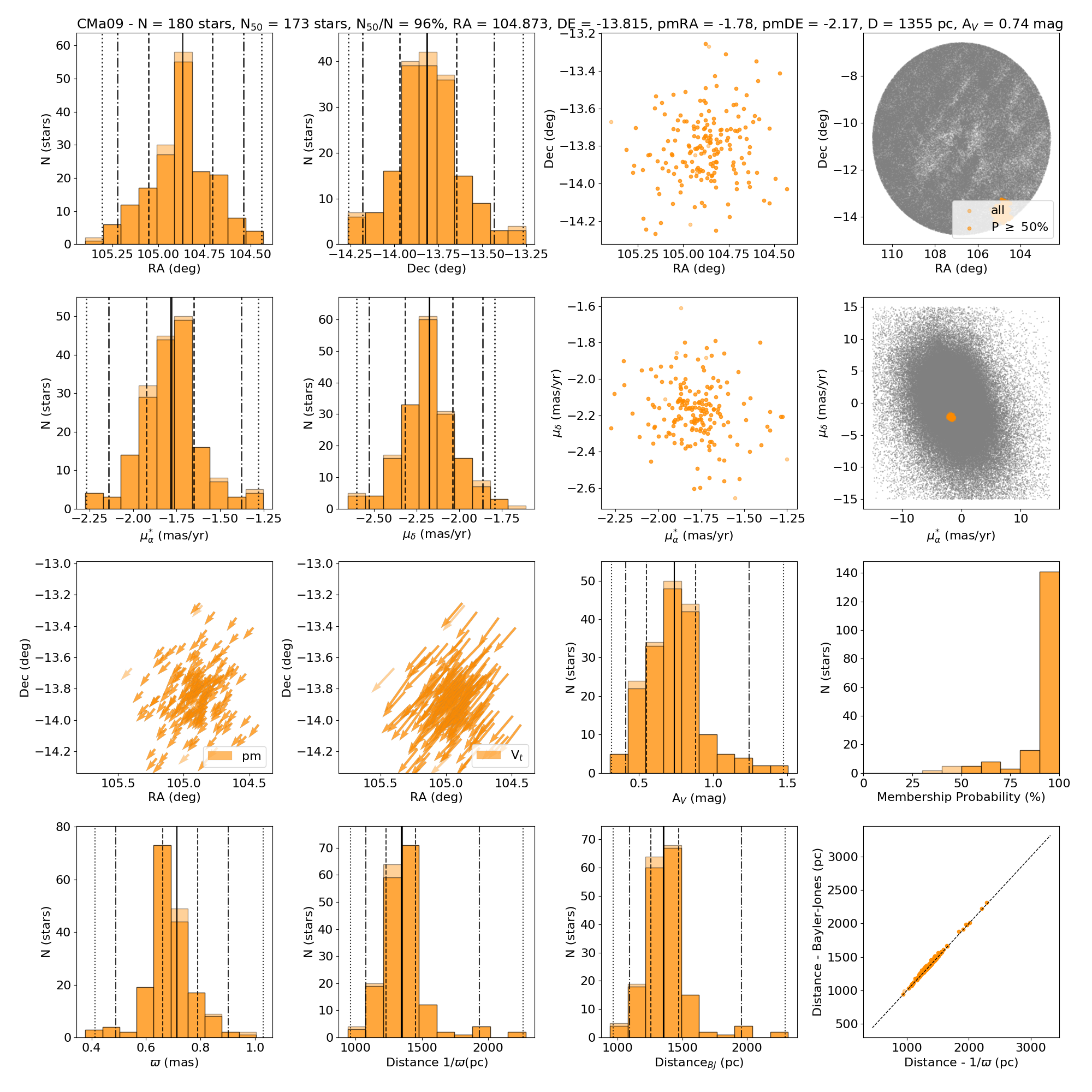}
\includegraphics[width=0.6\columnwidth, angle=0]{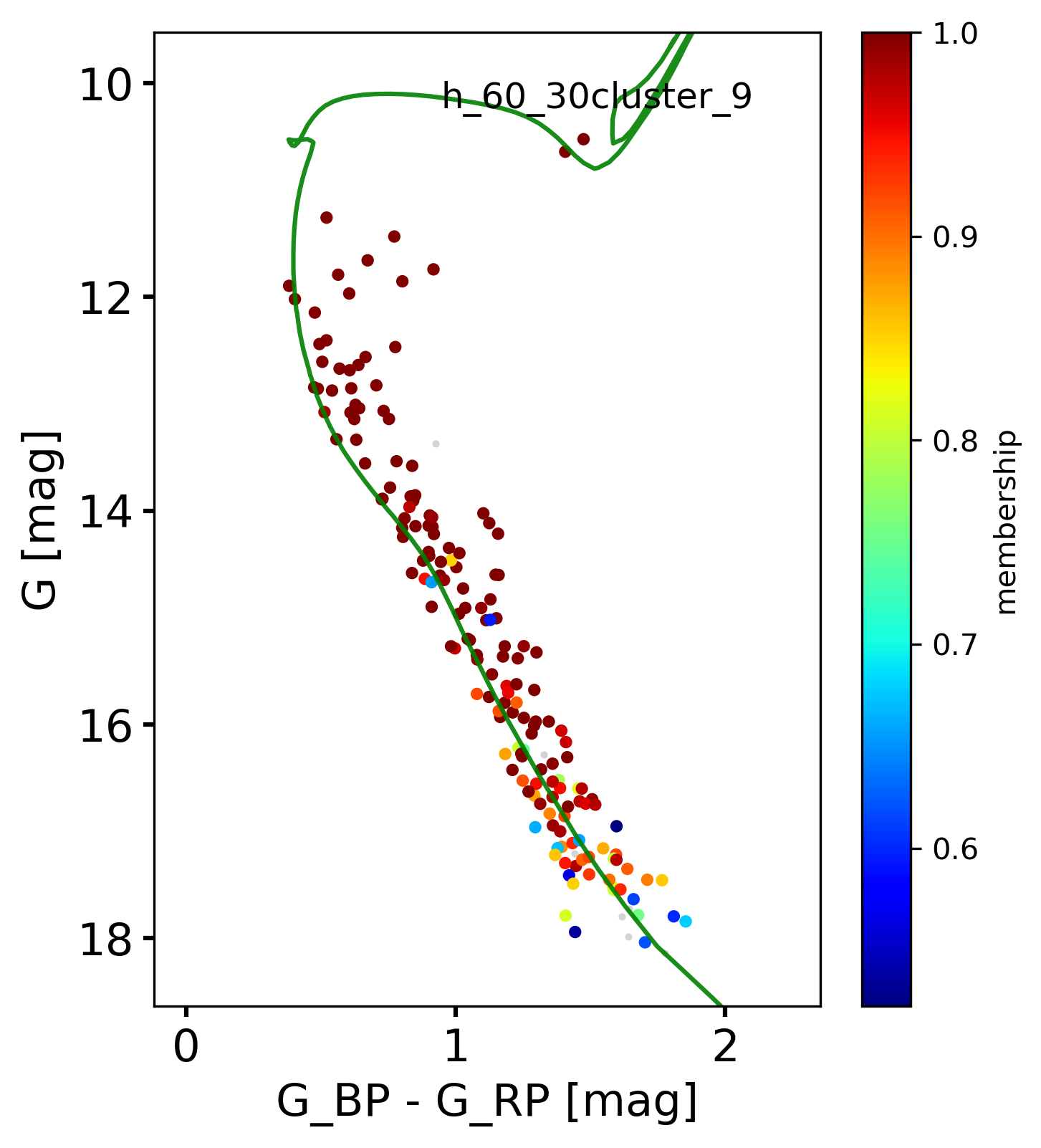}
\includegraphics[trim={0 1.4cm 0 0},clip, width=0.6\columnwidth, angle=0]{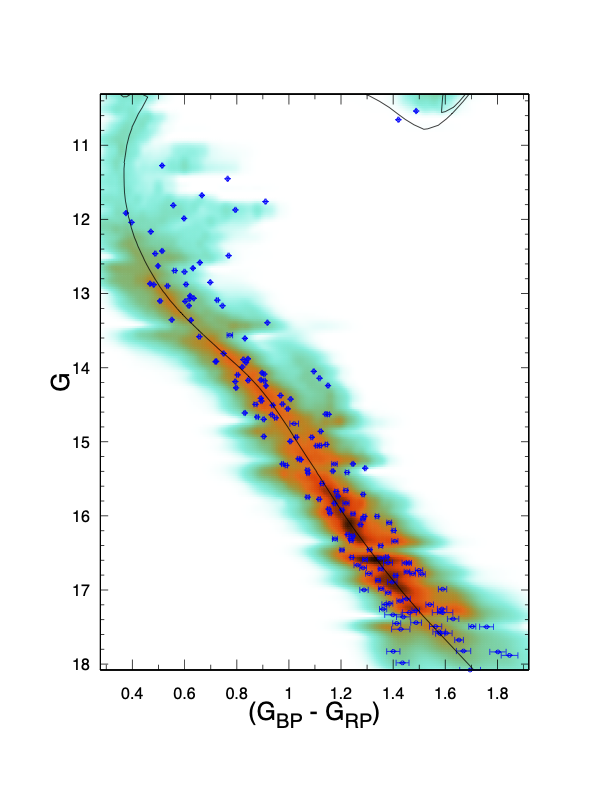}
\end{center}

\caption{The same of Fig. \ref{figA00} for CMa09.}

\label{figA09}

\end{figure*}



\begin{figure*}
\begin{center}

\includegraphics[width=1.8\columnwidth, angle=0]{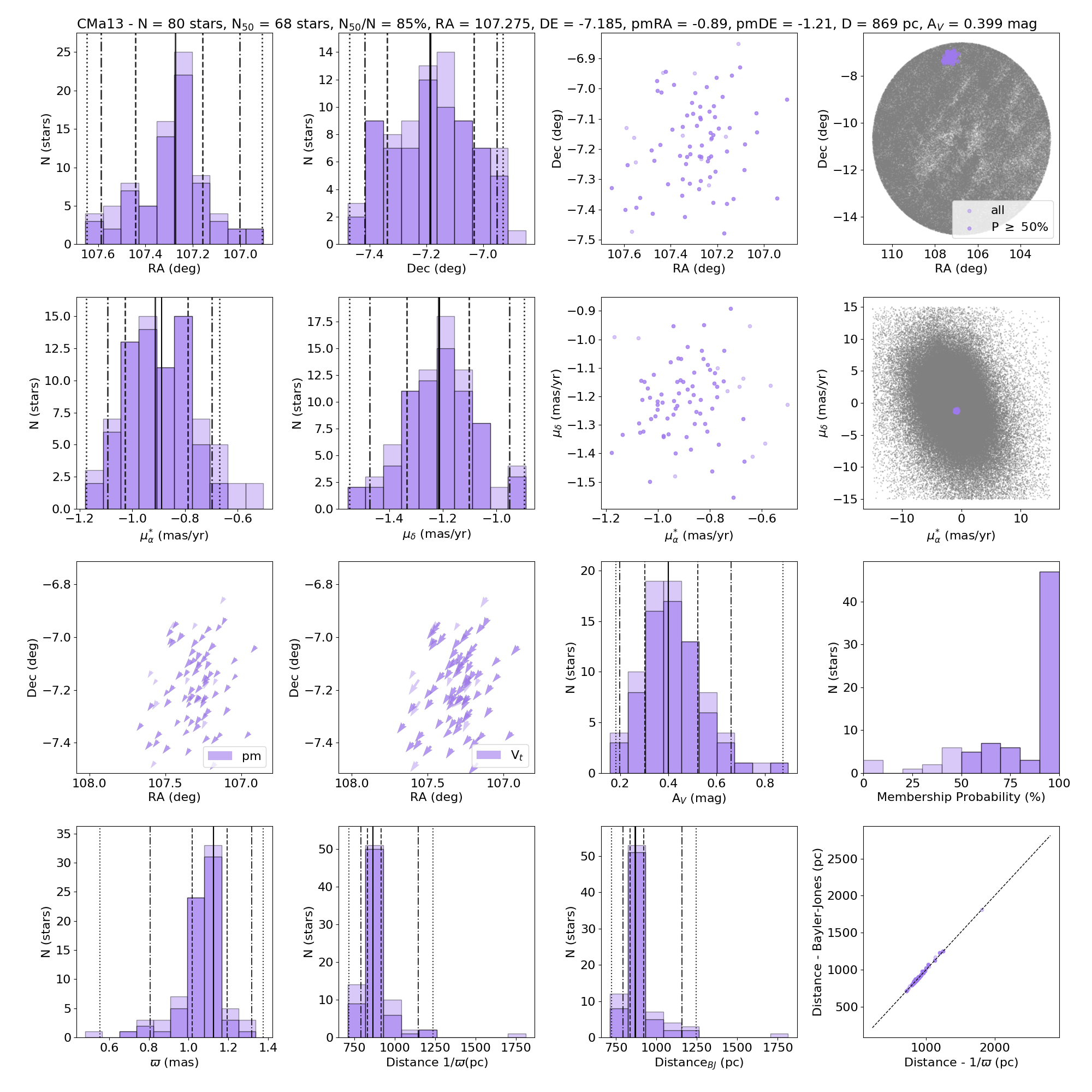}
\includegraphics[width=0.6\columnwidth, angle=0]{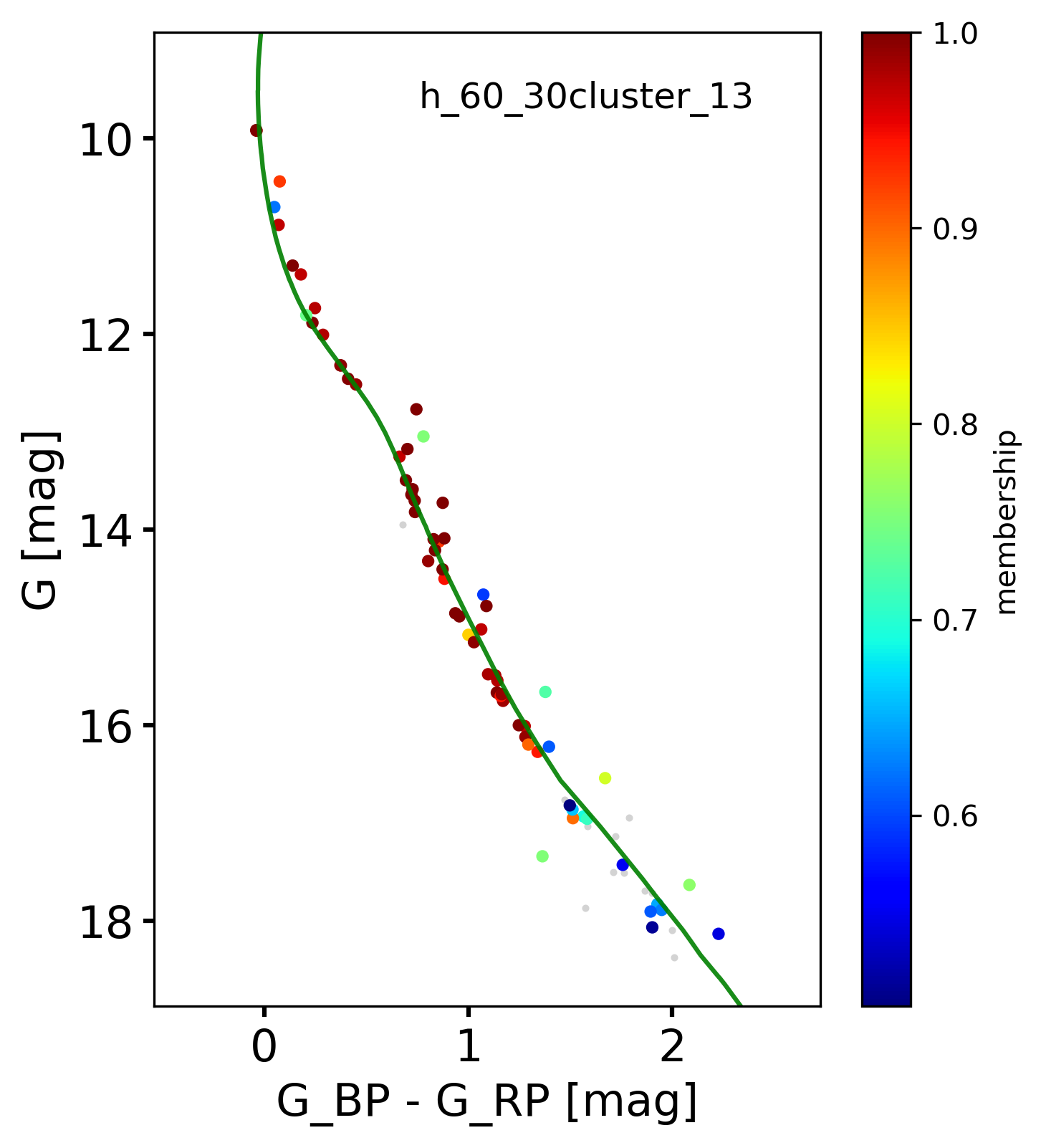}
\includegraphics[trim={0 1.4cm 0 0},clip, width=0.6\columnwidth, angle=0]{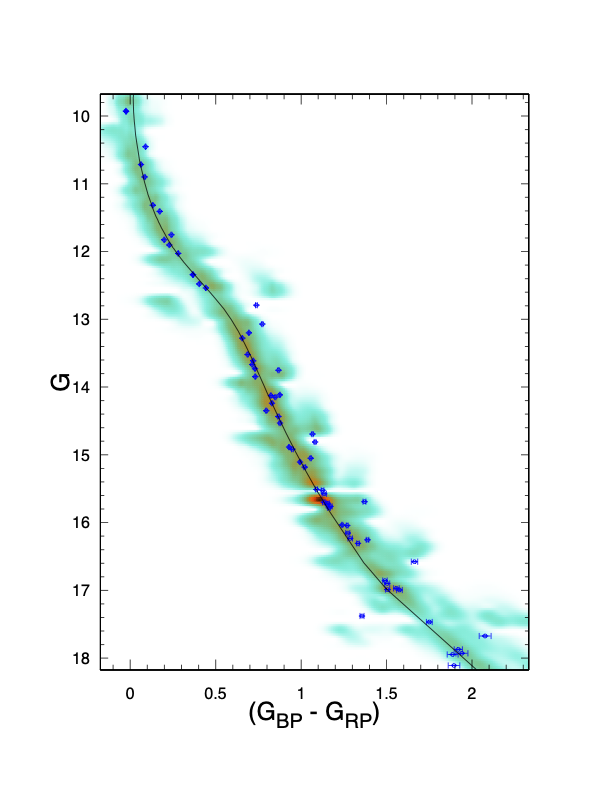}

\end{center}

\caption{The same of Fig. \ref{figA00} for CMa13.}

\label{figA13}

\end{figure*}



\begin{figure*}
\begin{center}

\includegraphics[width=1.8\columnwidth, angle=0]{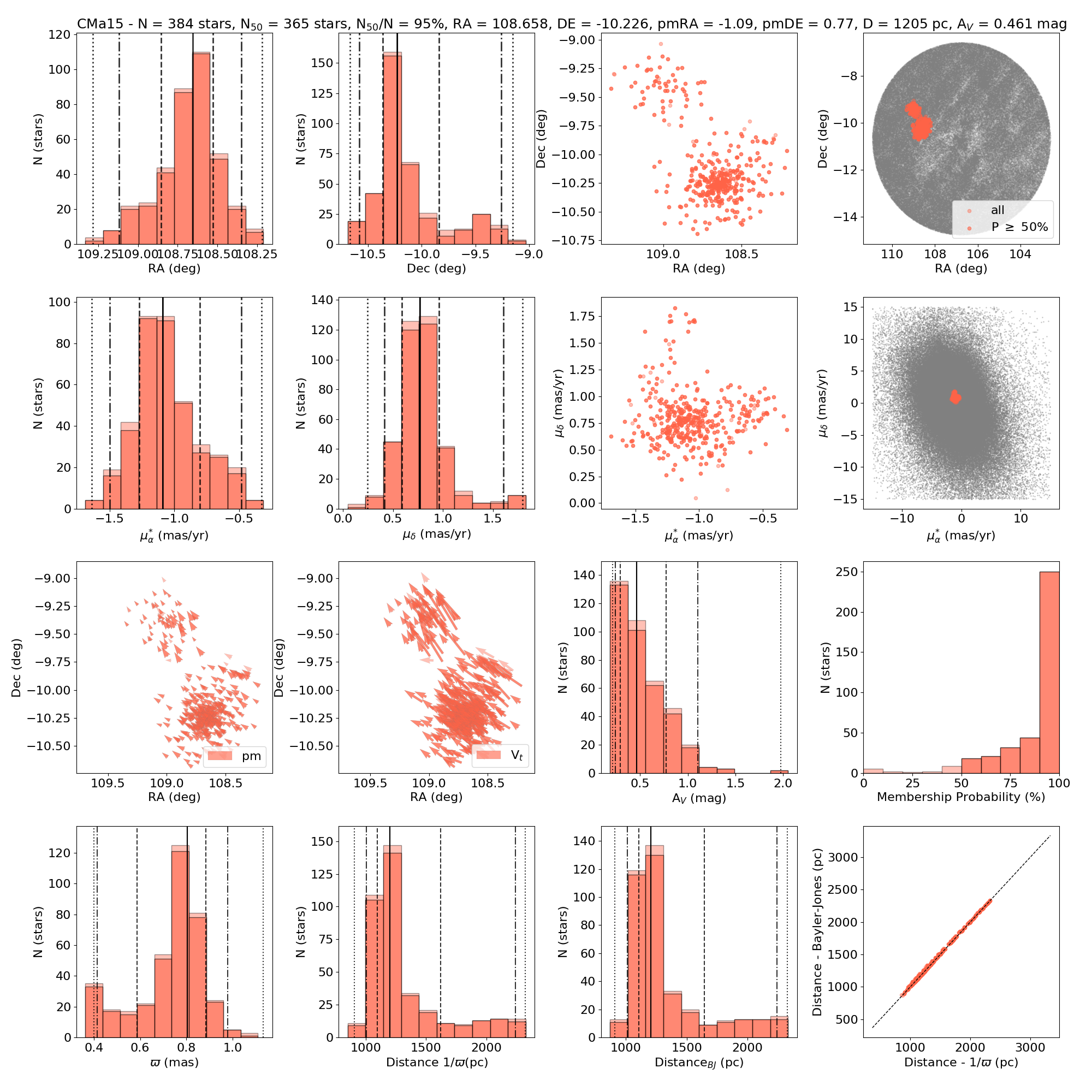}
\includegraphics[width=0.6\columnwidth, angle=0]{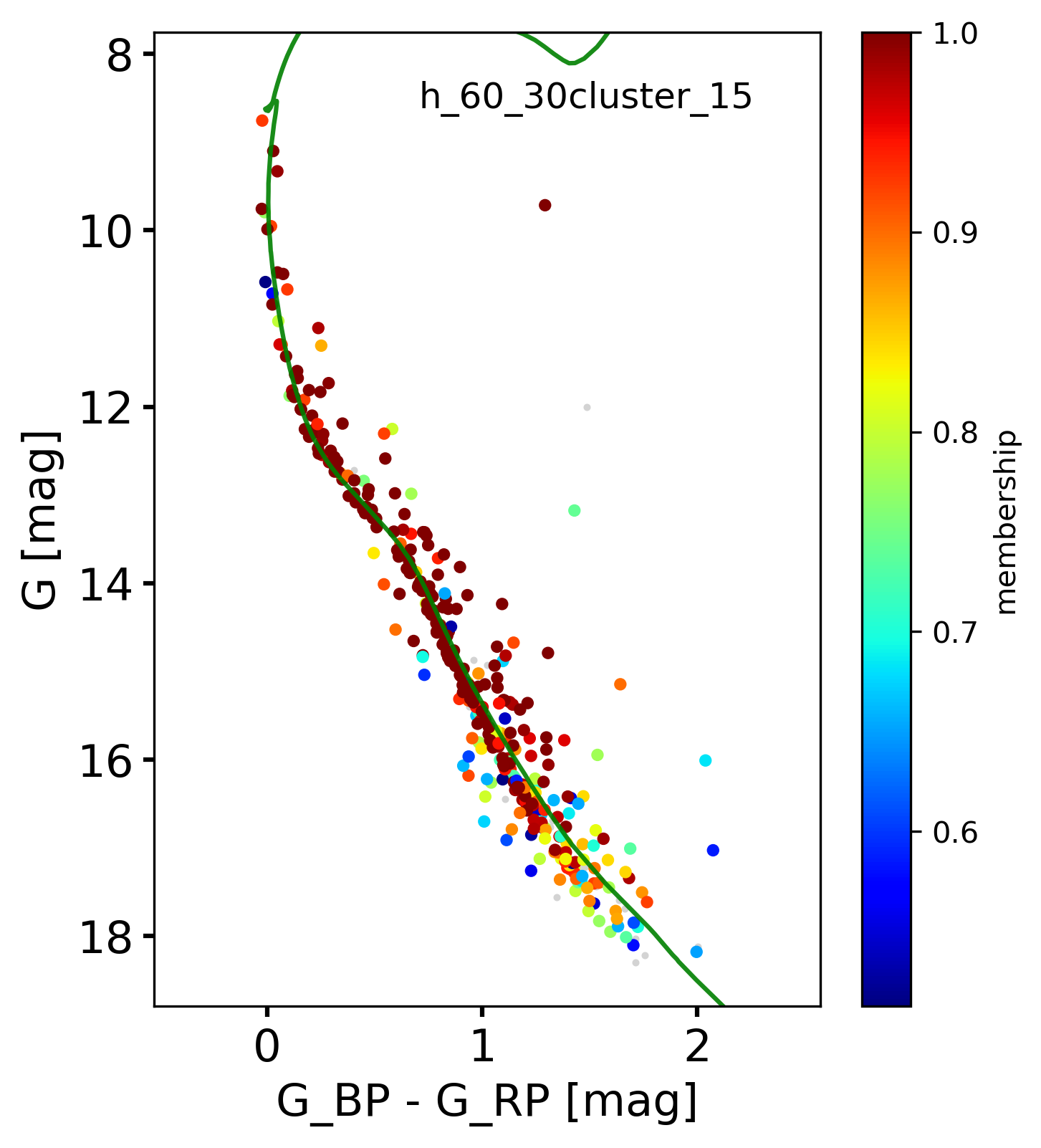}
\includegraphics[trim={0 1.4cm 0 0},clip, width=0.6\columnwidth, angle=0]{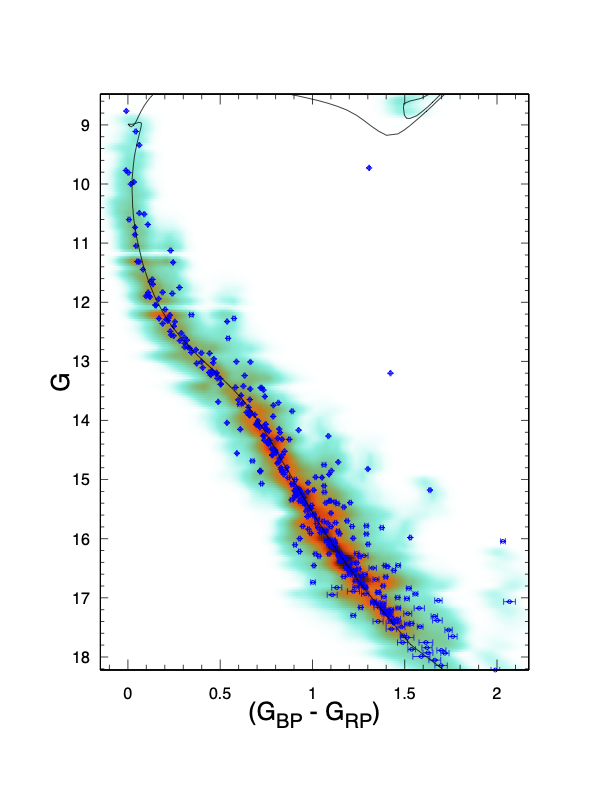}

\end{center}

\caption{The same of Fig. \ref{figA00} for CMa15.}

\label{figA15}

\end{figure*}



\begin{figure*}
\begin{center}

\includegraphics[width=1.8\columnwidth, angle=0]{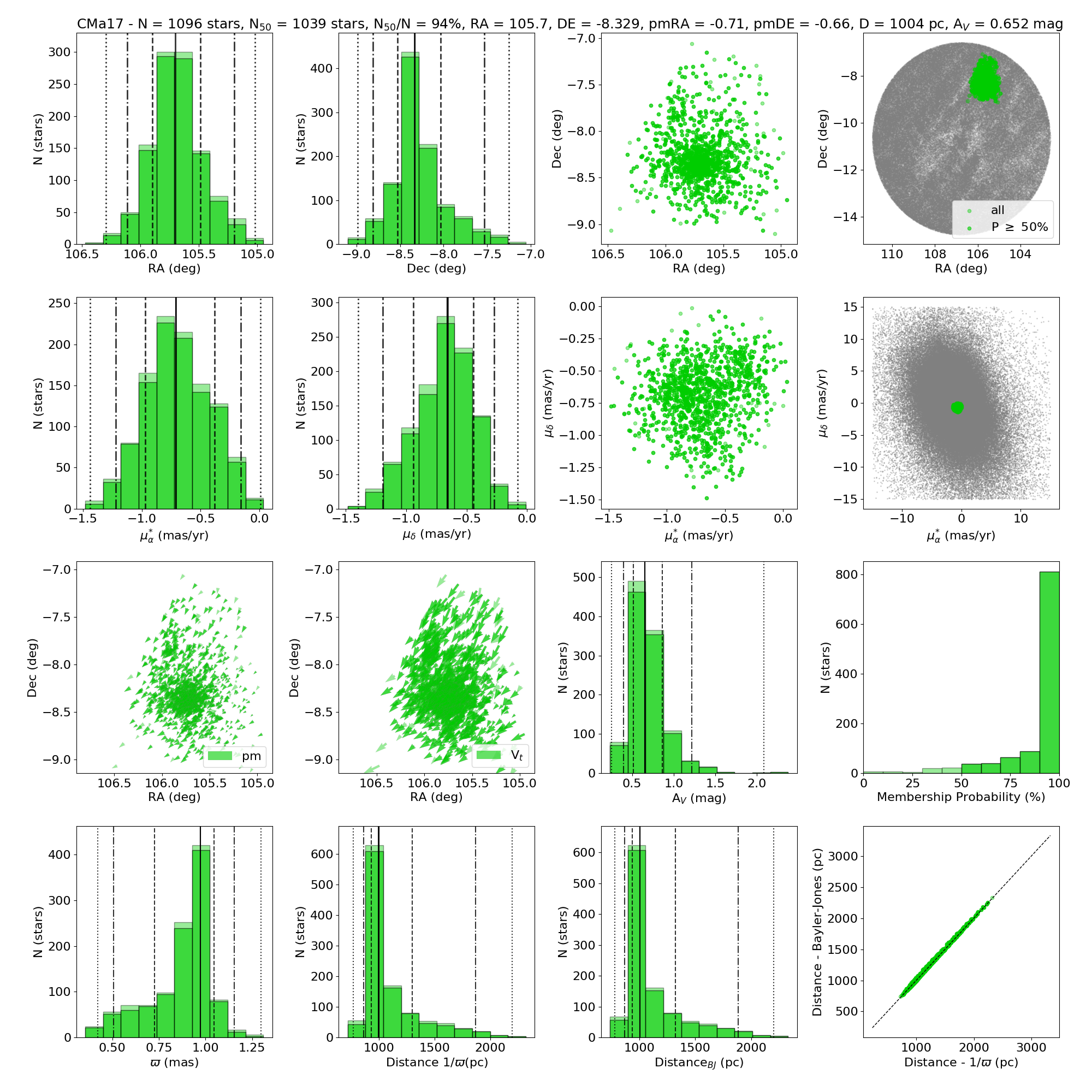}
\includegraphics[width=0.6\columnwidth, angle=0]{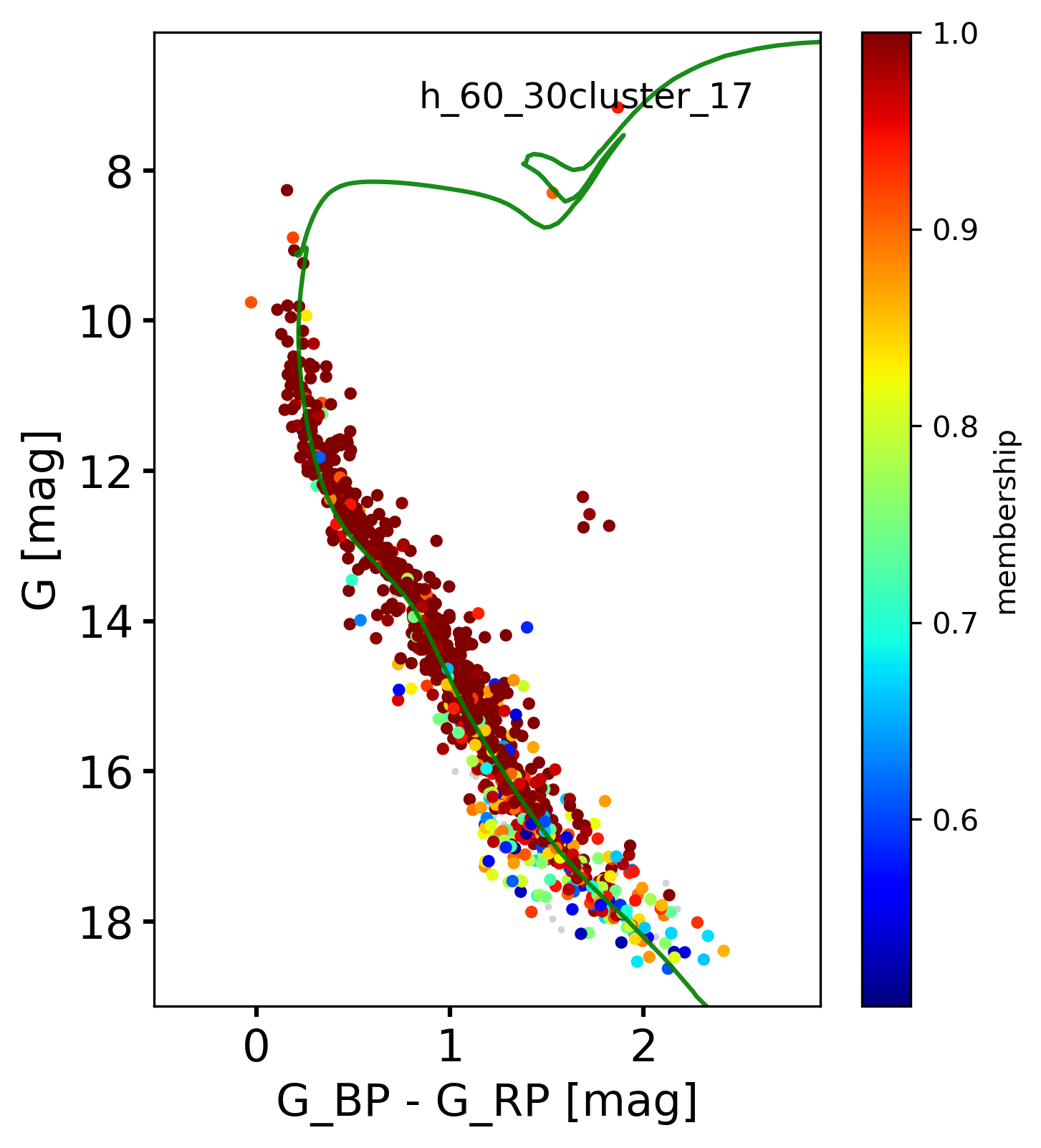}
\includegraphics[trim={0 1.4cm 0 0},clip, width=0.6\columnwidth, angle=0]{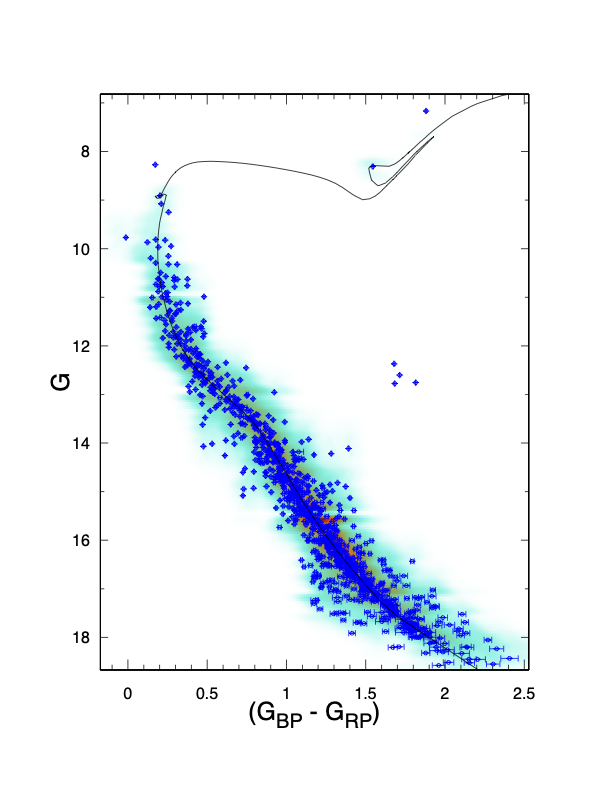}

\end{center}

\caption{The same of Fig. \ref{figA00} for CMa17.}

\label{figA17}

\end{figure*}



\begin{figure*}
\begin{center}

\includegraphics[width=1.8\columnwidth, angle=0]{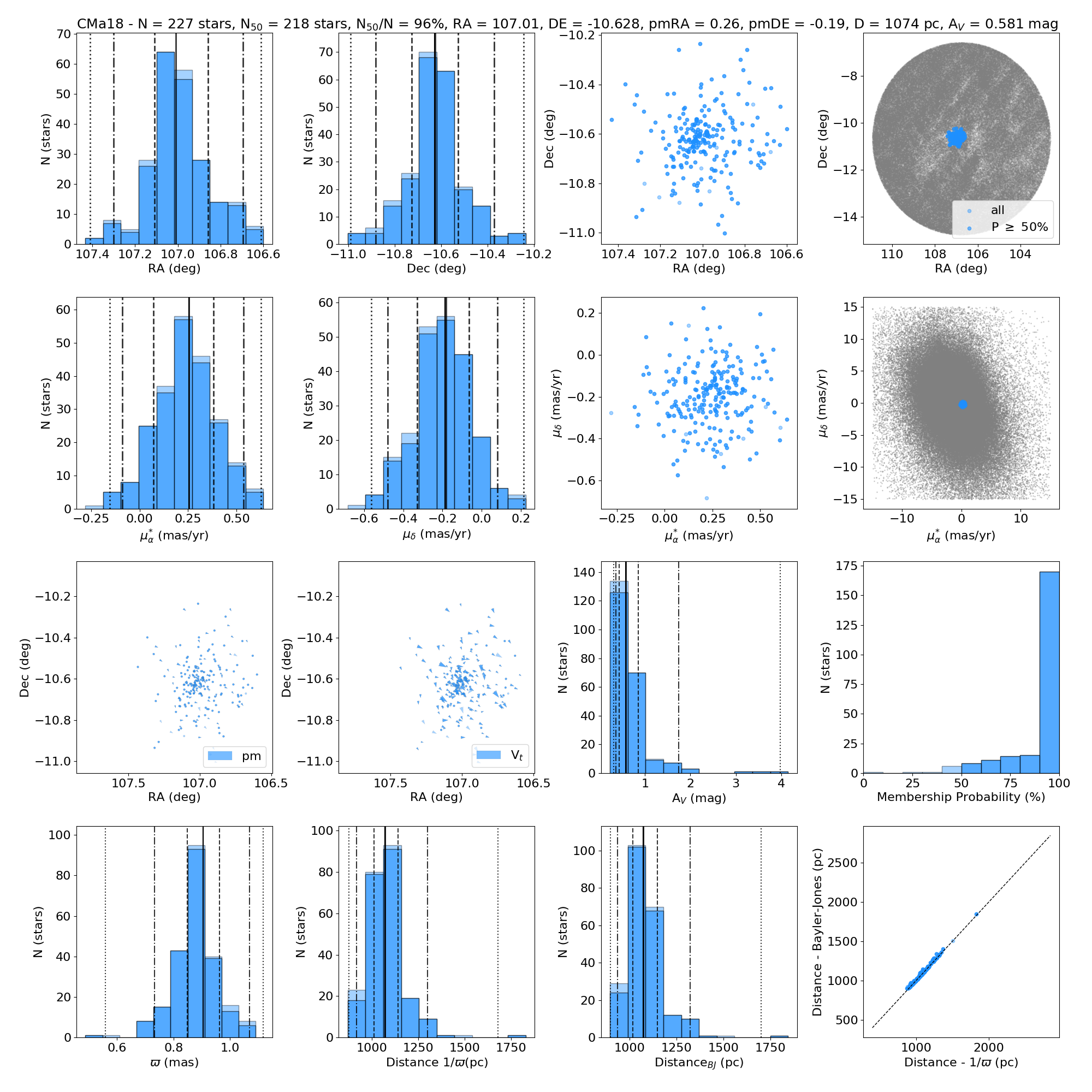}
\includegraphics[width=0.6\columnwidth, angle=0]{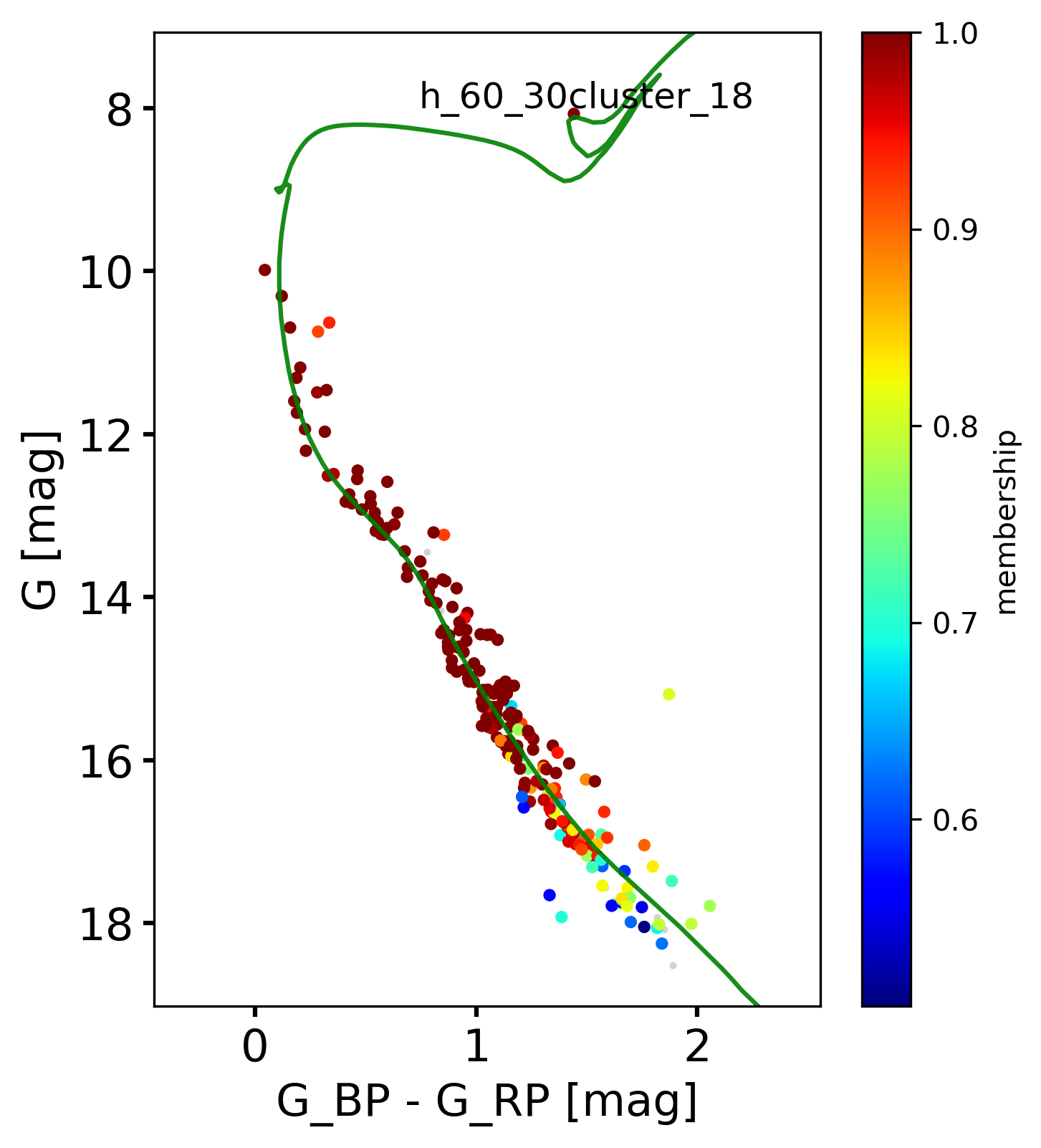}
\includegraphics[trim={0 1.4cm 0 0},clip, width=0.6\columnwidth, angle=0]{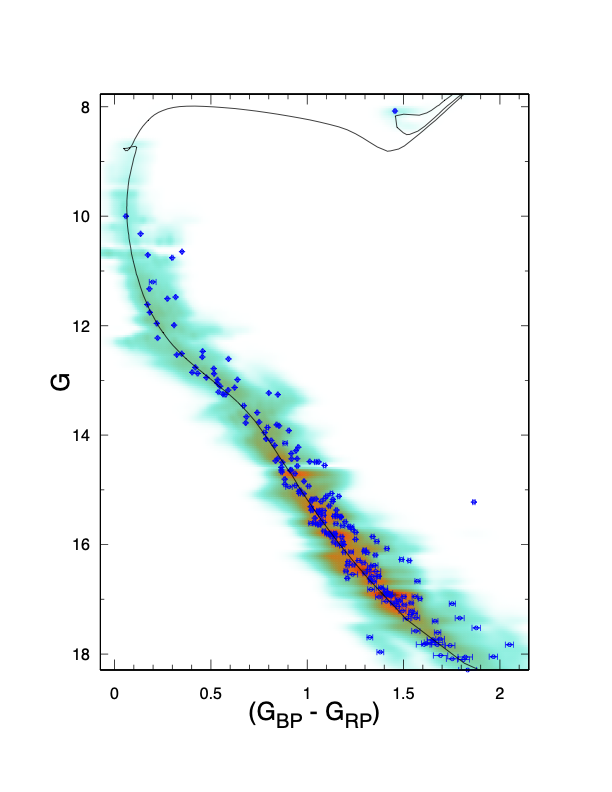}
\end{center}

\caption{The same of Fig. \ref{figA00} for CMa18.}

\label{figA18}

\end{figure*}



\begin{figure*}
\begin{center}

\includegraphics[width=1.8\columnwidth, angle=0]{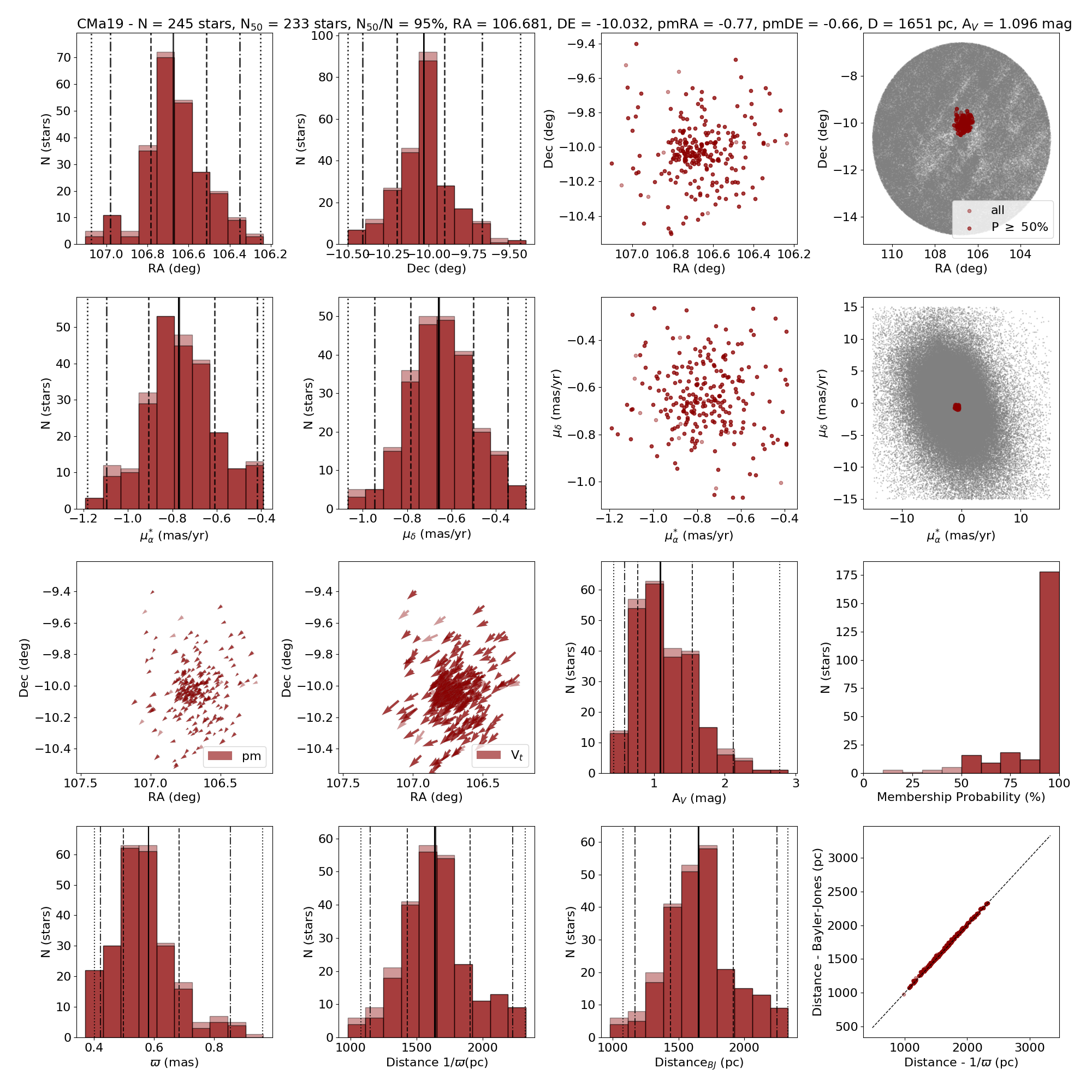}
\includegraphics[width=0.6\columnwidth, angle=0]{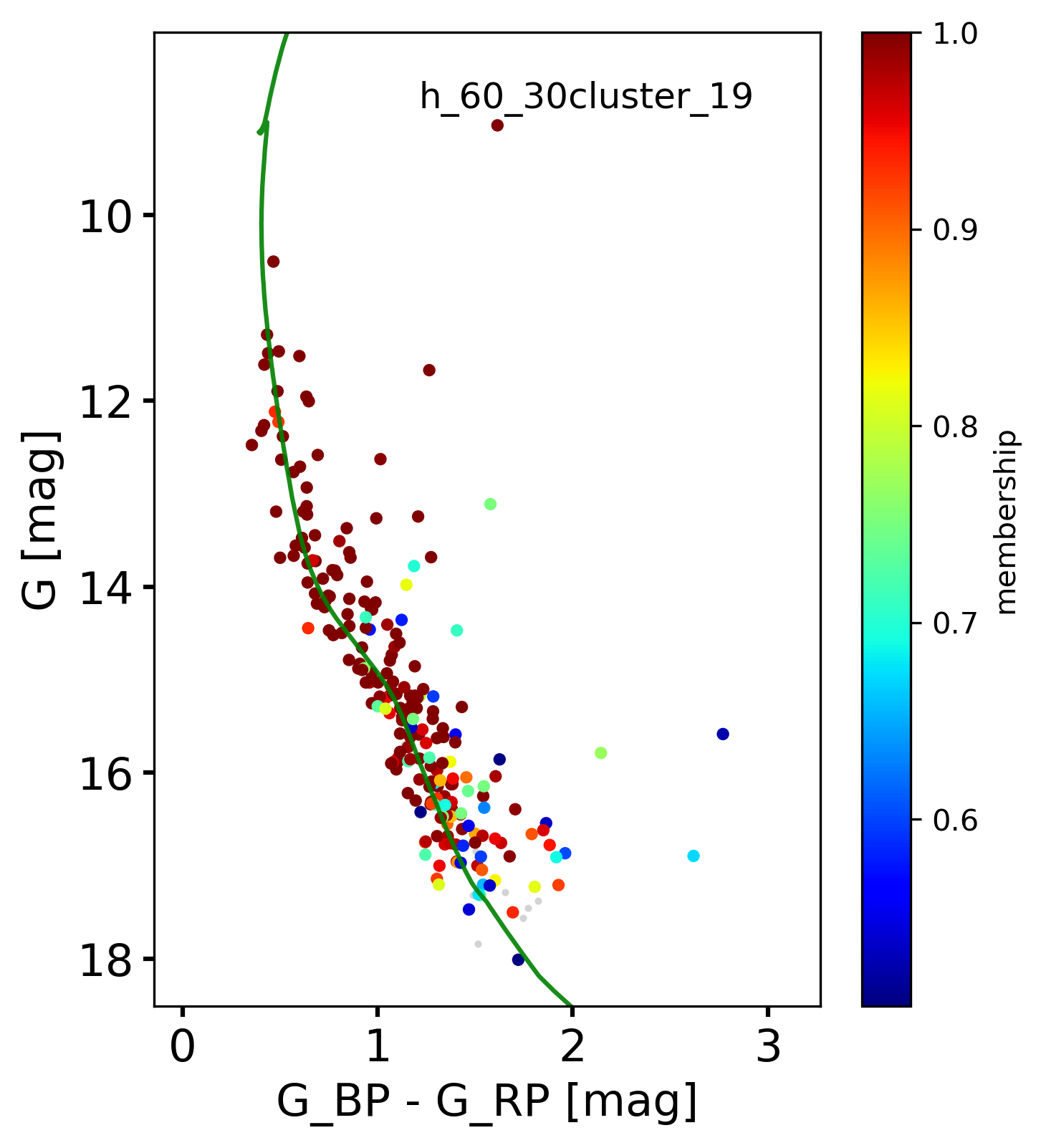}
\includegraphics[trim={0 1.4cm 0 0},clip, width=0.6\columnwidth, angle=0]{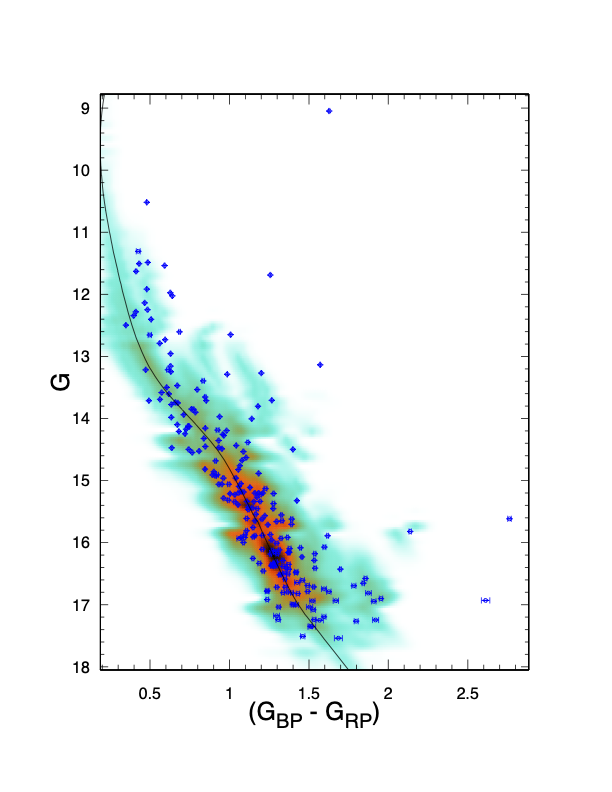}

\end{center}

\caption{The same of Fig. \ref{figA00} for CMa19.}

\label{figA19}

\end{figure*}



\begin{figure*}
\begin{center}

\includegraphics[width=1.8\columnwidth, angle=0]{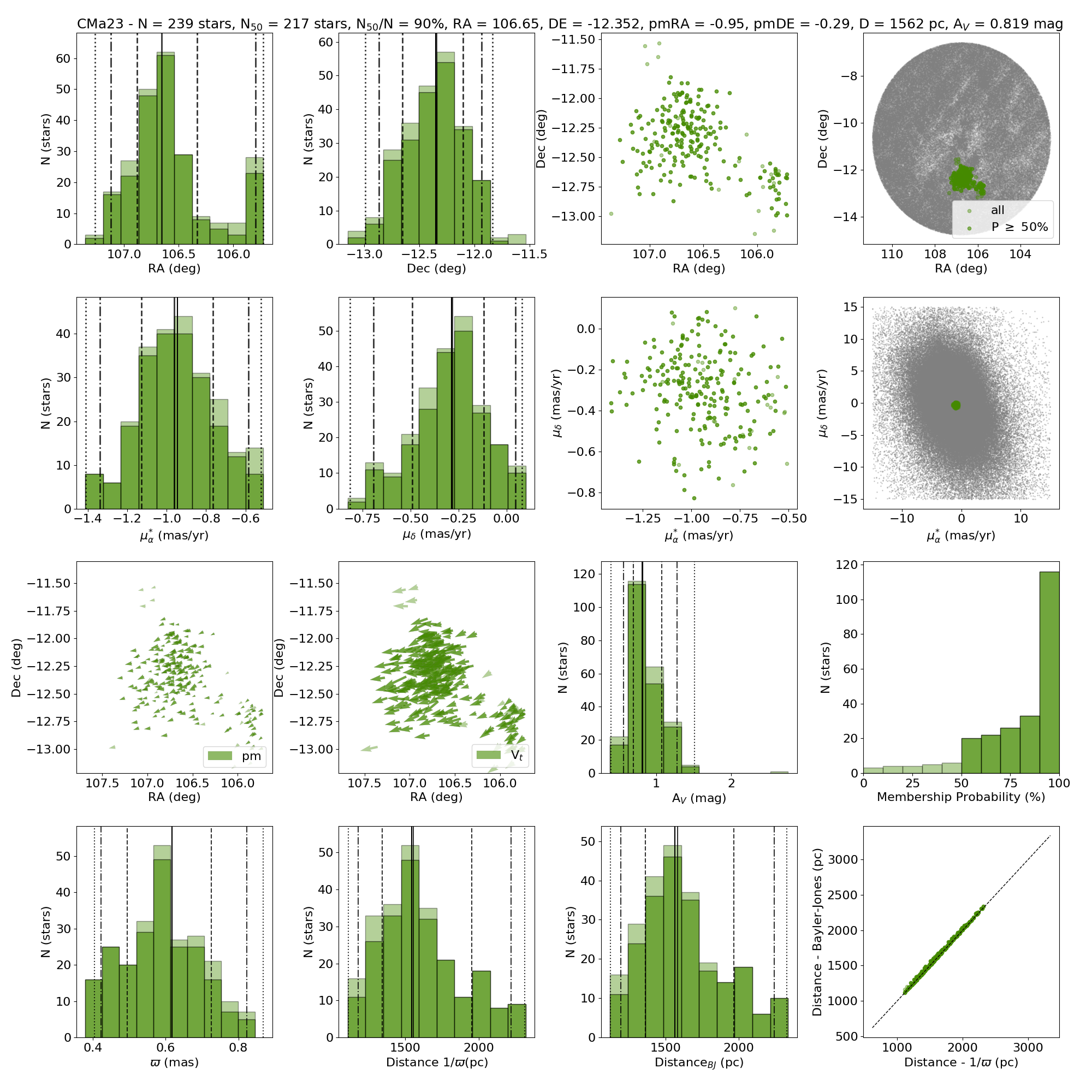}
\includegraphics[width=0.6\columnwidth, angle=0]{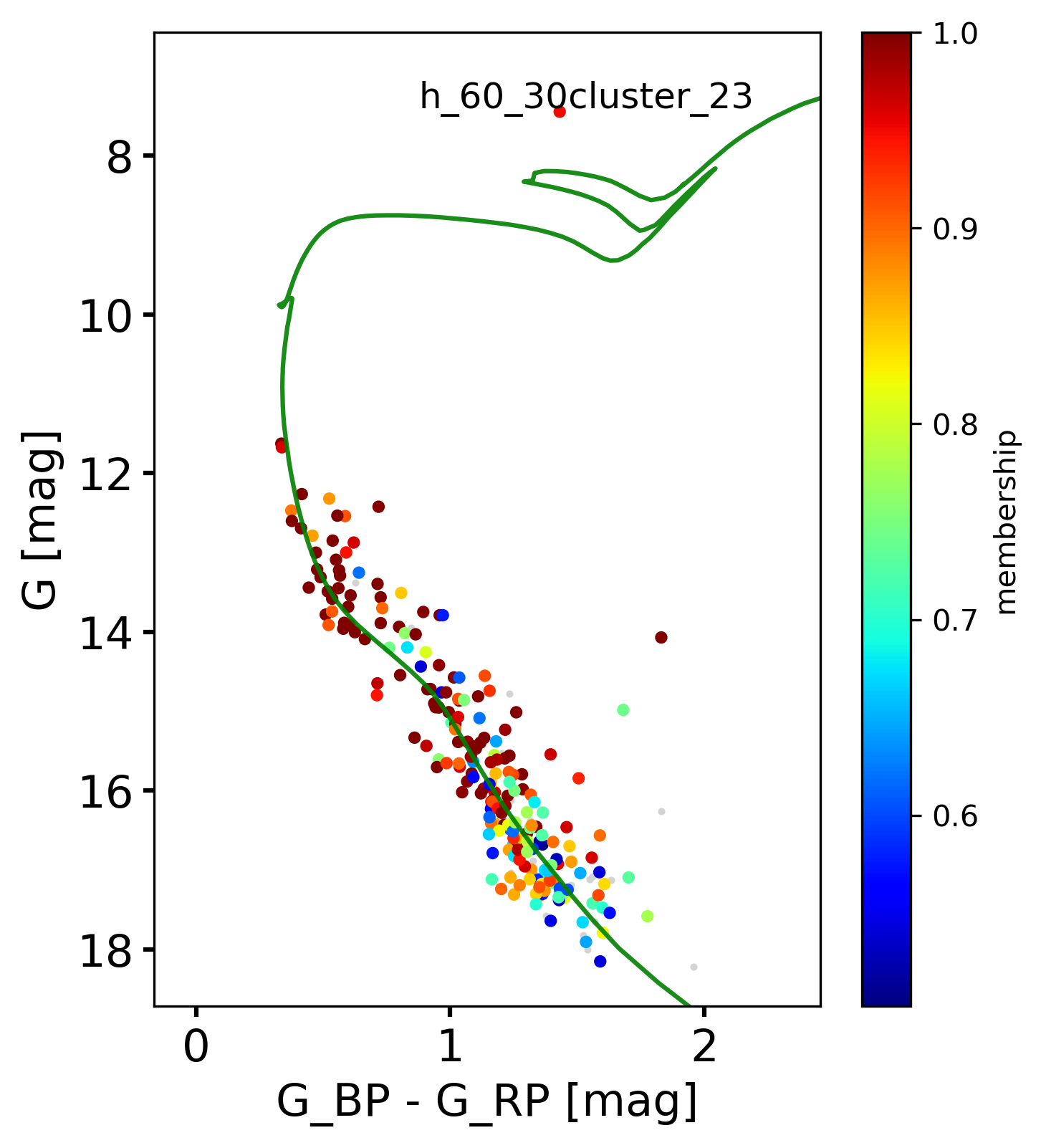}
\includegraphics[trim={0 1.4cm 0 0},clip, width=0.6\columnwidth, angle=0]{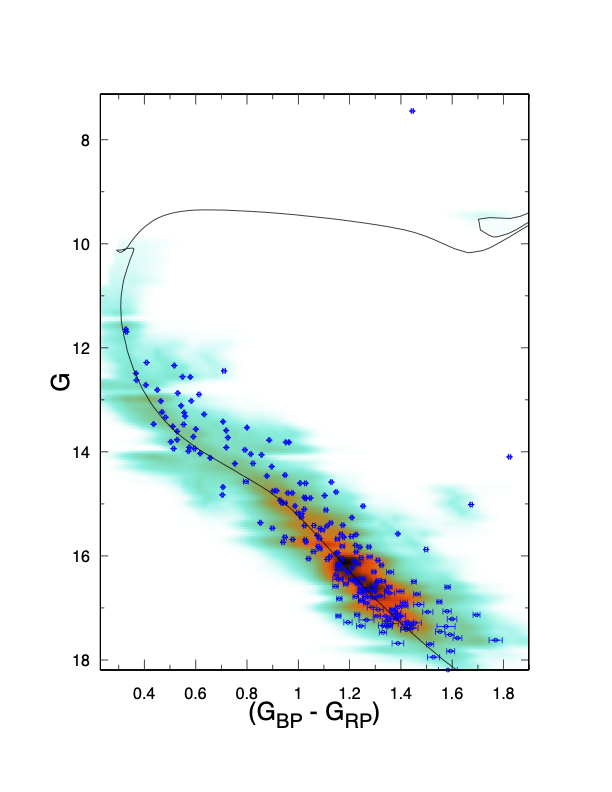}

\end{center}

\caption{The same of Fig. \ref{figA00} for CMa23.}

\label{figA23}

\end{figure*}


\bsp	
\label{lastpage}
\end{document}